
\input jnl

\def\comp{{\rm C}\llap{\vrule height7.1pt width1pt depth-.4pt\phantom t}}

\def\Fint{\rlap{$\Biggl\rfloor$}\Biggl\lceil}
\def\preprintno#1{
 \rightline{\rm #1}}    

\def\title                      
  {\null\vskip 3pt plus 0.2fill
   \beginlinemode \doublespace \raggedcenter \bf}

\def\author                     
  {\vskip 3pt plus 0.2fill \beginlinemode
   \singlespace \raggedcenter}

\def\affil                      
  {\vskip 3pt plus 0.1fill \beginlinemode
   \oneandahalfspace \raggedcenter \sl}

\def\abstract                   
  {\vskip 3pt plus 0.3fill \beginparmode
   \doublespace \narrower ABSTRACT: }

\def\endtitlepage               
  {\endpage                     
   \body}

\def\body                       
  {\beginparmode}               

\def\head#1{                    
  \filbreak\vskip 0.5truein     
  {\immediate\write16{#1}
   \raggedcenter \uppercase{#1}\par}
   \nobreak\vskip 0.25truein\nobreak}

\def\refto#1{$|{#1}$}           

\def\references                 
  {\head{References}            
   \beginparmode
   \frenchspacing \parindent=0pt \leftskip=1truecm
   \parskip=8pt plus 3pt \everypar{\hangindent=\parindent}}

\gdef\refis#1{\indent\hbox to 0pt{\hss#1.~}}    

\gdef\journal#1, #2, #3, 1#4#5#6{               
    {\sl #1~}{\bf #2}, #3, (1#4#5#6)}           

\def\refstylenp{                
  \gdef\refto##1{ [##1]}                                
  \gdef\refis##1{\indent\hbox to 0pt{\hss##1)~}}        
  \gdef\journal##1, ##2, ##3, ##4 {                     
     {\sl ##1~}{\bf ##2~}(##3) ##4 }}

\def\refstyleprnp{              
  \gdef\refto##1{ [##1]}                                
  \gdef\refis##1{\indent\hbox to 0pt{\hss##1)~}}        
  \gdef\journal##1, ##2, ##3, 1##4##5##6{               
    {\sl ##1~}{\bf ##2~}(1##4##5##6) ##3}}

\def\endreferences{\body}

\singlespace
\preprintno{UFIFT-93-20}
\preprintno{November, 1993}

\title Reduced Hamiltonians
\author J. A. Rubio$^*$
\author and
\author R. P. Woodard$^{\dagger}$
\affil
Department of Physics
University of Florida
Gainesville, FL 32611

\abstract \singlespace
We resurrect a standard construction of analytical mechanics dating
from the last century. The technique allows one to pass from any dynamical
system whose first order evolution equations are known, and whose bracket
algebra is not degenerate, to a system of canonical variables and a non-zero
Hamiltonian that generates their evolution. We advocate using this method to
infer a canonical formalism, as a prelude to quantization, for systems in which
the naive Hamiltonian is constrained to vanish. The construction agrees with
the
usual results for gauge theories and can be applied as well to gravity, {\it
even when the spatial manifold is closed.} As an example, we construct such a
reduced Hamiltonian in perturbation theory around a flat background on the
manifold $T^3 \times R$. The resulting Hamiltonian is positive semidefinite and
agrees with the A.D.M. energy in the limit that deviations from flat space
remain
localized as the toroidal radii become infinite. We also obtain closed form
expressions for the reduced Hamiltonians of two minisuperspace truncations.
Although our results are classical they can be formally quantized to give the
naive functional formalism. This is not only an effective starting point for
calculations, it also seems to provide a formulation of the quantum theory
which
is non-perturbative, at least in principle. The marriage we advocate between
the
old technique and canonical quantization seems to have profound implications
for
quantum gravity, especially as regards the conservation of energy, statistical
mechanics, and the problem of time.
\vskip .5cm
\noindent Int. Class. for Physics: 0460.
\vskip 2cm
\noindent $^*$ E-mail: RUBIO@UFHEPA.PHYS.UFL.EDU.

\noindent $^{\dagger}$ E-mail: WOODARD@UFHEPA.PHYS.UFL.EDU.

\endtitlepage
\doublespace

\centerline{\bf 1. Introduction}

It is traditional to develop the canonical formalism from a Lagrangian. One
first identifies variables whose Poisson bracket algebra is canonical, then the
Hamiltonian is constructed and used to generate first order evolution
equations.
However, the imposition of constraints can result in a system of dynamical
variables whose bracket algebra is not canonical and whose time evolution,
while
completely determined, is not generated by a known Hamiltonian. This
possibility
is especially relevant to systems in which gravity is dynamical and the spatial
manifold is closed because then the naive Hamiltonian is constrained to vanish
[1]. The ubiquity of canonical quantization and its evident failure in this
case
has led to much puzzlement over how these perfectly acceptable classical
systems
should be quantized [2]. We wish here to advocate a straightforward solution:
impose the constraints to obtain directly a set of first order evolution
equations and a bracket algebra for the reduced variables, then construct a
canonical formalism which reproduces this system. The resulting reduced
canonical formalism can then be quantized as usual.

Our motive in suggesting this step is the highly unsatisfactory situation which
currently prevails in the formulation of canonical quantum gravity. Those who
approach this field from the perspective of other disciplines have long been
frustrated by its failure to obey certain obvious correspondence limits. There
are at least four problems of this nature:
\item{(1)} The Paradox of Second Coordinatization --- In classical gravitation
we know that fixing the lapse and shift determines the coordinate system up to
deformations of the initial value surface. We also think we understand pretty
well how to infer physics from the behavior of the metric field in these
coordinates. In {\it quantum} gravity on a closed spatial manifold we are told
that even after fixing the lapse and the shift we must determine what time is
all over again. We are also told that physical information can only be gleaned
from observables which are manifestly coordinate invariant, that we don't
possess any such observables, and that even if we did it would not be possible
to compute their expectation values because an inner product cannot be given
until after the meaning of time has been clarified. How can changing $\hbar$
from zero so thoroughly confuse the way we infer physics from the field
variables?
\item{(2)} The Paradox of Dynamics --- A similarly striking issue concerns the
limit in which gravity becomes non-dynamical. Since the Hamiltonians of pure
matter theories are not typically zero, even on closed spatial manifolds, these
theories can be quantized canonically. But when gravity is made dynamical on a
spatially closed manifold it is asserted that we no longer know even what the
inner product is, no matter how weak the gravitational interaction might be
relative to other forces. Note that the alleged difficulty is unrelated to the
non-renormalizability of Einstein's theory; it would occur even in an
ultraviolet
finite theory of quantum gravity. How can changing Newton's constant from zero
affect the basic structure of quantum mechanics?
\item{(3)} The Paradox of Topology --- Although topology is not a continuous
parameter the comparison between manifolds with differing topologies is
disquieting. For spatially open manifolds there is no obstacle to canonical
quantization because the naive Hamiltonian can be non-zero [3,4]. Of course
there
remains the highly non-trivial problem of finding a theory whose dynamics are
consistent, but there is no confusion about basic issues such as the meaning of
time or how to define inner products. When the spatial manifold is closed it is
asserted that we no longer understand these issues. Yet there is no local
experiment which can distinguish a closed space from an open one, as witness
the
fact that either possibility might describe our own universe. How can
removing a
point from the other end of the universe affect our ability to provide a
quantum
mechanical description of local observations which have a vanishingly small
probability for even being in causal contact with the boundary?
\item{(4)} The Paradox of Stability --- Finally, there is the peculiarity
encountered in passing from pure quantum mechanics to statistical mechanics
on a
spatially closed manifold when gravity is dynamical. Since the energy is
constrained to be zero it follows that all states are degenerate. For example,
the total energy needed to add a particle-anti particle pair to any state is
zero because the negative gravitational interaction energy cancels the positive
energy of the pair. Simple considerations of entropy seem then to require that
the microcanonical ensemble should be concentrated around what would otherwise
be thought of as very highly excited states. To see this note that for every
``empty'' state there are a countably infinite number of degenerate states
which
contain a single particle-anti particle pair moving apart with various momenta.
There are even {\it more} states which contain two pairs, and more yet which
contain three, etc. So what prevents such a universe from evaporating into a
maelstrom of pairs?

We resolve the paradox of second coordinatization by denying that there is any
fundamental distinction between classical and quantum measurement beyond that
usually imposed by the uncertainty principle. What time means in quantum
gravity
is determined by fixing the lapse and the shift, just as in the classical
theory. Any question that can be answered classically by studying a functional
of the metric with fixed initial value data can be studied as well in the
quantum theory using the expectation value of this same functional in the
presence of a state whose probability is concentrated around the classical
initial value data. We can base observations on a non-invariant such as the
metric because the gauge has been fixed: {\it any} quantity becomes invariant
when it is defined in a unique coordinate system.

We resolve the paradoxes of dynamics and of topology by showing that for any
constrained system which can be reduced --- that is, for which the residual
gauge freedom can be fixed and the constraints imposed --- there is a canonical
choice of reduced dynamical variables and a non-zero Hamiltonian which
generates
their time evolution. In fact there are many such Hamiltonians, each
corresponding to a different identification of canonical variables in the
reduced dynamical system. For gravity these Hamiltonians seem to have the
property that the closed space versions go over to the known open space ones
when the coordinate volume is taken to infinity in such a way that the initial
value data are only locally disturbed from a background which obeys the
appropriate asymptotic conditions.\footnote{*}{For [3] this background would be
flat space, for [4] it would be the de Sitter geometry. Note that topological
obstructions --- for example, the inability to impose a flat metric on
$S^3$ ---
need have no physical relevance because we might be able to use half of the
infinite coordinate volume to approach the asymptotic geometry of the open
manifold and the other half to deviate so as to reconcile the topology of the
closed manifold.} When gravity is coupled to matter the non-zero reduced
Hamiltonians seem to agree with the corresponding pure matter Hamiltonians
in the limit that Newton's constant vanishes.

Although the Hamiltonians we are discussing generate time evolution they are
not
generally conserved. Outside of a few special cases the generally conserved
energy really {\it is} zero, so the universe can potentially suffer from the
instability described in (4). We see a mostly empty universe today because the
balance implicit in the $H = 0$ constraint is maintained by means of global,
negative energy modes, and the rate at which these modes can be excited is
highly suppressed by causality and by the weakness of the gravitational
interaction. There has simply not been enough time for the universe to
evaporate
into pairs. Depending on the topology and the causal structure of the initial
state there may never be enough time for this to happen.

Section 2 presents the construction by means of which any known system of first
order evolution equations and (not necessarily canonical) brackets can be used
to identify a set of canonical variables and a Hamiltonian that generates their
time evolution. This is not original work; the construction was given about a
century ago [5,6]. What does seem to be original is the idea of applying it
generally to gravity. In taking this step we are following the work in $2+1$
dimensions of Moncrief [7], Hosoya and Nakao [8], and Carlip [9]. We believe
that their constructions are special cases of the general technique; we believe
the same is true of the lovely formalism for open, asymptotically flat spaces
which was devised by Arnowitt, Deser and Misner (A.D.M.) [3, and references
therein].

In section 3 we apply the method to a standard gauge in scalar electrodynamics.
This is difficult to do in gravity because one can seldom obtain explicit
solutions for the constraints. However, the method is simple enough to carry
out
in perturbation theory or when the dynamics is suitably truncated. In section 4
we give a perturbative construction for a flat space background on the manifold
$T^3 \times R$; in section 5 we apply the method to several models of
minisuperspace. We show in section 6 that if the unreduced theory exists then
quantization along the lines we advocate results in the usual functional
formalism. In this form the technique is not only simple to exploit for the
purposes of perturbative calculations, it also provides a definition of the
theory that is valid beyond perturbation theory, at least in principle. Section
7 summarizes our resolutions to the aforementioned paradoxes and discusses the
implications our view has for statistical mechanics and the conservation of
energy.

\centerline{\bf 2. The Construction}

Consider a set of $2N$ dynamical variables, $\{v^i(t): R \longrightarrow R^{
2N}\}$.\footnote{*}{To emphasize that the dynamical variables are functions of
both the canonical positions and their conjugate momenta we have adopted the
neutral symbol, ``$v^i$,'' for ``variable.''} Suppose that the time evolution
of
these variables is determined by first order equations of the form:
$${\dot v}^i(t) = f^i(v,t) \eqno(2.1)$$
where the $f^i$'s are known functions of the $v^i$'s and possibly also of time.
We do not assume that the $v^i$'s are necessarily canonical but rather that
they
obey the following bracket algebra:
$$\Bigl\{v^i(t),v^j(t)\Bigr\} = J^{ij}(v,t) \eqno(2.2)$$
where the bracket matrix $J^{ij}$ is antisymmetric, invertible and obeys the
Jacobi identity. We shall denote the inverse bracket matrix by the symbol,
``$J_{ij}$'':
$$J_{ij} \thinspace J^{jk} = \delta_i^{~k} = J^{kj} \thinspace J_{ji}
\eqno(2.3)$$
Taking the time derivative of $(2.2)$ and substituting $(2.1)$ reveals the
following
relation:
$$J^{ik} \thinspace f^j_{~,k} - J^{jk} \thinspace f^i_{~,k} = J^{ij}_{~~,k}
\thinspace f^k + {\partial J^{ij} \over \partial t} \eqno(2.4a)$$
where a comma denotes differentiation. The analogous relation for the inverse
is:
$$J_{ik} \thinspace f^k_{~,j} - J_{jk} \thinspace f^k_{~,i} = - J_{ij,k}
\thinspace f^k - {\partial J_{ij} \over \partial t} \eqno(2.4b)$$

Since brackets obey the Jacobi identity we have the following differential
relation for the elements of the bracket matrix:
$$J^{i \ell} \thinspace J^{jk}_{~~,\ell} + J^{j\ell} \thinspace
J^{ki}_{~~,\ell}
+ J^{k\ell} \thinspace J^{ij}_{~~,\ell} = 0 \eqno(2.5a)$$
The analogous result for the inverse is:
$$J_{ij,k} + J_{jk,i} + J_{ki,j} = 0 \eqno(2.5b)$$
This last relation is reminiscent of the Bianchi identities in electromagnetism
and allows us to write the inverse bracket matrix as the curl of a ``vector
potential," $\phi_i$:
$$J_{ij} = \phi_{j,i} - \phi_{i,j} \eqno(2.6)$$
Just as in electrodynamics $\phi_i$ is undetermined up to the gradient of a
scalar function; similarly, one may need several coordinate patches if the
dynamical variables map into some space other than $R^{2N}$. A convenient
representation for the vector potential is:
$$\phi_i(v,t) = - \int_0^1 d\tau \thinspace \tau \thinspace J_{ij}\Bigl(v_0 +
\tau {\Delta v},t\Bigr) \thinspace {\Delta v}^j \eqno(2.7)$$
where $v^i_0$ is any fixed point and ${\Delta v}^i \equiv v^i - v_0^i$. Note
that under a change in the dynamical variables $J^{ij}$ transforms
contravariantly while $J_{ij}$ transforms covariantly; $\phi_i$ transforms
covariantly up to the addition of a gradient.

The simplest possibility for reconstructing a Hamiltonian is that the evolution
equations are integrable. The condition for this is that the bracket matrix
should be free of explicit time dependence:
$$\exists H(v,t) \ni {\dot v}(t) = \{v^i,H\} \qquad {\rm iff} \qquad {\partial
J^{ij} \over \partial t} = 0 \eqno(2.8)$$
For the proof note that, if it exists, the Hamiltonian is determined up to a
function of time by the equations:
$${\partial H \over \partial v^i} = J_{ij} \thinspace f^j \eqno(2.9)$$
Since we must have $H_{,ij} = H_{,ji}$ the integrability condition is:
$$\eqalignno{0 &= \Bigl(J_{ik} \thinspace f^k\Bigr)_{,j} - \Bigl(J_{jk}
\thinspace f^k\Bigr)_{,i} &(2.10a) \cr
&= -\Bigl(J_{ki,j} + J_{jk,i}\Bigr) \thinspace f^k + \Bigl(J_{ik} \thinspace
f^k_{~,j} - f^k_{~,i} \thinspace J_{jk}\Bigr) &(2.10b) \cr
&= - {\partial J_{ij} \over \partial t} &(2.10c) \cr}$$
where we have used $(2.4b)$ and $(2.5b)$ to reach the final line. The theorem
is
proved upon noting that the bracket matrix is invertible. When the
integrability
condition is met it is simple to check that:
$$H(v,t) - H(v_0,t) = \int_0^1 d\tau \thinspace {\Delta v}^i \thinspace J_{ij}
\Bigl(v_0 + \tau {\Delta v}\Bigr) \thinspace f^j\Bigl(v_0 + \tau {\Delta v},t
\Bigr) \eqno(2.11)$$
gives an explicit representation for the Hamiltonian.

To identify canonical variables in the local neighborhood of any point $v_0^i$
we construct $2N$ functions $Q^a(v,t)$ and $P_b(v,t)$ ($a,b = 1, 2, \dots , N$)
which are instantaneously invertible at any time and which obey:
$$\phi_i(v,t) = P_a(v,t) \thinspace {\partial Q^a \over \partial v^i}(v,t)
\eqno(2.12)$$
Note that invertibility implies:
$$\pmatrix{\delta^a_{~b} & 0 \cr 0 & \delta_a^{~b} \cr} =
\pmatrix{\frac{\partial
Q^a}{\partial v^i} \frac{\partial v^i}{\partial Q^b} & \frac{\partial Q^a}{
\partial v^i} \frac{\partial v^i}{\partial P_b} \cr \frac{\partial P_a}
{\partial
v^i} \frac{\partial v^i}{\partial Q^b} & \frac{\partial P_a}{\partial v^i}
\frac{\partial v^i}{\partial P_b} \cr} \eqno(2.13a)$$
$$\delta^i_{~j} = {\partial v^i \over \partial Q^a} \thinspace {\partial Q^a
\over \partial v^j} + {\partial v^i \over \partial P_a} \thinspace {\partial
P_a \over \partial v^j} \eqno(2.13b)$$
The problem of showing that such functions exist is known as {\it Pfaff's
problem} in honor of the German mathematician J. F. Pfaff [10]. This problem
was
solved long ago by G. Frobenius and J. G. Darboux [11,12,13].
That the transformation:
$$q^a(t) = Q^a\Bigl(v(t),t\Bigr) \qquad , \qquad p_a(t) = P_a\Bigl(v(t),t\Bigr)
\eqno(2.14)$$
gives canonical variables is straightforward to see. First, substitute $(2.12)$
into $(2.6)$ to obtain:
$$J_{ij} = {\partial P_a \over \partial v^i} \thinspace {\partial Q^a \over
\partial v^j} - {\partial Q^a \over \partial v^i} \thinspace {\partial P_a
\over \partial v^j} \eqno(2.15)$$
Now use this relation and $(2.13a)$ to show that the bracket matrix ``raises''
indices on the transformation matrices:
$$\eqalignno{{\partial v^i \over \partial Q^a} &= J^{ij} \thinspace J_{jk}
\thinspace {\partial v^k \over \partial Q^a} &(2.16a) \cr
&= J^{ij} \thinspace \Bigl({\partial P_b \over \partial v^j} \thinspace
{\partial Q^b \over \partial v^k} - {\partial Q^b \over \partial v^j}
\thinspace
{\partial P_b \over \partial v^k}\Bigr) \thinspace {\partial v^k \over \partial
Q^a} &(2.16b) \cr
&= J^{ij} \thinspace {\partial P_a \over \partial v^j} &(2.16c) \cr}$$
$$\eqalignno{{\partial v^i \over \partial P_a} &= J^{ij} \thinspace J_{jk}
\thinspace {\partial v^k \over \partial P_a} &(2.17a) \cr
&= J^{ij} \thinspace \Bigl({\partial P_b \over \partial v^j} \thinspace
{\partial Q^b \over \partial v^k} - {\partial Q^b \over \partial v^j}
\thinspace
{\partial P_b \over \partial v^k}\Bigr) \thinspace {\partial v^k \over \partial
P_a} &(2.17b) \cr
&= - J^{ij} \thinspace {\partial Q^a \over \partial v^j} &(2.17c) \cr}$$
Finally, take the brackets using $(2.2)$, $(2.13a)$, $(2.14)$, $(2.16)$ and
$(2.17)$:
$$\{q^a,q^b\} = {\partial Q^a \over \partial v^i} \thinspace J^{ij} \thinspace
{\partial Q^b \over \partial v^j} = - {\partial Q^a \over \partial v^i}
\thinspace {\partial v^i \over \partial P_b} = 0 \eqno(2.18a)$$
$$\{q^a,p_b\} = {\partial Q^a \over \partial v^i} \thinspace J^{ij} \thinspace
{\partial P_b \over \partial v^j} = {\partial Q^a \over \partial v^i}
\thinspace
{\partial v^i \over \partial Q^b} = \delta_a^{~b} \eqno(2.18b)$$
$$\{p_a,p_b\} = {\partial P_a \over \partial v^i} \thinspace J^{ij} \thinspace
{\partial P_b \over \partial v^j} = {\partial P_a \over \partial v^i}
\thinspace
{\partial v^i \over \partial Q^b} = 0 \eqno(2.18c)$$
A constructive --- and slightly different --- procedure for finding canonical
coordinates is credited to the French mathematician Darboux and can be found in
the text by Arnold [14]. The older derivation is discussed in the text by
Whittaker [15].

It is possible to give explicit formulae for $N=1$. For this case we can write
down the inverse bracket matrix:
$$\pmatrix{J_{11} & J_{12} \cr J_{21} & J_{22} \cr} = \pmatrix{0 & -\frac1{
J^{12}} \cr \frac1{J^{12}} & 0 \cr} \eqno(2.19)$$
and we further know that the function $J^{12}(v,t)$ is non-zero. The most
convenient ``gauge'' for the vector potential is an axial one:
$$\phi_1(v^1,v^2,t) = \int_0^{v^2} \thinspace {ds \over J^{12}(v^1,s,t)}
\eqno(2.20a)$$
$$\phi_2(v^1,v^2,t) = 0 \eqno(2.20b)$$
Comparison with $(2.12)$ reveals that $Q$ cannot depend upon $v^2$ and that the
choice of $Q$ uniquely determines $P$. The simplest choice is:
$$Q(v^1,v^2,t) = v^1 \eqno(2.21a)$$
$$P(v^1,v^2,t) = \int_0^{v^2} \thinspace {ds \over J^{12}(v^1,s,t)}
\eqno(2.21b)$$
Note that although the general construction is only local --- and so must be
built up in patches --- the formulae for $N=1$ are valid globally if the domain
of $(v^1,v^2)$ is $R^2$.

Of course the mapping to canonical variables can only be unique up to canonical
transformations --- even in the Lagrangian formalism. This shows up slightly in
the integration constants which come from solving $(2.12)$; for the case of
$N=1$
above we had the freedom to choose $Q$ to be any function of $v^2$ and $t$
which
was instantaneously invertible on $v^2$. By far the larger source of variation
is the freedom to add the gradient of a scalar to $\phi_i$. {}From the
instantaneous relation:
$$\phi_i \thinspace dv^i = p_a \thinspace dq^a \eqno(2.22)$$
we see that the ``gauge transformation," $\phi_i \longrightarrow \phi_i -
\partial_i \theta$, induces a canonical transformation whose generating
function is $-\theta$.

Once the brackets are canonical they are also independent of time, so we can
get
a Hamiltonian in the new coordinate system by using $(2.11)$. The evolution
equations in these coordinates are:
$${\dot q}^a(t) = f^a\Bigl(q(t),p(t),t\Bigr) \eqno(2.23a)$$
$${\dot p}_a(t) = f_a\Bigl(q(t),p(t),t\Bigr) \eqno(2.23b)$$
where we define:
$$f^a\Bigl(Q(v,t),P(v,t),t\Bigr) \equiv {\partial Q^a \over \partial v^i}(v,t)
\thinspace f^i(v,t) + {\partial Q^a \over \partial t}(v,t) \eqno(2.24a)$$
$$f_a\Bigl(Q(v,t),P(v,t),t\Bigr) \equiv {\partial P_a \over \partial v^i}(v,t)
\thinspace f^i(v,t) + {\partial P_a \over \partial t}(v,t) \eqno(2.24b)$$
Substituting into $(2.11)$ gives:
$$\eqalign{H&(q,p,t) - H(q_0,p_0,t) \cr
&= \int_0^1 d\tau \thinspace \Biggl\{{\Delta p}_a \thinspace f^a\Bigl(q_0 +
\tau {\Delta q}, p_0 + \tau {\Delta p},t\Bigr) - {\Delta q}^a \thinspace f_a
\Bigl(q_0 + \tau {\Delta q}, p_0 + \tau {\Delta p},t\Bigr)\Biggr\} \cr}
\eqno(2.25)$$
where $(q_0,p_0)$ is any point in phase space and we define ${\Delta q}^a
\equiv
q^a - q_0^a$ and ${\Delta p}_a \equiv p_a - p_{0a}$.

Note that the Hamiltonian $(2.25)$ only generates the evolution $(2.23)$ of the
canonical coordinates. Its bracket with $v^i(t)$ does {\it not} generally give
${\dot v}^i(t)$. If we call the inverse transformation $V^i(q,p,t)$ then the
relation:
$$v^i(t) = V^i\Bigl(q(t),p(t),t\Bigr) \eqno(2.26)$$
implies that the original dynamical variables acquire only a portion of their
time dependence from that of the canonical coordinates:
$$\eqalignno{{\dot v}^i(t) &= {\partial V^i \over \partial q^a} \thinspace
{\dot q}^a + {\partial V^i \over \partial p_a} \thinspace {\dot p}_a +
{\partial
V^i \over \partial t} &(2.27a) \cr
&= \Bigl\{v^i(t),H\Bigr\} + {\partial V^i \over \partial t} &(2.27b) \cr}$$
We have already seen that the final term in $(2.27b)$ must be non-zero if the
$J^{ij}$'s contain explicit time dependence.

Far from being a problem, this extra source of time dependence is a blessing.
It
is what keeps the evolution of the original dynamical variables fixed if we
alter what is meant by the canonical variables. As long as there is a clear
procedure for inferring physics from the $v^i(t)$'s it does not matter that we
can choose the canonical variables differently. This is a subtle point and a
crucially important one. It is also counterintuitive because any choice of
canonical coordinates {\it does} offer a complete description of the degrees of
freedom available to the system. The point is that what these degrees of
freedom
mean physically changes as we change our representation of them.

The possibility for this sort of subtlety is always present in physics because
time dependent canonical transformations can be made in any classical theory
and, with proper attention to operator ordering, on the quantum level as well.
For example, in perturbative quantum field theory there is a well known
transformation, called as the ``interaction representation,'' which takes one
to a set of fields whose time evolution and commutation relations are free.
Consider the theory of a scalar field, $\phi(t,{\vec x})$, whose action has the
following form:
$$S[\phi] = \int d^4x \thinspace \Bigl\{\frac12 \partial_{\mu} \phi \thinspace
\partial^{\mu} \phi - \frac12 m^2 \thinspace \phi^2\Bigr\} + S_I[\phi]
\eqno(2.28)$$
where $S_I[\phi]$ is an ultralocal interaction. Since $S_I[\phi]$ is free of
derivatives the momentum canonically conjugate to $\phi(t,{\vec x})$ is just
its
time derivative. The only non-vanishing equal time bracket is:
$$\Bigl\{\phi(t,{\vec x}),{\dot \phi}(t,{\vec y})\Bigr\} = \delta^3({\vec x} -
{\vec y}) \eqno(2.29)$$
The field equation:
$$(\square - m^2) \thinspace \phi + {\delta S_I[\phi] \over \delta \phi} = 0
\eqno(2.30)$$
can be integrated to give a new field which obeys the Klein-Gordon equation:
$$\Phi[\phi] \equiv \phi + {1 \over \square - m^2} \thinspace {\delta S_I[\phi]
\over \delta \phi} \eqno(2.31)$$
If we define the inverse of $\square - m^2$ so that it and its first derivative
vanish at $t=0$ then we have also:
$$\Phi(0,{\vec x}) = \phi(0,{\vec x}) \eqno(2.32a)$$
$${\dot \Phi}(0,{\vec x}) = {\dot \phi}(0,{\vec x}) \eqno(2.32b)$$
It follows that $\Phi(t,{\vec x})$ not only {\it evolves} like a free field, it
also obeys the same bracket algebra. We do not conclude that all such theories
are free, even perturbatively, because we insist upon inferring physics from
the
original variables. The only difference between this standard situation and the
previous discussion of this section is that even the original scalar is
canonical whereas we do not make this assumption about the $v^i$'s.

Why the imposition of constraints tends to result in a non-canonical bracket
matrix is obvious to anyone who has ever constructed a Dirac bracket [16]. Even
though the gauge and constrained variables are irrelevant to physics, the
unreduced canonical formalism is an artificial construct in which the
unphysical
variables play an essential role in keeping the brackets canonical under time
evolution. After reduction these variables are no longer independent and this
changes the bracket algebra. Unless what we call the reduced variables are
chosen very carefully they will not have a canonical bracket algebra. This does
not mean that other choices are ``wrong.'' In the sense of providing a complete
description of physics a non-canonical set of variables can be as ``right'' as
a
canonical set. Non-canonical reduced variables can even be highly preferred on
account of bearing a simpler relation than any canonical variables to the
original degrees of freedom from which physics is inferred.

In the next section we use scalar electrodynamics to provide an explicit
example
of how reduction can yield a non-canonical bracket matrix. For now we can
understand the issues in a simple but somewhat contrived fashion through a
model
of two coupled oscillators whose Lagrangian is:
$$L = \frac12 m \thinspace \Bigl({\dot q}_1^2 + {\dot q}_2^2\Bigr) - \frac12
m \omega^2 \thinspace \Bigl(\frac54 q_1^2 + q_1 q_2 + \frac54 q_2^2\Bigr)
\eqno(2.33)$$
The resulting canonical formalism will serve as our model for an unreduced
canonical formalism, in spite of the fact that the Lagrangian possesses no
continuous symmetries and gives rise to no constraint equations. We shall make
up for the absence of constraints by imposing them {\it ad hoc}.

It is straightforward to show that the canonical variables of the coupled
oscillator system have the following time evolution:
$$\eqalign{q_1(t) = \frac12 \Bigl({\widehat q}_1 &+ {\widehat q}_2\Bigr)
\thinspace \cos\Bigl(\frac32 \omega t\Bigr) + \frac12 \Bigl({\widehat q}_1 -
{\widehat q}_2\Bigr) \thinspace \cos\Bigl(\frac12 \omega t\Bigr) \cr
&+ \frac1{3 m \omega} \Bigl({\widehat p}_1 + {\widehat p}_2\Bigr) \thinspace
\sin\Bigl(\frac32 \omega t\Bigr) + \frac1{m \omega} \Bigl({\widehat p}_1 -
{\widehat p}_2\Bigr) \thinspace \sin\Bigl(\frac12 \omega t\Bigr) \cr}
\eqno(2.34a)$$
$$\eqalign{p_1(t) = - \frac34 m \omega \Bigl({\widehat q}_1 &+ {\widehat q}_2
\Bigr) \thinspace \sin\Bigl(\frac32 \omega t\Bigr) - \frac14 m \omega
\Bigl({\widehat q}_1 - {\widehat q}_2\Bigr) \thinspace \sin\Bigl(\frac12 \omega
t\Bigr) \cr
&+ \frac12 \Bigl({\widehat p}_1 + {\widehat p}_2\Bigr) \thinspace \cos\Bigl(
\frac32 \omega t\Bigr) + \frac12 \Bigl({\widehat p}_1 - {\widehat p}_2 \Bigr)
\thinspace \cos\Bigl(\frac12 \omega t\Bigr) \cr} \eqno(2.34b)$$
$$\eqalign{q_2(t) = \frac12 \Bigl({\widehat q}_1 &+ {\widehat q}_2\Bigr)
\thinspace \cos\Bigl(\frac32 \omega t\Bigr) + \frac12 \Bigl(-{\widehat q}_1 +
{\widehat q}_2\Bigr) \thinspace \cos\Bigl(\frac12 \omega t\Bigr) \cr
&+ \frac1{3 m \omega} \Bigl({\widehat p}_1 + {\widehat p}_2\Bigr) \thinspace
\sin\Bigl(\frac32 \omega t\Bigr) + \frac1{m \omega} \Bigl(-{\widehat p}_1 +
{\widehat p}_2\Bigr) \thinspace \sin\Bigl(\frac12 \omega t\Bigr) \cr}
\eqno(2.34c)$$
$$\eqalign{p_2(t) = - \frac34 m \omega \Bigl({\widehat q}_1 &+ {\widehat q}_2
\Bigr) \thinspace \sin\Bigl(\frac32 \omega t\Bigr) - \frac14 m \omega \Bigl(
-{\widehat q}_1 + {\widehat q}_2\Bigr) \thinspace \sin\Bigl(\frac12 \omega t
\Bigr) \cr
&+ \frac12 \Bigl({\widehat p}_1 + {\widehat p}_2\Bigr) \thinspace \cos\Bigl(
\frac32 \omega t\Bigr) + \frac12 \Bigl(-{\widehat p}_1 + {\widehat p}_2 \Bigr)
\thinspace \cos\Bigl(\frac12 \omega t\Bigr) \cr} \eqno(2.34d)$$
In these formulae we have denoted the initial values by a hat. It is these
initial values that are the independent degrees of freedom of this system; in
particular, only they have independent Poisson brackets. It is straightforward
to verify that if the initial variables have Poisson brackets then the time
evolution given in $(2.34)$ preserves this feature. That is, the only non-zero
equal time brackets are:
$$\Bigl\{q_1(t),p_1(t)\Bigr\} = 1 = \Bigl\{q_2(t),p_2(t)\Bigr\} \eqno(2.35)$$
We emphasize that the bracket algebras for different times are not
independently
specifiable. Since the variables at time $t$ are uniquely determined in terms
of
their initial values, the Poisson bracket algebra at time $t$ is determined by
time evolution and by the initial bracket algebra.

The model at this stage should be thought of as analogous to temporal gauge QED
or Yang-Mills, or to synchronous gauge gravity. A constraint in these models is
a relation between the initial value variables, and such constraints are
reduced
by imposing a gauge condition on the initial value surface. It is only in
exceptional cases that we can find gauge conditions which are preserved under
time evolution. Initial value gauge conditions always imply {\it some} relation
between the later canonical variables, but very seldom the same relation.
Unless
we choose what we call the variables of the reduced theory to compensate for
this change, the reduced brackets will become non-canonical because at any
instant they are the Dirac brackets associated with different gauge conditions.

Let us suppose that the reduction of our oscillator model is accomplished by
setting ${\widehat q}_2 = 0 = {\widehat p}_2$. Let us further suppose that we
take as our reduced variables, $v^1(t) \equiv q_1(t)$ and $v^2(t) \equiv
p_1(t)$. Since the constraint and its conjugate gauge condition affect only the
initial values we see from $(2.34)$ that the reduced variables have the
following evolution:
$$v^1(t) = \frac12 {\widehat q}_1 \thinspace \cos\Bigl(\frac32 \omega t\Bigr)
+ \frac12 {\widehat q}_1 \thinspace \cos\Bigl(\frac12 \omega t\Bigr) + \frac1{3
m \omega} {\widehat p}_1 \thinspace \sin\Bigl(\frac32 \omega t\Bigr) + \frac1{m
\omega} {\widehat p}_1 \thinspace \sin\Bigl(\frac12 \omega t\Bigr)\eqno(2.36a)
$$
$$v^2(t) = -\frac34 m \omega \thinspace {\widehat q}_1 \thinspace \sin\Bigl(
\frac32 \omega t\Bigr) - \frac14 m \omega \thinspace {\widehat q}_1 \thinspace
\sin\Bigl(\frac12 \omega t\Bigr) + \frac12 {\widehat p}_1 \thinspace \cos\Bigl(
\frac32 \omega t\Bigr) + \frac12 {\widehat p}_1 \thinspace \cos\Bigl(\frac12
\omega t\Bigr) \eqno(2.36b)$$
It follows that the only non-zero equal time Poisson bracket is:
$$\Bigl\{v^1(t),v^2(t)\Bigr\} = \frac43 \cos^2\Bigl(\frac12 \omega t\Bigr)
- \frac13 \cos^2\Bigl(\omega t\Bigr) \eqno(2.37)$$
Although it is canonical at $t=0$, it is not so later on, and it can even pass
through zero!

If we ignore the fact that the bracket matrix becomes non-invertible at
$\omega t
\approx (2n+1) \pi \pm .12 \pi$ then the construction gives the following
canonical variables:
$$q(t) \equiv v^1(t) \eqno(2.38a)$$
$$p(t) \equiv {3 \thinspace v^2(t) \over 4 \cos^2\Bigl(\frac12 \omega t\Bigr)
- \cos^2\Bigl(\omega t\Bigr)} \eqno(2.38b)$$
The associated Hamiltonian is:
$$\eqalign{H(t,q,p) = &\Bigl[\frac43 \cos^2\Bigl(\frac12 \omega t\Bigr) -
\frac13 \cos^2\Bigl(\omega t\Bigr) \Bigr] \thinspace {p^2 \over 2 m} \cr
&\qquad + \frac12 m \omega^2 \thinspace q^2 \thinspace {\cos^2\Bigl(\frac12
\omega t\Bigr) + \frac14 \cos^2\Bigl(\omega t\Bigr) \over \Bigl[\frac43 \cos^2
\Bigl(\frac12 \omega t\Bigr) - \frac13 \cos^2\Bigl(\omega t\Bigr) \Bigr]^2} \cr
}
\eqno(2.39)$$
Note that although this Hamiltonian generates time evolution for $q(t)$ and
$p(t)$, it does not do so for either the reduced variables or for the original
ones. The variable $v^2(t)$ acquires additional time dependence from relation
$(2.38b)$, and the original variables have the following expressions in terms
of
$q(t)$ and $p(t)$:
$$q_1(t) = q(t) \eqno(2.40a)$$
$$p_1(t) = \Bigl[\frac43 \cos^2\Bigl(\frac12 \omega t\Bigr) - \frac13 \cos^2
\Bigl(\omega t\Bigr)\Bigr] \thinspace p(t) \eqno(2.40b)$$
$$q_2(t) = {-2 \sin\Bigl(\frac12 \omega t\Bigr) \thinspace \sin\Bigl(\frac32
\omega t\Bigr) \over 4 \cos^2\Bigl(\frac12 \omega t\Bigr) - \cos^2\Bigl(\omega
t\Bigr)} \thinspace q(t) + \frac43 \sin^2\Bigl(\frac12 \omega t\Bigr)
\thinspace
\sin\Bigl(\omega t\Bigr) \thinspace {p(t) \over m \omega} \eqno(2.40c)$$
$$p_2(t) = {-3 \cos^2\Bigl(\frac12 \omega t\Bigr) \thinspace \sin\Bigl(\omega t
\Bigr) \over 4 \cos^2\Bigl(\frac12 \omega t\Bigr) - \cos^2\Bigl(\omega t\Bigr)}
\thinspace m \omega q(t) + \frac23 \sin\Bigl(\frac 12 \omega t\Bigr) \thinspace
\sin\Bigl(\frac32 \omega t\Bigr) \thinspace p(t) \eqno(2.40d)$$

Although the preceding canonical formalism is valid, it is not simple because
the reduced variables were badly chosen. A much more convenient choice is:
$$v^1(t) \equiv q_1(t) + q_2(t) = {\widehat q}_1 \thinspace \cos\Bigl(\frac32
\omega t\Bigr) + \frac2{3 m \omega} \thinspace {\widehat p}_1 \thinspace
\sin\Bigl(\frac32 \omega t\Bigr) \eqno(2.41a)$$
$$v^2(t) \equiv p_1(t) + p_2(t) = - \frac32 m \omega \thinspace {\widehat q}_1
\thinspace \sin\Bigl(\frac32 \omega t\Bigr) + {\widehat p}_1 \thinspace
\cos\Bigl(\frac32 \omega t\Bigr) \eqno(2.41b)$$
The resulting reduced bracket algebra is canonical:
$$\Bigl\{v^1(t),v^2(t)\Bigr\} = 1 \eqno(2.42a)$$
and the Hamiltonian is time independent:
$$H = \frac1{2 m} \Bigl(v^2\Bigr)^2 + \frac98 m \omega^2 \Bigl(v^1\Bigr)^2
\eqno(2.42b)$$
This Hamiltonian generates the evolution of $v^1(t)$ and $v^2(t)$ but not that
of the original variables. They acquire additional time dependence through the
relations:
$$q_1(t) =  \Bigl[2 \cos^2\Bigl(\frac12 \omega t\Bigr) - \cos^2\Bigl(\omega t
\Bigr)\Bigr] \thinspace v^1(t) - \frac43 \sin^2\Bigl(\frac12 \omega t\Bigr)
\thinspace \sin\Bigl(\omega t\Bigr) \thinspace {v^2(t) \over m \omega}
\eqno(2.43a)$$
$$p_1(t) =  \cos^2\Bigl(\frac12 \omega t\Bigr) \thinspace \sin\Bigl(\omega t
\Bigr) \thinspace m \omega v^1(t) + \Bigl[\frac23 \cos^2\Bigl(\frac12 \omega t
\Bigr) + \frac13 \cos^2\Bigl(\omega t\Bigr)\Bigr] \thinspace v^2(t)
\eqno(2.43b)$$
$$q_2(t) =  \Bigl[2 \sin^2\Bigl(\frac12 \omega t\Bigr) - \sin^2\Bigl(\omega t
\Bigr)\Bigr] \thinspace v^1(t) + \frac43 \sin^2\Bigl(\frac12 \omega t\Bigr)
\thinspace \sin\Bigl(\omega t\Bigr) \thinspace {v^2(t) \over m \omega}
\eqno(2.43c)$$
$$p_2(t) = - \cos^2\Bigl(\frac12 \omega t\Bigr) \thinspace \sin\Bigl(\omega t
\Bigr) \thinspace m \omega v^2(t) + \Bigl[\frac23 \sin^2\Bigl(\frac12 \omega t
\Bigr) + \frac13 \sin^2\Bigl(\omega t\Bigr)\Bigr] \thinspace v^1(t)
\eqno(2.43d)$$

Another convenient choice is:
$$v^1(t) \equiv q_1(t) - q_2(t) = {\widehat q}_1 \thinspace \cos\Bigl(\frac12
\omega t\Bigr) + \frac2{m \omega} \thinspace {\widehat p}_1 \thinspace
\sin\Bigl(\frac12 \omega t\Bigr) \eqno(2.44a)$$
$$v^2(t) \equiv p_1(t) - p_2(t) = - \frac12 m \omega \thinspace {\widehat q}_1
\thinspace \sin\Bigl(\frac12 \omega t\Bigr) + {\widehat p}_1 \thinspace
\cos\Bigl(\frac12 \omega t\Bigr) \eqno(2.44b)$$
As before, the resulting reduced bracket algebra is canonical:
$$\Bigl\{v^1(t),v^2(t)\Bigr\} = 1 \eqno(2.45a)$$
and the Hamiltonian is time independent:
$$H = \frac1{2 m} \Bigl(v^2\Bigr)^2 + \frac18 m \omega^2 \Bigl(v^1\Bigr)^2
\eqno(2.45b)$$
This Hamiltonian generates the evolution of $v^1(t)$ and $v^2(t)$. The original
variables acquire additional time dependence through the relations:
$$q_1(t) =  \Bigl[\frac23 \cos^2\Bigl(\frac12 \omega t\Bigr) +\frac13 \cos^2
\Bigl(\omega t\Bigr)\Bigr] \thinspace v^1(t) + \frac43 \sin^2\Bigl(\frac12
\omega t\Bigr) \thinspace \sin\Bigl(\omega t\Bigr) \thinspace {v^2(t) \over m
\omega} \eqno(2.46a)$$
$$p_1(t) = - \cos^2\Bigl(\frac12 \omega t\Bigr) \thinspace \sin\Bigl(\omega t
\Bigr) \thinspace m \omega \thinspace v^1(t) + \Bigl[2 \cos^2\Bigl(\frac12
\omega t\Bigr) - \cos^2\Bigl(\omega t\Bigr)\Bigr] \thinspace v^2(t)
\eqno(2.46b)$$
$$q_2(t) = \Bigl[-\frac23 \sin^2\Bigl(\frac12 \omega t\Bigr) - \frac13 \sin^2
\Bigl(\omega t\Bigr)\Bigr] \thinspace v^1(t) + \frac43 \sin^2\Bigl(\frac12
\omega t\Bigr) \thinspace \sin\Bigl(\omega t\Bigr) \thinspace {v^2(t) \over m
\omega} \eqno(2.46c)$$
$$p_2(t) = - \cos^2\Bigl(\frac12 \omega t\Bigr) \thinspace \sin\Bigl(\omega t
\Bigr) \thinspace m \omega \thinspace v^1(t) + \Bigl[-2 \sin^2\Bigl(\frac12
\omega t\Bigr) + \sin^2\Bigl(\omega t\Bigr)\Bigr] \thinspace v^2(t)
\eqno(2.46d)$$

Three points deserve mention. First, although the latter two formulations are
significantly simpler than the first, all three give precisely the same
evolution for $q_i(t)$ and $p_i(t)$. As long as we infer physics from these
variables the seemingly different theories are in fact identical. Physicists
are
so conditioned to regard the Hamiltonian as an observable that this point can
not be overemphasized. A glance at the first Hamiltonian $(2.39)$ suggests a
harmonic oscillator with time dependent mass and frequency:
$$m(t) = {m \over \frac43 \cos^2\Bigl(\frac12 \omega t\Bigr) - \frac13 \cos^2
\Bigl(\omega t\Bigr)} \eqno(2.47a)$$
$$\omega^2(t) = \Bigl[\cos^2\Bigl(\frac12 \omega t\Bigr) + \frac14 \cos^2\Bigl(
\omega t\Bigr)\Bigr] \thinspace \omega^2 \eqno(2.47b)$$
The second Hamiltonian $(2.42b)$ is that of an oscillator of mass $m$ and
frequency $\frac32 \omega$, while the third Hamiltonian $(2.45b)$ is that of an
oscillator with mass $m$ and frequency $\frac12 \omega$. In fact the system is
none of these things, it is rather a set of coupled oscillators which has been
subjected to a constraint. This only becomes apparent by ignoring the
Hamiltonian and concentrating instead upon the $q_i(t)$'s and the $p_i(t)$'s.

Note that although the reduced canonical variables provide a complete and
minimal representation of the system's dynamical degrees of freedom, there is
nothing wrong with using as observables the overcomplete representation
provided
by the original variables. These will include some pure gauge degrees of
freedom
and some constrained ones. Of course the former are unphysical but the latter
encode perfectly valid information, even if this information can be recovered
from the reduced variables. The longitudinal electric field offers a familiar
example. We really do not need this quantity since it can be recovered from a
knowledge of the positions of all the charges, but there is nothing wrong with
regarding it as an observable since it can be measured using the Lorentz force
law.

The second point is that the reduced canonical variables {\it do} have a role
to
play. In the quantum theory they tell us how to describe states, and how the
original variables act as operators upon these states. For example, suppose we
quantize the first representation of the coupled oscillator system. A state in
the Schr\"odinger picture position representation is described by a square
integrable wavefunction, $\psi(q,t)$. The time evolution of this wavefunction
is
generated by the Hamiltonian $(2.39)$. The operators $q_i(t)$ and $p_i(t)$ act
on
such a state through relations $(2.40)$, where $q(t) \thinspace \psi(q,t) = q
\thinspace \psi(q,t)$ and $p(t) \thinspace \psi(q,t) = - i {\partial \psi(q,t)
\over \partial q}$. Note that even in the Schr\"odinger picture the observables
$q_i(t)$ and $p_i(t)$ have time dependence. In the Heisenberg picture the state
is time independent and the evolution of the reduced canonical variables,
$q(t)$
and $p(t)$, is generated by $(2.39)$.

The final point concerns energy. We have stated that none of the Hamiltonians
--- $(2.39)$, $(2.42b)$ and $(2.45b)$ --- plays the usual role of the energy
as an
observable, so what does? The answer is the Hamiltonian of the original theory,
after the imposition of our ``gauge condition'' and ``constraint,'' ${\widehat
q}_2 = 0 = {\widehat p}_2$:
$$\eqalignno{E &= {{\widehat p}_1^2 \over 2 m} + {{\widehat p}_2^2 \over 2 m}
+ \frac12 m \omega^2 \thinspace \Bigl(\frac54 {\widehat q}_1^2 + {\widehat q}_1
\thinspace {\widehat q}_2 + \frac54 {\widehat q}_2^2\Bigr) &(2.48a) \cr
&\mathop{\longrightarrow}\limits_{{\widehat q}_2 = 0 = {\widehat p}_2} \quad
{{\widehat p}_2^2 \over 2 m} + \frac58 m \omega^2 \thinspace {\widehat q}_1^2
&(2.48b) \cr}$$
This energy does not generate time evolution for the reduced canonical
variables
in any of the three representations given above. For example, in the last
representation we can use relations $(2.34)$ and $(2.46)$ to express the
energy as:
$$\eqalign{E\Bigl(v^1,v^2,t\Bigr) = \Bigl[\frac9{16} &+ \frac7{16} \cos(\omega
t)\Bigr] \thinspace \frac12 m \omega^2 \thinspace \Bigl(v^1\Bigr)^2 \cr
&- \frac{15}8 \sin(\omega t) \thinspace \omega \thinspace v^1 \thinspace v^2
+ \Bigl[\frac{17}8 - \frac{15}8 \cos(\omega t)\Bigr] \thinspace {\Bigl(v^2
\Bigr)^2 \over 2 m} \cr} \eqno(2.49)$$
{}From $(2.40a)$ it follows that:
$$\Bigl\{E,v^1(t)\Bigr\} = - \frac{15}8 \sin(\omega t) \thinspace \omega
\thinspace v^1(t) + \Bigl[\frac{17}8 - \frac{15}8 \cos(\omega t)\Bigr]
\thinspace
{v^2(t) \over m} \neq {v^2(t) \over m} = {\dot v}^1(t) \eqno(2.50a)$$
$$\Bigl\{E,v^2(t)\Bigr\} = - \Bigl[\frac9{16} + \frac7{16} \cos(\omega t)\Bigr]
\thinspace m \omega^2 \thinspace v^1(t) + \frac{15}8 \sin(\omega t) \thinspace
\omega \thinspace v^2(t) \neq - \frac14 m \omega^2 \thinspace v^1(t) = {\dot
v}^2(t) \eqno(2.50b)$$
In spite of the fact that the physical energy $E$ does not generate time
evolution, it {\it is} conserved:
$${d E \over dt} \equiv {\partial E \over \partial v^1} \thinspace {\dot v}^1
+ {\partial E \over \partial v^2} \thinspace {\dot v}^2 + {\partial E \over
\partial t} = 0 \eqno(2.51)$$
This potential for a disagreement between the physical energy and the
Hamiltonian
which generates time evolution is a general feature of reduced canonical
systems.

\vskip 1cm
\centerline{\bf 3. Correspondence With Gauge Theories}

The purpose of this section is to study a simple model in which the various
features of the construction and the context in which it takes place can be
clearly understood. The model is scalar QED in flat space, the Lagrangian for
which is:
$${\cal L} = -\frac14 \thinspace F_{\mu \nu} \thinspace F^{\mu \nu} + \Bigl(
\partial_{\mu} - i e \thinspace A_{\mu}\Bigr) \thinspace \phi^* \thinspace
\Bigl(\partial^{\mu} + i e \thinspace A^{\mu}\Bigr) \thinspace \phi \eqno(3.1)
$$
Here $F_{\mu \nu} \equiv \partial_{\mu} A_{\nu} - \partial_{\nu} A_{\mu}$ is
the
field strength tensor, $\phi$ is a complex scalar field, $e$ is the
electromagnetic coupling constant, and we are using a spacelike metric. The
Lagrangian is invariant under a gauge transformation:
$$A_{\mu}(x) \mapsto A_{\mu}(x) - \partial_{\mu} \theta(x) \eqno(3.2a)$$
$$\phi(x) \mapsto \exp\Bigl[i e \thinspace \theta(x)\Bigr] \thinspace \phi(x)
\eqno(3.2b)$$
which is parametrized by an arbitrary real scalar function, $\theta(x)$. By
imposing temporal gauge:
$$A_0(x) = 0 \eqno(3.3)$$
one obtains a constrained canonical system which still possesses invariance
under
gauge transformations parametrized by time independent functions $\theta({\vec
x})$. We will show how this system can be reduced to a completely gauge fixed
system containing only dynamical degrees of freedom, and we will explain why
the
resulting Poisson brackets can fail to be canonical. We will then change
variables to obtain canonical Poisson brackets and infer the Hamiltonian which
generates time evolution for the reduced system.

It is a simple matter to show that in temporal gauge the momenta canonically
conjugate to $A_i$, $\phi$ and $\phi^*$ are, respectively:
$$E_i(t,{\vec x}) = {\dot A}_i(t,{\vec x}) \eqno(3.4a)$$
$$\pi(t,{\vec x}) = {\dot \phi}^*(t,{\vec x}) \eqno(3.4b)$$
$$\pi^*(t,{\vec x}) = {\dot \phi}(t,{\vec x}) \eqno(3.4c)$$
What it means to be ``canonical'' is that the non-zero equal time Poisson
brackets are:
$$\Bigl\{A_i(t,{\vec x}),E_j(t,{\vec y})\Bigr\} = \delta_{ij} \thinspace
\delta^3({\vec x} - {\vec y}) \eqno(3.5a)$$
$$\Bigl\{\phi(t,{\vec x}),\pi(t,{\vec y})\Bigr\} = \delta^3({\vec x} -
{\vec y})
\eqno(3.5b)$$
$$\Bigl\{\phi^*(t,{\vec x}),\pi^*(t,{\vec y})\Bigr\} = \delta^3({\vec x} -
{\vec y}) \eqno(3.5c)$$
The Hamiltonian is:
$$H = \int d^3x \thinspace \Bigl\{\frac12 \thinspace E_i \thinspace E_i +
\frac14 \thinspace F_{ij} \thinspace F_{ij} + \pi^* \thinspace \pi + \Bigl(
\partial_i - i e \thinspace A_i\Bigr) \thinspace \phi^* \thinspace \Bigl(
\partial_i + i e \thinspace A_i\Bigr) \thinspace \phi\Bigr\} \eqno(3.6)$$
It is straightforward to check that this functional generates time evolution.
That is, the Poisson brackets of the Hamiltonian with the canonical coordinates
give relations $(3.4a)$ through $(3.4c)$, and the Poisson brackets of the
conjugate
momenta give the canonical formulation of the dynamical Euler-Lagrange
equations:
$${\dot E}_i = \Bigl\{E_i,H\Bigr\} = \partial_j \thinspace F_{ji} + i e
\thinspace \phi^* \thinspace \Bigl(\partial_i + i e \thinspace A_i\Bigr)
\thinspace \phi - i e \thinspace \phi \thinspace \Bigl(\partial_i - i e
\thinspace A_i\Bigr) \thinspace \phi^* \eqno(3.7a)$$
$${\dot \pi} = \Bigl\{\pi,H\Bigr\} = \Bigl(\partial_i - i e \thinspace A_i
\Bigr)
\thinspace \Bigl(\partial_i - i e \thinspace A_i\Bigr) \thinspace \phi^*
\eqno(3.7b)$$
$${\dot \pi}^* = \Bigl\{\pi,H\Bigr\} = \Bigl(\partial_i + i e \thinspace A_i
\Bigr)\thinspace \Bigl(\partial_i + i e \thinspace A_i\Bigr) \thinspace \phi
\eqno(3.7c)$$
The non-dynamical Euler-Lagrange equation --- the one obtained by variation
with
respect to $A_0$ --- is not realized through the definition of time evolution.
It must be imposed as a constraint:
$$\partial_i \thinspace E_i + ie \thinspace \Bigl(\pi \thinspace \phi - \pi^*
\thinspace \phi^*\Bigr) = 0\eqno(3.8)$$
Of course the left hand side of this constraint is also the generator of
infinitesimal, time-independent gauge transformations.

Although this system possesses a time independent gauge symmetry, it is still
complete in the sense that the equations of evolution uniquely determine the
fields at any time in terms of those on an initial value surface. To economize
the notation let us refer to a general field as $\psi_a(t,{\vec x})$ and an
initial value configuration as ${\widehat \psi}_a({\vec x})$. The evolution
equations $(3.4)$ and $(3.7)$ uniquely determine the former in terms of the
latter. Even though we cannot exhibit this relation as we could for the
coupled
oscillator system of the previous section --- cf. relations $(2.34)$ --- we
can still represent it functionally:
$$\psi_a(t,{\vec x}) = \Psi\Bigl[{\widehat \psi}~\Bigr]_a(t,{\vec x})
\eqno(3.9)$$
Just as with the coupled oscillator system discussed in the previous section,
it
is the ${\widehat \psi}_a({\vec x})$'s that represent the dynamical degrees of
freedom of the unconstrained system. Only their Poisson brackets are
independent, for example:
$$\Bigl\{\psi_a(t,{\vec x}),\psi_b(t,{\vec y})\Bigr\} = \int d^3u \thinspace
\int d^3v \thinspace {\delta \Psi\Bigl[{\widehat \psi}~\Bigr]_a(t,{\vec x})
\over \delta {\widehat \psi}_c({\vec u})} \thinspace \Bigl\{{\widehat \psi}_c(
{\vec u}), {\widehat \psi}_d({\vec v})\Bigr\} \thinspace {\delta \Psi\Bigl[{
\widehat \psi}~\Bigr]_b(t,{\vec y}) \over \delta {\widehat \psi}_d({\vec v})}
\eqno(3.10)$$
The equal time bracket algebra remains canonical because of the way time
evolution acts; in our notation, because of the special way the functionals
$\Psi\Bigl[{\widehat \psi}~\Bigr]_a(t,{\vec x})$ depend upon the initial value
configurations.

The constraint $(3.8)$ represents a relation between the
${\widehat \psi}_a({\vec
x})$'s. The theory is reduced by identifying a gauge condition on the
${\widehat
\psi}_a({\vec x})$'s that can be imposed by a unique, field dependent
transformation of the residual symmetry group. Together the constraint and this
gauge condition serve to eliminate a conjugate pair of the ${\widehat \psi}_a({
\vec x})$'s. This does not change the way time evolution makes the fields
depend
upon their initial configurations, it only fixes the values of a conjugate pair
of these initial configurations. The result is to change the equal time bracket
algebra $(3.10)$ since the Poisson brackets of the initial configurations are
replaced with Dirac brackets.

After reduction the full set of $\psi_a(t,{\vec x})$'s provides an overcomplete
description of the system at any time. Although it is always {\it possible} to
identify a minimal set of reduced variables whose equal time bracket algebra is
canonical, making such an identification is not necessarily easy and we often
settle for a set of reduced variables whose bracket algebra is not degenerate
but also not canonical. (If the bracket algebra becomes degenerate it means
that
the reduced variables do not provide a complete description of physics. That
is,
we cannot use the reduced set of $\psi_a$'s to recover the reduced ${\widehat
\psi}_a$'s which are the true dynamical degrees of freedom.) This set is
typically the evolution of whatever initial configurations remain after
reduction. For example, in reducing the coupled oscillator system of the last
section we eliminated ${\widehat q}_2$ and ${\widehat p}_2$, leaving ${\widehat
q}_1$ and ${\widehat p}_1$. The natural first choice for reduced variables was
accordingly $q_1(t)$ and $p_1(t)$. We then discovered that it would have been
better to choose either $q_1(t) + q_2(t)$ and $p_1(t) + p_2(t)$ or $q_1(t) -
q_2(t)$ and $p_1(t) - p_2(t)$ because these variables have a canonical equal
time bracket algebra while the original choice does not. We shall now witness a
similar phenomenon in reducing scalar electrodynamics.

A natural gauge condition compatible with $(3.8)$ is:
$$\partial_i \thinspace A_i(0,{\vec x}) = 0 \eqno(3.11)$$
{}From this and relation $(3.4a)$ we infer the following evolution for the
divergence of the vector potential:
$$\partial_i \thinspace A_i(t,{\vec x}) = - i e \thinspace \int_0^t ds
\thinspace \Bigl[\pi(s,{\vec x}) \thinspace \phi(s,{\vec x}) -
\pi^*(s,{\vec x})
\thinspace \phi^*(s,{\vec x})\Bigr] \eqno(3.12)$$
Of course $(3.11)$ does not completely fix the gauge; it is still possible to
make
time independent, harmonic transformations. This freedom can be eliminated with
a surface condition. A typical choice for the latter would be setting the
normal
component of $A_i(0,{\vec x})$ to zero on the ``surface at infinity,'' which
could realized as the limit of successively larger spheres. Rather than
burdening the formalism with a cumbersome limiting procedure we will instead
subsume the surface condition into the asymptotic fall-off usually assumed for
the gauge invariant part of the vector potential in order to make the magnetic
field energy finite.

There is a very close relation between fixing the harmonic gauge freedom and
dividing the vector potential --- and all other vector fields --- into
``transverse'' and ``longitudinal'' components. With the just-stated convention
the transverse and longitudinal parts of a vector field $f_i(t,{\vec x})$ which
obeys the asymptotic fall-off condition are is defined as follows:
$$f^T_i(t,{\vec x}) \equiv f_i(t,{\vec x}) - {\partial \over \partial x^i} \int
d^3y \thinspace G({\vec x};{\vec y}) \thinspace \partial_j f_j(t,{\vec y})
\eqno(3.13a)$$
$$f^L_i(t,{\vec x}) \equiv {\partial \over \partial x^i} \int d^3y \thinspace
G({\vec x};{\vec y}) \thinspace \partial_j f_j(t,{\vec y}) \eqno(3.13b)$$
where the Green's function is, $G({\vec x};{\vec y}) \equiv - \Bigl(4 \pi \Vert
{\vec x} - {\vec y} \Vert\Bigr)^{-1}$.

Since the longitudinal components of the electric field and the vector
potential
are entirely constrained it is natural to take the transverse components to be
among the reduced dynamical variables, along with the scalar field and its
conjugate momentum. Although the Dirac brackets of these variables are
initially
canonical:
$$\Bigl\{A^T_i(0,{\vec x}),E^T_j(0,{\vec y})\Bigr\} = \delta_{ij} \thinspace
\delta^3({\vec x} - {\vec y}) + \frac{\partial}{\partial x^i} \frac{\partial}{
\partial y^j} \thinspace G({\vec x};{\vec y}) \eqno(3.14a)$$
$$\Bigl\{\phi(0,{\vec x}),\pi(0,{\vec y})\Bigr\} = \delta^3({\vec x} -
{\vec y})
\eqno(3.14b)$$
$$\Bigl\{\phi^*(0,{\vec x}),\pi^*(0,{\vec y})\Bigr\} = \delta^3({\vec x} -
{\vec y}) \eqno(3.14c)$$
time evolution makes the equal time bracket algebra non-canonical. We cannot
exhibit the relation in closed form the way we did for the coupled oscillator
system of the previous section but we can evaluate the second time derivative
at
$t=0$ easily enough. This is because reduction affects only the initial value
fields; the time evolution equations are unchanged from $(3.4)$ and $(3.7)$.
We can
therefore use the evolution equations to reduce any number of time derivatives
of an initial bracket to brackets of the initial value fields, which can then
be
evaluated using $(3.14)$. For example, we compute:
$$\eqalignno{\Bigl(\frac{\partial}{\partial t}&\Bigr)^2 \thinspace \Bigl\{\phi(
t,{\vec x}),\pi(t,{\vec y})\Bigr\}\Bigl\vert_{t=0} = \cr
&= \frac{\partial}{\partial t} \thinspace \Biggl(\Bigl\{\pi^*(t,{\vec x}),
\pi(t,
{\vec y})\Bigr\} + \Bigl\{\phi(t,{\vec x}),(\partial_i - i e \thinspace A_i)^2
\thinspace \phi^*(t,{\vec y})\Bigr\}\Biggr)\Bigl\vert_{t=0} &(3.15a) \cr
&= \Bigl\{(\partial_i + i e \thinspace A_i)^2 \thinspace \phi(0,{\vec x}),
\pi(0,{\vec y})\Bigr\} + \Bigl\{\pi^*(0,{\vec x}),(\partial_i - i e \thinspace
A_i)^2 \thinspace \phi^*(0,{\vec y})\Bigr\} \cr
&+ \Bigl\{\pi^*(0,{\vec x}),(\partial_i - i e \thinspace A_i)^2 \thinspace
\phi^*(0,{\vec y})\Bigr\} + \Bigl\{\phi(0,{\vec x}),(\partial_i - i e
\thinspace
A_i)^2 \thinspace \pi(0,{\vec y})\Bigr\} &(3.15b) \cr
&+ \Bigl\{\phi(0,{\vec x}),- i e \thinspace E_i \thinspace (\partial_i - i e
\thinspace A_i) \thinspace \phi^*(0,{\vec y})\Bigr\} + \Bigl\{\phi(0,{\vec x}),
-i e \thinspace (\partial_i - i e \thinspace A_i) \thinspace E_i \thinspace
\phi^*(0,{\vec y})\Bigr\} \cr
&= -e^2 \thinspace {\partial G({\vec y};{\vec x}) \over \partial y^i}
\thinspace
{\widehat \phi}({\vec x}) \thinspace \Bigl[\frac{\partial}{\partial y^i} - i e
\thinspace {\widehat A}_i^T({\vec y})\Bigr] \thinspace {\widehat \phi}^*({\vec
y}) \cr
&\qquad \qquad \qquad -e^2 \thinspace \Bigl[\frac{\partial}{\partial y^i} - i e
\thinspace {\widehat A}_i^T({\vec y})\Bigr] \thinspace \Biggl\{ {\widehat
\phi}^*({\vec y}) \thinspace {\partial G({\vec y};{\vec x}) \over \partial y^i}
\Biggr\} \thinspace {\widehat \phi}({\vec x}) &(3.15c) \cr}$$
Of course the failure of the $\phi$-$\pi$ bracket to remain canonical is due to
the fact that reduction makes the longitudinal components of the vector
potential and the electric field depend upon the charge fields.

One consequence of $(3.15)$ is that the bracket matrix contains explicit time
dependence in addition to being non-canonical. To see this note that the
bracket
matrix must be time dependent because the second derivative of one of its
components fails to vanish. If this time dependence were exclusively implicit
--- that is, if it derived only from the bracket matrix's dependence upon the
time dependent dynamical variables --- then the initial matrix elements would
depend upon the reduced dynamical variables. Because they do not we infer that
the bracket matrix must harbor explicit as well as implicit time dependence.

We saw in the discussion associated with $(2.8)$ that explicit time dependence
in
the bracket matrix precludes the existence of a Hamiltonian which generates
time
evolution for the reduced dynamical variables. This is a little strange because
$(3.6)$ {\it is} the conserved, gauge invariant energy functional for scalar
electrodynamics. However, it is easy to verify that the change reduction
effects
in the longitudinal electric field prevents even the initial brackets from
agreeing with $(3.4)$ and $(3.7)$:
$$\eqalign{\Bigl\{\phi(0,{\vec x}),H\Bigr\} = \pi^*&(0,{\vec x}) \cr
&+ e^2 \thinspace \int d^3y \thinspace \Bigl[\pi(0,{\vec y}) \thinspace
\phi(0,{\vec y}) - \pi^*(0,{\vec y}) \thinspace \phi^*(0,{\vec y})\Bigr]
G({\vec y};{\vec x}) \thinspace \phi(0,{\vec x}) \cr} \eqno(3.16a)$$
$$\eqalign{\Bigl\{\pi(0,{\vec x}),H\Bigr\} =\Bigl[&\frac{\partial}
{\partial x^i}
- i e \thinspace A_i^T(0,{\vec x})\Bigr] \thinspace \Bigl[\frac{\partial}{
\partial x^i} - i e \thinspace A_i^T(0,{\vec x})\Bigr] \thinspace \phi^*(0,
{\vec x}) \cr
&- e^2 \thinspace \int d^3y \thinspace \Bigl[\pi(0,{\vec y}) \thinspace
\phi(0,{\vec y}) - \pi^*(0,{\vec y}) \thinspace \phi^*(0,{\vec y})\Bigr]
G({\vec y};{\vec x}) \thinspace \pi(0,{\vec x}) \cr} \eqno(3.16b)$$
Note again that our result is stronger than just that $(3.6)$ fails to generate
time evolution: there does not {\it exist} a Hamiltonian which generates time
evolution for these reduced variables.

As was explained in the last section, the problem with finding a Hamiltonian
that generates time evolution derives from an unfortunate choice of the reduced
variables. We emphasize that there is nothing dynamically wrong with the choice
we made. It is the natural one, and it does provide a complete and minimal
description of the physics of scalar electrodynamics. Our choice is not even
particularly inconvenient for computation; we will see in section 6 that it can
be written in the functional formalism almost as simply as any other choice.
The
only problem comes if we insist on an explicit operator formalism in which time
evolution is generated by a Hamiltonian. The cure for this problem is a
necessarily time dependent field redefinition to canonical variables.

There are many sets of canonical variables but the simplest is surely that
which
is related to our set of reduced variables through a time dependent gauge
transformation. The gauge parameter is:
$$\theta(t,{\vec x}) = \int_0^t ds \thinspace \alpha_0(s,{\vec x})
\eqno(3.17a)$$
$$\alpha_0(t,{\vec x}) \equiv i e \thinspace \int d^3y \thinspace G({\vec x};
{\vec y}) \thinspace \Bigl\{\pi(t,{\vec y}) \thinspace \phi(t,{\vec y}) -
\pi^*(t,{\vec y}) \thinspace \phi^*(t,{\vec y})\Bigr\} \eqno(3.17b)$$
We will adopt the convention that the Greek or Latin letter which denotes one
of
the old variables goes over in the new variables to the corresponding Latin or
Greek letter respectively. The new variables are:
$$\alpha^T_i(t,{\vec x}) \equiv A^T_i(t,{\vec x}) \eqno(3.18a)$$
$$\epsilon^T_i(t,{\vec x}) \equiv E^T_i(t,{\vec x}) \eqno(3.18b)$$
$$f(t,{\vec x}) \equiv \exp\Bigl[i e \thinspace \theta(t,{\vec x})\Bigr]
\thinspace \phi(t,{\vec x}) \eqno(3.18c)$$
$$p(t,{\vec x}) \equiv \exp\Bigl[- i e \thinspace \theta(t,{\vec x})\Bigr]
\thinspace \Bigl\{\pi(t,{\vec x}) - i e \thinspace \alpha_0(t,{\vec x})
\thinspace \phi^*(t,{\vec x})\Bigr\} \eqno(3.18d)$$
$$f^*(t,{\vec x}) \equiv \exp\Bigl[-i e \thinspace \theta(t,{\vec x})\Bigr]
\thinspace \phi^*(t,{\vec x}) \eqno(3.18e)$$
$$p^*(t,{\vec x}) \equiv \exp\Bigl[i e \thinspace \theta(t,{\vec x})\Bigr]
\thinspace \Bigl\{\pi^*(t,{\vec x}) + i e \thinspace \alpha_0(t,{\vec x})
\thinspace \phi(t,{\vec x})\Bigr\} \eqno(3.18f)$$
Since the quantity $\alpha_0(t,{\vec x})$ is a gauge invariant it has the same
form in terms of $f$ and $p$ as $\phi$ and $\pi$:
$$\alpha_0(t,{\vec x}) = i e \int d^3y \thinspace G({\vec x};{\vec y})
\thinspace \Bigl\{p(t,{\vec y}) \thinspace f(t,{\vec y}) - p^*(t,{\vec y})
\thinspace f^*(t,{\vec y})\Bigr\} \eqno(3.19)$$
By differentiating these relations and then using $(3.4)$ and $(3.7)$ we infer
the
following evolution equations:
$${\dot \alpha}^T_i(t,{\vec x}) = \epsilon^T_i(t,{\vec x}) \eqno(3.20a)$$
$$\eqalign{{\dot \epsilon}^T_i(t,{\vec x}) = \frac{\partial}{\partial x^j}
\frac{\partial}{\partial x^j} \thinspace &\alpha^T_i(t,{\vec x}) + i e
\thinspace f^*(t,{\vec x}) \thinspace \Bigl[\partial_i + i e \thinspace
\alpha^T_i(t,{\vec x})\Bigr] \thinspace f(t,{\vec x}) \cr
&- i e \thinspace f(t,{\vec x}) \thinspace \Bigl[\partial_i - i e \thinspace
\alpha^T_i(t,{\vec x})\Bigr] \thinspace f^*(t,{\vec x}) \cr} \eqno(3.20b)$$
$${\dot f}(t,{\vec x}) = \pi^*(t,{\vec x}) - i e \thinspace \alpha_0(t,
{\vec x})
\thinspace f(t,{\vec x}) \eqno(3.20c)$$
$${\dot p}(t,{\vec x}) = i e \thinspace \alpha_0(t,{\vec x}) \thinspace p(t,
{\vec x}) + \Bigl[\frac{\partial}{\partial x^i} - i e \thinspace \alpha^T_i(t,
{\vec x})\Bigr] \thinspace \Bigl[\frac{\partial}{\partial x^i} - i e \thinspace
\alpha^T_i(t,{\vec x})\Bigr] \thinspace f^*(t,{\vec x}) \eqno(3.20d)$$
$${\dot f}^*(t,{\vec x}) = \pi(t,{\vec x}) + i e \thinspace \alpha_0(t,
{\vec x})
\thinspace f^*(t,{\vec x}) \eqno(3.20e)$$
$${\dot p}^*(t,{\vec x}) = - i e \thinspace \alpha_0(t,{\vec x}) \thinspace
p^*(t,{\vec x}) + \Bigl[\frac{\partial}{\partial x^i} + i e \thinspace
\alpha^T_i(t,{\vec x})\Bigr] \thinspace \Bigl[\frac{\partial}{\partial x^i} +
i e \thinspace \alpha^T_i(t,{\vec x})\Bigr] \thinspace f(t,{\vec x})
\eqno(3.20f)$$
Of course these relations are generated by the Hamiltonian $(3.6)$, which in
the
new variables has the form:
$$H = \int d^3x \thinspace \Bigl\{\frac12 \epsilon^T_i \thinspace \epsilon^T_i
+ \frac12 \partial_i \alpha^T_j \thinspace \partial_i \alpha^T_j + \frac12
\partial_i \alpha_0 \thinspace \partial_i \alpha_0 + p^* \thinspace p + \Bigl(
\partial_i - i e \thinspace \alpha^T_i\Bigr) \thinspace f^* \thinspace \Bigl(
\partial_i + i e \thinspace \alpha^T_i\Bigr) \thinspace f\Bigr\} \eqno(3.21)$$
The reason the old Hamiltonian can generate evolution for these variables and
not for our first choice of reduced dynamical variables is that the field
redefinition contains explicit time dependence through the time integration in
$(3.17a)$.

It should now be apparent --- and it can be, with some difficulty, checked
using
relations $(3.18)$ --- that the bracket algebra is canonical. That is, the
non-zero
equal time brackets are:
$$\Bigl\{\alpha^T_i(t,{\vec x}),\epsilon^T_j(t,{\vec y})\Bigr\} = \delta_{ij}
\thinspace \delta^3({\vec x} - {\vec y}) + \frac{\partial}{\partial x^i}
\frac{\partial}{\partial y^j} \thinspace G({\vec x};{\vec y}) \eqno(3.22a)$$
$$\Bigl\{f(t,{\vec x}),p(t,{\vec y})\Bigr\} = \delta^3({\vec x} - {\vec y})
\eqno(3.22b)$$
$$\Bigl\{f^*(t,{\vec x}),p^*(t,{\vec y})\Bigr\} = \delta^3({\vec x} - {\vec y})
\eqno(3.22c)$$
It should also be apparent that the canonical formalism we have constructed is
just that which follows from the invariant action $(3.1)$ by the imposition of
Coulomb gauge:
$$\partial_i A_i(t,{\vec x}) = 0 \eqno(3.23)$$
Relation $(3.19)$ comes from solving the constraint equation:
$$\partial_i \partial_i A_0(t,{\vec x}) = i e \thinspace \Bigl[{\dot \phi}^*(t,
{\vec x}) \thinspace \phi(t,{\vec x}) - {\dot \phi}(t,{\vec x}) \thinspace
\phi^*(t,{\vec x})\Bigr] \eqno(3.24)$$
subject to a surface condition at spatial infinity which fixes the freedom to
perform time dependent, harmonic gauge transformations.

\vskip 1cm
\centerline{\bf 4. Perturbative Gravity Around Flat Space On $T^3 \times R$}

\def \o#1{{\cal{O}}(\kappa^{#1})}

This section is divided into four parts. In the first we describe the canonical
formalism for gravity in a general closed spatial manifold. The second part
introduces the mode and tensor decompositions we shall use for $T^3 \times R$.
In the third part we apply this mode decomposition to perturbation theory
around
flat space. It is here that we impose the constraints and fix the gauge to
obtain the reduced theory. In the final part we obtain a reduced canonical
formalism and we show that our result agrees in the limit of infinite toroidal
radius and localized initial value data with that obtained by A.D.M. [3] for
open, asymptotically flat space.

\item{(1)}--- Description of Canonical Formalism.

We define the lapse $N^0$ and the shift $N^i$ via the invariant interval:
$$ds^2\,=\,-\,\left(N^0\right)^2\,dt^2\,+\,\gamma_{ij}\,
\left(\,dx^i\,+\,N^i\,dt\,\right)\,\left(\,dx^j\,+\,N^j\,dt\,\right) \;.
\eqno(4.1)$$
This implies the $4$--metric $g_{\mu\nu}$ and its inverse $g^{\mu\nu}$ are:
$$\eqalignno{
g_{\mu\nu}&=\pmatrix{-\,\left(N^0\right)^2\,+\,N^k\,N^l\,\gamma_{kl}\;&\;
N^k\,\gamma_{kj} \cr\cr \gamma_{ik}\,N^k&\gamma_{ij}} &(4.2) \cr\cr\cr
g^{\mu\nu}&={1 \over \left(N^0\right)^2}\,\pmatrix{-\,1&N^j\cr\cr
N^i\;&\,\left(N^0\right)^2\,\gamma^{ij}\,-\,N^i\,N^j} \;. &(4.3)}$$

\noindent The usual Hilbert action for gravity
$$\eqalignno{
S\,&=\,\int d^4x \left[ {1 \over \kappa^2}\,R\,\sqrt{-\,g} \right] &(4.4a) \cr
\noalign{\hbox{can be written in canonical form as:}}
S\,&=\,\int d^4x \left[ -\,\dot \pi^{ij}\,\gamma_{ij}\,-\,N^\mu\,{\cal H}_\mu
\right] \;. &(4.4b)}$$
An integration by parts was used to arrive at $(4.4b)$ from $(4.4a)$ and
the following definitions were used:
$$\eqalignno{
\pi^{ij}\,&\equiv\,{\sqrt{\gamma} \over 2 N^0 \kappa^2}\,
\left(\,\gamma^{ik}\,\gamma^{jl}\,-\,\gamma^{ij}\,\gamma^{kl}\,\right)
\,\left(\dot\gamma_{kl}\,-\,N_{k;l}\,-\,N_{l;k}\right) &(4.5a) \cr\cr
{\cal H}_0\,&\equiv\,{\kappa^2 \over \sqrt{\gamma}}\,\left(
\,\gamma_{ik}\,\gamma_{jl}\,-\,\frac12\,\gamma_{ij}\,\gamma_{kl}\,\right)
\,\pi^{ij}\,\pi^{kl}\,-\,{\sqrt{\gamma} \over \kappa^2}\,{\cal R} &(4.5b) \cr
\cr{\cal H}_i\,&\equiv\,-\,2\,\gamma_{ij}\,\pi^{jl}_{~~;l} \;. &(4.5c) }$$
In the previous expressions a semicolon indicates covariant differentiation on
the spatial sections using the connection compatible with the $3$--metric,
$\gamma_{ij}$; ${\cal R}$ is the Ricci scalar formed from $\gamma_{ij}$.

In these variables the Hamiltonian is,
$$
H\,=\,\int d^3x \,N^\mu\,{\cal H}_\mu\eqno(4.6)
$$
Variations of it with respect to $\pi^{ij}$ and $\gamma_{ij}$ give us the
evolution equations,
$$
\eqalignno{
\dot \gamma_{ij}\,=&\,{2\,\kappa^2 \over \sqrt{\gamma}}\,N^0\,
\left(\,\pi_{ij}\,-\,\frac12 \, \gamma_{ij}\,\pi\,\right)\,+\,
N_{i;j}\,+\,N_{j;i} &(4.7a)\cr\cr
\dot \pi^{ij}\,=&\,-\,{\sqrt{\gamma} \over \kappa^2}
\,N^0\,\left({\cal R}^{ij}\,-\,\frac12\,
\gamma^{ij}\,{\cal R}\,\right)\,+\,{ \kappa^2 \over 2\,\sqrt{\gamma}}
\,N^0 \,\gamma^{ij}\,\left(\,\pi^{lm}\,\pi_{lm}\,-\,\frac12\,\pi^2\right)\,
\cr\cr
&\,-\,{2\,\kappa^2 \over \sqrt{\gamma}}\,N^0 \,\left(\,\pi^{il}\,\pi_l^{~j}\,
-\,\frac12\,\pi\,\pi^{ij}\,\right)\,+\,{\sqrt{\gamma} \over\kappa^2}\,
\left(N^{0;ij}\,-\,\gamma^{ij}\,{N^{0;l}}_{l}\,\right)\cr\cr&\,+\,
\left(\,\pi^{ij}\,N^l\,\right)_{;l}\,-\,N^i_{~;l}\,\pi^{lj}\,
-\,N^j_{~;l}\,\pi^{li} &(4.7b)}
$$
\noindent while variation with respect to $N^\mu$ gives the constraint
equation,
$${\cal H}_\mu\,=\,0 \eqno(4.7c)$$
Of course $(4.7a)$ is just a restatement of the definition $(4.5a)$ of the
conjugate
momentum. Relations $(4.7b)$ are canonical versions of the six $g_{ij}$
Euler-Lagrange equations; the constraints $(4.7c)$ are linear combinations of
the
four $g_{\mu 0}$ equations.

We imagine the volume gauge to have been fixed by specifying the lapse and
shift, possibly as functionals of the 3-metric and its conjugate momentum. Such
a gauge condition eliminates the ability to perform diffeomorphisms which are
locally time dependent, as witness the fact that the Cauchy problem has a
unique
solution for fixed (and non-degenerate) lapse and shift. Just as with temporal
gauge in scalar electrodynamics, our gravitational gauge leaves a residual
symmetry of transformations which are completely characterized by their action
on the initial value surface, and by the condition that they do not affect the
lapse and shift.

Suppose we represent a general infinitesimal diffeomorphism, $x^{\mu} \mapsto
x^{\mu} + \theta^{\mu}(x)$, using the parameter $\theta^{\mu}(x)$. It is a
simple exercise to show that the 4-metric is changed by the following amount:
$$\delta_{\theta} \Bigl(g^{\mu\nu}\Bigr) = \theta^{\mu}_{~,\rho} \thinspace
g^{\rho \nu} + \theta^{\nu}_{~,\rho} \thinspace g^{\mu \rho} - g^{\mu \nu}_{~~,
\rho} \thinspace \theta^{\rho} \;. \eqno(4.8a)$$
$$\delta_{\theta} \Bigl(g_{\mu\nu}\Bigr) = - \theta^{\rho}_{~,\mu} \thinspace
g_{\rho \nu} - \theta^{\rho}_{~,\nu} \thinspace g_{\mu \rho} - g_{\mu \nu ,
\rho} \thinspace \theta^{\rho} \;. \eqno(4.8b)$$
By requiring that $\delta_{\Theta} g^{\mu 0} = 0$ we see that the residual
transformations, $\Theta\Bigl[{\widehat \theta}\Bigr](t,{\vec x})$, are
characterized by their initial values, ${\widehat \theta}^{\mu}({\vec x})$, and
by the following evolution equations:
$${\dot \Theta}^0 = \Theta^0_{~,j} \thinspace N^j - {N^0_{~,\rho} \over N^0}
\thinspace \Theta^{\rho} \eqno(4.9a)$$
$${\dot \Theta}^i = \Theta^i_{~,j} \thinspace N^j - \Theta^0_{~,j} \thinspace
\gamma^{ji} \thinspace \Bigl(N^0\Bigr)^2 + {N^0_{~,\rho} \over N^0} \thinspace
\Theta^{\rho} \thinspace N^i - N^i_{~,\rho} \thinspace \Theta^{\rho} \;.
\eqno(4.9b)$$
As with temporal gauge scalar electromagnetism, the constraints generate
residual symmetry transformations. That is, if we define,
$${\cal H}\left[{\widehat \theta}\right]\,\equiv\,\int d^3x ~ \Bigl\{{\widehat
\theta}^{~0}({\vec x}) \thinspace N^\mu(0,{\vec x}) \thinspace {\cal H}_\mu(0,
{\vec x}) + {\widehat \theta~}^i({\vec x}) \thinspace {\cal H}_i(0,{\vec x})
\Bigr\} \;, \eqno(4.10)$$
then explicit calculation shows that
$$\eqalignno{\Bigl\{{\widehat \gamma_{ij}}, {\cal H}\Bigl[{\widehat \theta}
\Bigr]\Bigr\} &= {\widehat \theta}^{~k}_{~~;i} \thinspace
{\widehat \gamma}_{kj}
+ {\widehat \theta}^{~k}_{~~;j} \thinspace {\widehat \gamma}_{ik} + {\dot
{\widehat \gamma}_{ij}} \thinspace {\widehat \theta}^{~0} + {\widehat \theta}^{
{}~0}_{~~,i} \thinspace {\widehat \gamma}_{jk} \thinspace {\widehat N}^k +
{\widehat \theta}^{~0}_{~~,j} \thinspace {\widehat \gamma}_{ik} \thinspace
{\widehat N}^k  &(4.11a) \cr
&= - \delta_{\Theta[{\widehat \theta}~]} \Bigl({\widehat \gamma}_{ij}\Bigr)
&(4.11b) \cr}$$
$$\eqalignno{\Bigl\{{\widehat \pi^{ij}} , {\cal H}\Bigl[{\widehat \theta}\Bigr]
\Bigr\} &= {\dot {\widehat \pi}}^{ij} \thinspace {\widehat \theta}^{~0} +
{\sqrt{\widehat \gamma} \over \kappa^2} \thinspace \Bigl({\widehat \theta}^{~0;
ij} \thinspace {\widehat N}^0 + {\widehat \theta}^{~0,i} \thinspace {\widehat
N}^{0,j} + {\widehat \theta}^{~0,j} \thinspace {\widehat N}^{0,i}\Bigr) \cr
&-{\sqrt{\widehat \gamma} \over \kappa^2} \thinspace {\widehat \gamma}^{ij}
\thinspace \Bigl({\widehat \theta}^{~0;k}_{~~~~k} \thinspace {\widehat N}^{0}
+ 2 {\widehat \theta}^{~0,k} \thinspace {\widehat N}^0_{~~,k}\Bigr) - \Bigl(
{\widehat N}^i \thinspace {\widehat \pi}^{~kj} + {\widehat N}^j \thinspace
{\widehat \pi}^{ik}\Bigr) \thinspace {\widehat \theta}^{~0}_{~~,k} \cr
& + {\widehat \pi}^{~ij} \thinspace {\widehat N}^k\thinspace {\widehat \theta}^
{~0}_{~~,k} - {\widehat \theta}^{~i}_{~~;k} \thinspace {\widehat \pi}^{~kj} -
{\widehat \theta}^{~j}_{~~;k} \thinspace {\widehat \pi}^{~ik} + ({\widehat
\pi}^{~ij} \thinspace {\widehat \theta}^{~k})_{;k} &(4.12a) \cr
&= - \delta_{\Theta[{\widehat \theta}~]} \Bigl({\widehat \pi}^{ij}\Bigr) \;.
&(4.12b) \cr}$$
Note that this is not a definition. The left hand sides of $(4.11)$ and
$(4.12)$
are defined by $(4.10)$, $(4.5b)$ and $(4.5c)$, while the identifications on
the right hand side are made by applying $(4.8)$ to the canonical coordinates
and
taking any time derivatives from $(4.9)$.

\item{(2)}--- Mode Decomposition on $T^3 \times R$.

Now that the canonical formalism for a general space with closed spatial
sections has been described it will be specialized to the treatment of $T^3
\times R$. The coordinate ranges are $t \in R$ and $0\le x^i < L$. The points
$x^i=0$ and $x^i=L$ are identified. Any function $f(t,{\vec x})$ can be
decomposed in modes in the following way:
$$f(t{,}\vec x)\,=\,L^{-3/2}\, \sum\limits^\infty_{n_1=-\infty}
\,\sum\limits^\infty_{n_2=-\infty}\,\sum\limits^\infty_{n_3=-\infty}\,
\exp\Bigl[i \frac{2 \pi}{L} {\vec n} \cdot {\vec x}\Bigr] \thinspace
{\widetilde f}(t{,}\vec n) \eqno(4.13a)$$
$${\widetilde f}(t,{\vec n}) \equiv (2 \pi)^{-3} \thinspace L^{-3/2} \thinspace
\int_0^L dx_1 \int_0^L dx_2 \int_0^L dx_3 \thinspace \exp\Bigl[-i \frac{2 \pi}{
L} {\vec n} \cdot {\vec x}\Bigr] \thinspace f(t{,}\vec x) \eqno(4.13b)$$
Note that when $f(t,{\vec x})$ is real we have $\widetilde f^*(t{,}\vec n)\,=\,
\widetilde f(t{,}-\vec n)$.

In representing tensors such as $\gamma_{ij}$ and $\pi^{ij}$ it is convenient
to decompose the index structure in a way that depends on the mode number. Let
us define the 3-momentum, the transverse projection operator and the
longitudinal inversion operator as follows:
$$\vec k \equiv {2\pi \over L} \vec n \eqno(4.14a)$$
$$T_{ij} \equiv \delta_{ij}\,-\,{k_i\,k_j \over k^2} \eqno(4.14b)$$
$$L_{ij} \equiv \delta_{ij}\,-\,{k_i\,k_j \over 2 k^2} \eqno(4.14c)$$
For ${\vec k} \neq 0$ one can decompose any symmetric $2$--tensor into three
component pieces:
$$\widetilde f_{j k}\,=\,\widetilde f^{~tt}_{j k}\,+\,\widetilde f^{~t}_{j k}\,
+\, i \Bigl( \widetilde f_j\,k_k\,+\, \widetilde f_k\,k_j\Bigr) \eqno(4.15a)$$
$$\widetilde f^{~tt}_{ij}\,\equiv \,\left(\,T_{ik}\,T_{jl}\,- \,\frac12
T_{ij}\,T_{kl} \,\right)\,\widetilde f_{kl} \eqno(4.15b)$$
$$\widetilde f^{~t}_{ij}\,\equiv \, \frac12 \, T_{ij} \, T_{kl} \,
\widetilde f_{kl} \equiv \,\frac12\,T_{ij}\, {\widetilde f}^{~t} \eqno(4.15c)$$
$$\widetilde f_j\,\equiv \, -\,{i \over k^2}\,L_{jk} \, \widetilde f_{j k}
\eqno(4.15d)$$
Note that for each ${\vec k} \neq 0$ there are two independent transverse
traceless components ${\widetilde f}^{~tt}_{ij}$, three longitudinal components
${\widetilde f}_i$, and one independent transverse component
${\widetilde f}^{~t
}$. Of course for $\vec k = 0$ all components satisfy the transversality
condition. We therefore decompose the zero mode tensor into five transverse
traceless components and one trace:
$$\widetilde f_{ij}(t,0)\,=\,\widetilde f^{~tt}_{ij}(t,0)\,+\,{1 \over 3}\,
\delta_{ij}\widetilde f^{~tr}(t) \eqno(4.16)$$
We can carry the decomposition over into position space through the inverse
transform as follows:
$$f_{ij}\,=\,\frac13\,f^{tr}\,\delta_{ij}\,+\,f^{tt}_{ij}\,+\,f^t_{ij}\,+\,
\left(\,f_{i,j}\,+\,f_{j,i}\,\right)\eqno(4.17a)$$
$$f^{tr}(t) \equiv L^{-3/2} \thinspace {\widetilde f}^{~tr}(t) \eqno(4.17b)$$
$$f^{tt}_{ij}(t,{\vec x}) \equiv L^{-3/2}\, \sum_{{\vec n}} \exp\Bigl[i
\frac{2 \pi}{
L} {\vec n} \cdot {\vec x}\Bigr] \thinspace \widetilde f^{~tt}_{ij}(t{,}\vec n)
\eqno(4.17c)$$
$$f^{t}_{ij}(t,{\vec x}) \equiv L^{-3/2}\, \sum_{{\vec n} \neq 0} \exp\Bigl[i
\frac{2\pi}{L} {\vec n} \cdot {\vec x}\Bigr] \thinspace \widetilde
f^{~t}_{ij}(t{,}\vec n) \eqno(4.17d)$$
$$f_i(t,{\vec x}) \equiv L^{-3/2}\, \sum_{{\vec n} \neq 0} \exp\Bigl[i \frac{2
\pi}{L} {\vec n} \cdot {\vec x}\Bigr] \thinspace \widetilde f_i(t{,}\vec n)
\eqno(4.17e)$$
Note that the longitudinal and transverse components contain no spatial zero
modes while $f^{tr}$ is {\it all} zero mode. The transverse traceless
components alone contain both zero and non-zero modes.

\item{(3)} --- Perturbing Around Flat Space

Since $\eta_{\mu\nu}$ is a solution of Einstein's equations in $T^3 \times R$
we
can perturb around flat space, $g_{\mu\nu}\,=\,\eta_{\mu\nu}\,+\,\kappa\,h_{\mu
\nu}$. (We define the constant $\kappa^2 \equiv 16 \pi G$.) The corresponding
expansions for the various canonical variables are:
$$\eqalignno{\gamma_{ij}\,=&\,\delta_{ij}\,+\,\kappa\,h_{ij}&(4.18a)\cr
\pi^{ij}\,=&\,{1 \over \kappa}\,p^{ij}&(4.18b)\cr
N^0\,=&\,1\,+\,\kappa\,n^0 &(4.18c)\cr
N^i\,=&\,0\,+\,\kappa\,n^i &(4.18d)\cr}$$
We refer to $h_{ij}$, $p^{ij}$, $n^0$ and $n^i$ collectively as the {\it weak
fields}. By convention the background metric is used to raise and lower indices
on the weak fields. Since the background metric in this case is
$\eta_{\mu \nu}$
it is irrelevant whether the spatial indices of weak fields are up or down, and
raising a temporal index merely flips the sign. Note that the placement of
$\kappa$'s in $(4.18a)$ and $(4.18b)$ implies that $h_{ij}$ and $p^{ij}$ have
the
same bracket or commutation relations as $\gamma_{ij}$ and $\pi^{ij}$.

If we expand the equations of time evolution, $(4.7a)$ and $(4.7b)$, and then
segregate according to tensor components, the following equations result:
$${\dot h}^{tt}_{ij} = 2 p^{tt}_{ij} + {\cal O}(\kappa) \eqno(4.19a)$$
$${\dot p}^{tt}_{ij} = \frac12 \nabla^2 \thinspace h^{tt}_{ij} + {\cal O}(
\kappa) \eqno(4.19b)$$
$${\dot h}^t = 0 + {\cal O}(\kappa) \eqno(4.20a)$$
$${\dot p}^t = - 2 \nabla^2 \thinspace n^0 + {\cal O}(\kappa) \eqno(4.20b)$$
$$\nabla^2 \thinspace {\dot h}_i + {\dot h}_{j,ji} = - 2 p^t_{~,i} + \nabla^2
\thinspace n_i + n_{j,ji} + {\cal O}(\kappa) \eqno(4.21a)$$
$$\nabla^2 \thinspace {\dot p}_i + {\dot p}_{j,ji} = 0 + {\cal O}(\kappa)
\eqno(4.21b)$$
$${\dot h}^{tr} = - p^{tr} + {\cal O}(\kappa) \eqno(4.22a)$$
$${\dot p}^{tr} = 0 + {\cal O}(\kappa) \eqno(4.22b)$$
In these relations we have implicitly regarded the various weak fields as being
of order one. This is not really correct because not all the fields are
independent. Even in a theory without local symmetries we could use the
equations of time evolution to express the weak fields at any time as
functionals of the initial weak fields. It is traditional in this case to
develop perturbative solutions as though the initial value configurations are
of
order one in the coupling constant. The scheme is more complicated in a theory
which possesses local symmetries because then one must impose a volume gauge
condition in order to define a canonical formalism. Further, the canonical
formalism so obtained possesses a set of constraints upon the initial value
configurations and also, typically, a local but time independent residual
symmetry. This residual symmetry is fixed by imposing gauge conditions on the
initial weak field configurations. In our case we shall find it convenient to
imagine that the surface gauge conditions are of order one, but we shall allow
for the possibility of higher order terms in the volume gauge conditions. The
constraints are solved perturbatively on the initial value surface to express
the initial values of the constrained fields as power series expansions in
functionals of the initial values of the unconstrained fields, regarding the
latter as of order one. One then solves the perturbative equations of time
evolution as for a theory without constraints but remembering that not all the
initial configurations are of order one, and that the volume gauge conditions
may also supply higher order terms.

The four constraints can be expanded as follows in powers of the weak fields:
$$\eqalignno{{\cal H}_0 &= \frac1{\kappa} \Bigl(h_{,ii} - h_{ij,ij}\Bigr) +
\Bigl(\frac12 h h_{,k} - \frac12 h h_{jk,j} - h_{,j} h_{jk} - h_{ij} h_{ij,k}
+ h_{ki} h_{ij,j} + h_{ij} h_{kj,i}\Bigr)_{,k} \cr
&+ \Bigl(p_{ij} p_{ij} - \frac12 p^2\Bigr) + \Bigl(-\frac14 h_{,i} h_{,i} +
\frac12 h_{,i} h_{ij,j} + \frac14 h_{ij,k} h_{ij,k} - \frac12 h_{ij,k} h_{kj,i}
\Bigr) + {\cal O}(\kappa) \qquad &(4.23a) \cr}$$
$${\cal H}_i = -\frac2{\kappa} \thinspace p_{ij,j} - 2 \Bigl(h_{ij} p_{jk}
\Bigr)_{,k} + h_{jk,i} \thinspace p_{jk} + {\cal O}(\kappa) \eqno(4.23b)$$
Substitution of the tensor decomposition $(4.17)$ reveals that the ${\cal H}_0$
constraint determines the weak field $h^t$:
$$\nabla^2 h^t = \kappa \thinspace {\cal Q}_0\Bigl[h^{tt},p^{tt};h^t,p^t;h,p;
h^{tr},p^{tr}\Bigr] \eqno(4.24a)$$
$$\eqalignno{{\cal Q}_0 \equiv \Bigl(-\frac12 h h_{,k} + \frac12 h &h_{jk,j} +
h_{,j} h_{jk} + h_{ij} h_{ij,k} - h_{ki} h_{ij,j}- h_{ij} h_{kj,i}\Bigr)_{,k}
+ \Bigl(-p_{ij} p_{ij} + \frac12 p^2\Bigr) \cr
&+ \Bigl(\frac14 h_{,i} h_{,i} - \frac12 h_{,i} h_{ij,j} - \frac14 h_{ij,k}
h_{ij,k} + \frac12 h_{ij,k} h_{kj,i}\Bigr) + {\cal O}(\kappa) &(4.24b) \cr}$$
Similarly, the ${\cal H}_i$ constraint gives an equation for the weak field
$p_i$:
$$\nabla^2 p_i + p_{j,ji} = \kappa \thinspace {\cal Q}_i
\Bigl[h^{tt},p^{tt};h^t,
p^t;h,p;h^{tr},p^{tr}\Bigr] \eqno(4.25a)$$
$${\cal Q}_i \equiv - \Bigl(h_{ij} p_{jk}\Bigr)_{,k} + \frac12 h_{jk,i} p_{jk}
+ {\cal O}(\kappa) \eqno(4.25b)$$
We can solve perturbatively for $h^t$ and $p_i$ because these weak field
components contain no zero modes and the Laplacian is therefore a negative
definite operator. However, we must first subtract off the zero mode parts of
the sources ${\cal Q}_{\mu}$. For any function $f(t,{\vec x})$ we define its
non-zero mode part as:
$$f^{\rm NZ}(t,{\vec x}) \equiv f(t,{\vec x}) - L^{-3} \int_0^L dy_1 \int_0^L
dy_2 \int_0^L dy_3 \thinspace f(t,{\vec y}) \eqno(4.26)$$
To solve for $h^t$ and $p_i$ one simply inverts the Laplacian on the non-zero
mode sectors of $(4.24a)$ and $(4.25a)$:
$$h^t = {\kappa \over \nabla^2} \thinspace {\cal Q}_0^{\rm NZ}\Bigl[h^{tt},
p^{tt};h^t,p^t;h,p;h^{tr},p^{tr}\Bigr] \eqno(4.27a)$$
$$p_i = {\kappa \over \nabla^2} \thinspace L_{ij} \thinspace
{\cal Q}_j^{\rm NZ}
\Bigl[h^{tt},p^{tt};h^t,p^t;h,p;h^{tr},p^{tr}\Bigr] \eqno(4.27b)$$
and then substitutes the resulting equations to re-express any $h^t$'s or
$p_i$'s which appear in the sources. For example the first iteration gives:
$$h^t = {\kappa \over \nabla^2} \thinspace {\cal Q}_0^{\rm NZ}\Bigl[h^{tt},
p^{tt};\frac{\kappa}{\nabla^2}{\cal Q}_0^{\rm NZ},p^t;h,\frac{\kappa}{\nabla^2}
L_{ij}{\cal Q}_j^{\rm NZ};h^{tr},p^{tr}\Bigr] \eqno(4.28a)$$
$$p_i = {\kappa \over \nabla^2} \thinspace L_{ij} \thinspace
{\cal Q}_j^{\rm NZ}
\Bigl[h^{tt},p^{tt};\frac{\kappa}{\nabla^2}{\cal Q}_0^{\rm NZ},p^t;h,\frac{
\kappa}{\nabla^2}L_{kl} {\cal Q}_l^{\rm NZ};h^{tr},p^{tr}\Bigr] \eqno(4.28b)$$
Of course there are still $h^t$'s and $p_i$'s inside the new sources --- though
space presents us from displaying it explicitly --- but whereas these fields
might appear at order $\kappa$ on the right hand side of $(4.27)$ they cannot
appear before order $\kappa^2$ on the right hand side of $(4.28)$. Because each
iteration moves them to a higher order in $\kappa$ we can obtain in this way an
asymptotic series solution as a functional of $h^{tt}$, $p^{tt}$, $p^t$, $h_i$,
$h^{tr}$ and $p^{tr}$.

Although we have just seen that the constraints completely determine $h^t$ and
$p_i$ it is not quite true that constraining $h^t$ and $p_i$ completely
enforces
the constraints. There remain the zero modes. One can see by direct integration
that although the zero mode constraints are free of terms linear in the weak
fields they are not trivial at the next order, even when $h^t$ and $p_i$ are
set
to their constrained values:
$$\eqalignno{&\int d^3x \thinspace {\cal H}_0 = \int d^3x \thinspace{\cal Q}_0
&(4.29a) \cr
&= \int d^3x \thinspace \Biggl\{-\frac16 (p^{tr})^2 + p_{ij}^{tt} \thinspace
p_{ij}^{tt} + 2 p_{i,j} \thinspace p_{i,j} - 2 p_{i,i} \thinspace p^t + \frac14
h^{tt}_{ij,k} \thinspace h^{tt}_{ij,k} + \frac18 h^t_{,i} \thinspace h^t_{,i}
\Biggr\} + {\cal O}(\kappa) \qquad \qquad &(4.29b) \cr
&\mathop{\longrightarrow}\limits_{\scriptstyle h^t, p_i} - \frac16 L^3
\thinspace (p^{tr})^2 + \int d^3x \thinspace \Biggl\{p_{ij}^{tt} \thinspace
p_{ij}^{tt} + \frac14 h^{tt}_{ij,k} \thinspace h^{tt}_{ij,k}\Biggr\} + \kappa
\thinspace {\cal C}_0\Bigl[h^{tt},p^{tt};h,p^t;h^{tr},p^{tr}\Bigr]
&(4.29c) \cr}$$
$$\eqalignno{\int d^3 \thinspace {\cal H}_i &= \int d^3x \thinspace {\cal Q}_i
&(4.30a) \cr
&= \int d^3x \thinspace \Biggl\{p_{jk}^{tt} \thinspace h_{jk,i}^{tt} + \frac12
p^t \thinspace h^t_{,i} + p_{j,j} \thinspace h_{k,ki} - 2 p_{j,k} \thinspace
h_{j,ki}\Biggr\} + {\cal O}(\kappa) &(4.30b) \cr
&\mathop{\longrightarrow}\limits_{\scriptstyle h^t, p_i} \int d^3x \thinspace
p_{jk}^{tt} \thinspace h_{jk,i}^{tt} + \kappa \thinspace {\cal C}_i
\Bigl[h^{tt},p^{tt};h,p^t;h^{tr},p^{tr}\Bigr] &(4.30c) \cr}$$
(The functionals ${\cal C}_{\mu}\Bigl[h^{tt},p^{tt};h,p^t;h^{tr},p^{tr}\Bigr]$
are of
cubic order and higher in the remaining weak fields.) A consequence is that
there are solutions to the linearized field equations which can not be
perturbatively corrected to give asymptotic solutions to the full field
equations. This phenomenon is known as {\it linearization instability}, and it
afflicts gravitational perturbation theory whenever the background possesses
Killing vectors\footnote{*}{The number of constraints which lack linear terms
is equal to the number of Killing vectors. We have four because only the four
translations give global Killing vectors for flat space on $T^3 \times R$.
Lorentz transformations --- which also give Killing vectors for flat space on
$R^3 \times R$ --- do not respect the identification of $T^3 \times R$.} on a
spatially closed manifold [17,18,19].

The linearization instability is sometimes regarded as a non-trivial obstacle
to
the development of perturbation theory. This is not correct. We need merely to
restrict to those linearized solutions which satisfy the first non-trivial
parts
of the four zero mode constraints and then develop systematic corrections as
usual. Because our strategy is different for the global Hamiltonian constraint
$(4.29)$ than for the global momentum constraint $(4.30)$ we shall discuss them
separately.

At quadratic order in the remaining independent weak fields the global
Hamiltonian constraint is the difference of two manifestly positive quantities.
This means we can solve it explicitly as follows:
$$p^{tr} = \pm \sqrt{6} L^{-3/2} \thinspace \Biggl[\int d^3x \thinspace \Bigl(
p_{ij}^{tt} p_{ij}^{tt} + \frac14 h_{ij,k}^{tt} h_{ij,k}^{tt}\Bigr) + \kappa
\thinspace {\cal C}_0\Bigl[h^{tt},p^{tt};h,p^t;h^{tr},p^{tr}\Bigr]\Biggr]^{
\frac12} \eqno(4.31)$$
The issue of choosing the sign in $(4.31)$ commands considerably more attention
than it deserves. The constraint equation does not fix it, and either choice is
allowed classically --- the positive sign corresponds to a contracting universe
while the negative sign gives an expanding universe. If we are to avoid
imposing extraneous conditions, and especially if quantum mechanics is to
recover classical results in the correspondence limit then we must include {\it
both} signs. This is achieved by using a multi-component wavefunction(al):
$$\Psi \longrightarrow \Biggl({\Psi^+ \atop \Psi^-}\Biggr) \eqno(4.32)$$
The action of the the wholly constrained operator $p^{tr}$ on $\Psi^{\pm}$ is
defined by $(4.31)$ with the plus or minus sign respectively. A purely
contracting universe would be represented by $\Psi^+=0$ whereas a purely
expanding universe would have $\Psi^-=0$. We shall see in section~6 that the
straightforward application of the Faddeev-Popov technique for gauge fixing
results in the absolute value of an operator which causes the inner product to
segregate into a manifestly positive contribution from each component.
\footnote{
*}{We wish to suggest that the same procedure be applied {\it whenever}
discrete
choices must be made in solving constraints.}

There is also the issue of perturbatively iterating $(4.31)$ to achieve an
asymptotic series solution which is free of dependence upon $p^{tr}$. We first
define the zeroth order energy:
$$E_0 \equiv \int d^3x \thinspace \Biggl\{p^{tt}_{ij} \thinspace p^{tt}_{ij} +
\frac14 h^{tt}_{ij,k} \thinspace h^{tt}_{ij,k}\Biggr\} \eqno(4.33)$$
and then expand the square root:
$$\eqalignno{p^{tr} &= \pm \sqrt{6} L^{-3/2} \thinspace \sqrt{E_0} \thinspace
\Biggl\{1 + \frac12 \kappa {{\cal C}_0 \over E_0} - \sum_{n=2}^{\infty}
\thinspace {(2n-3)!! \over n!} \thinspace \Biggl(-{\kappa {\cal C}_0 \over 2
E_0}\Biggr)^n\Biggr\} &(4.34a) \cr
&\equiv \pm \sqrt{6} L^{-3/2} \thinspace \Biggl\{ \sqrt{E_0} + \kappa
\thinspace
{\cal S}_0\Bigl[h^{tt},p^{tt};h,p^t;h^{tr},p^{tr}\Bigr]\Biggr\} &(4.34b) \cr}$$
Assuming that it is fair to regard the ratio ${\cal C}_0/E_0$ as of first plus
higher orders in the weak fields, we then obtain an asymptotic series solution
by iteration. For example, the first iteration gives:
$$p^{tr} = \pm \sqrt{6} L^{-3/2} \thinspace \Biggl\{\sqrt{E_0} + \kappa
\thinspace {\cal S}_0\Bigl[h^{tt},p^{tt};h,p^t;h^{tr},\pm \sqrt{6} L^{-3/2}
\Bigl\{\sqrt{E_0} + \kappa{\cal S}_0\Bigr\}\Bigr]\Biggr\} \eqno(4.35)$$
As with $h^t$ and $p_i$, successive iterations push to ever higher orders any
dependence of the right hand side upon $p^{tr}$.

Of course the iterative solution for $p^{tr}$ will result in nonsense, even
perturbatively, if $E_0$ can be made to vanish without ${\cal C}_0$ vanishing
at least as rapidly. It turns out that this cannot happen for three reasons.
First, we will shortly see that $h_i$ and $p^t$ can be gauged to zero. Second,
$E_0$ is a sum of squares of all the remaining variables except for $h^{tr}$
and
the zero modes of $h^{tt}$. Finally, the dependence ${\cal C}_0$ inherits from
${\cal H}_0$ implies that each of its terms must vanish at least quadratically
with $p^{tt}_{ij}$ and/or the non-zero modes of $h^{tt}_{ij}$. To see this last
point note from substituting the weak field expansions $(4.18)$ into $(4.5b)$
that
${\cal H}_0$ consists of terms quadratic in $p_{ij}$ with any number of
$h_{ij}$'s and other terms which are free of $p_{ij}$ but contain at least one
differentiated $h_{ij}$. Upon integration over $T^3$ each of the pure $h_{ij}$
terms must contain at least {\it two} non-zero modes of $h_{ij}$. Constraining
$h^t$ and $p_i$ to zero can result in terms which have any even power of the
remaining components of $p_{ij}$ but it cannot result in odd powers of the
momentum nor can it introduce pure $h_{ij}$ terms which fail to possess at
least
two non-zero modes. Upon gauging $p^t$ to zero we see that every term in ${\cal
C}_0$ must either contain a positive even power of $p^{tt}_{ij}$ or $p^{tr}$,
or
it must contain at least two non-zero modes of $h^{tt}_{ij}$. It follows that
whenever $E_0$ vanishes ${\cal C}_0$ must vanish at least as rapidly, so the
ratio ${\cal C}_0/E_0$ can be legitimately regarded as of order one and higher
in the weak fields.

The three global momentum constraints cannot be imposed this way because we see
from $(4.30c)$ that their quadratic parts are not differences of squares. Our
strategy is therefore to leave them as constraints upon the classical initial
value data or, in the quantum theory, upon the space of states. We can get away
with this for three reasons. First, their imposition is not necessary in order
to construct a reduced canonical formalism with a non-zero Hamiltonian. This
was
obviously {\it not} the case for the global Hamiltonian constraint. Second, the
global momentum constraints remove no negative energy modes, unlike the global
Hamiltonian constraint. Finally, the symmetry generated by the global momentum
constraints consists of constant spatial translations on $T^3$. Since these
form
a compact group there is no need to gauge fix them in the functional formalism.
Additional support for the viability of not attempting to reduce the global
momentum constraints can be found in the close parallel with the
$N ={\widetilde
N}$ constraint of closed bosonic string theory in lightcone gauge [20].

We turn now to the issue of gauge fixing. Since we wish in the end to compare
our results with those of A.D.M. [3] we shall of course need to follow them in
the
choice of gauge. Their perspective was slightly different from ours: whereas we
impose the volume gauge by choosing the lapse and shift, and then fix (most of)
the residual symmetry with gauge conditions on the initial value surface,
A.D.M.
impose a volume gauge condition on the weak fields $h_{ij}$ and $p^{ij}$ and
then use the evolution equations for the frozen variables to determine the
lapse and the shift. We can obtain the same result by merely choosing our lapse
and shift, and our auxiliary surface conditions, so as to agree with A.D.M. The
distinction between the two methods is important only to Faddeev-Popov gauge
fixing in the quantum theory which was developed years after A.D.M. wrote. In
our notation the conditions favored by A.D.M. [3] are:\footnote{*}{The
component
fields $h_i$ and $p^t$ used by A.D.M. actually contain zero modes, unlike
ours, and
these zero modes were given non-zero values determined by the asymptotically
flat boundary conditions [3].}
$$h_i(t,{\vec x}) = 0 \eqno(4.36a)$$
$$p^t(t,{\vec x}) = 0 \eqno(4.36b)$$
We shall accordingly begin by showing that the residual symmetry allows the
perturbative imposition ${\widehat h}_i = 0$ and ${\widehat p}^{~t} = 0$. We
then argue that $n^0(t,{\vec x})$ and $n^i(0,{\vec x})$ can be chosen so as to
perturbatively enforce the A.D.M. condition $(4.36)$.

To properly organize the notion of a perturbative transformation we insert a
factor of $\kappa$ into the infinitesimal transformation parameter, $\theta^{
\mu} \equiv \kappa \thinspace \tau^{\mu}$. By substituting this and the
perturbative expansions $(4.18)$, and by iterating $(4.11)$ we obtain the
following
expressions for the non-infinitesimal but perturbatively small transformations
of ${\widehat h}_{ij}$ and ${\widehat p}_{ij}$:
$$\eqalignno{{\widehat h}_{ij}'  - {\widehat h}_{ij} &= -
({\widehat \tau}_{i,j}
+ {\widehat \tau}_{j,i}) \cr
&+ \kappa \Bigl(\frac12 ({\widehat \tau}_{k,i} \thinspace {\widehat \tau}_k)_{,
j} + {\widehat \tau}_{k,i} \thinspace {\widehat \tau}_{k,j} + \frac12
({\widehat
\tau}_{k,j} \thinspace {\widehat \tau}_k)_{,i} - {\widehat \tau}_{k,i}
\thinspace {\widehat h}_{kj} - {\widehat \tau}_{k,j} \thinspace {\widehat h}_{i
k} - {\widehat h}_{ij,k} \thinspace {\widehat \tau}_k \cr
&\qquad \qquad - {\widehat \tau}^{~0}_{~~,i} \thinspace {\widehat n}_j -
{\widehat \tau}^{~0}_{~,j} \thinspace {\widehat n}_i + {\widehat \tau}^0_{~,ij}
\thinspace {\widehat \tau}^0 - 2 {\widehat p}_{ij} \thinspace {\widehat \tau}^0
+ \delta_{ij} \thinspace {\widehat p} \thinspace {\widehat \tau}^0\Bigr)
+ {\cal O}(\kappa^2) &(4.37a)\cr
&\equiv - ({\widehat \tau}_{i,j} + {\widehat \tau}_{j,i}) + \kappa \thinspace
{\cal H}_{ij}\Bigl[{\widehat h}^{tt},{\widehat p}^{~tt};
{\widehat h}^t,{\widehat
p}^{~t};{\widehat h},{\widehat p};{\widehat h}^{tr},{\widehat p}^{~tr};
{\widehat
\tau}^0,{\widehat \tau}\Bigr] &(4.37b) \cr}$$
$${\widehat p}_{ij}{~'} - {\widehat p}_{ij} = - ({\widehat \tau}^{0}_{~,ij}
- \delta_{ij} \thinspace {\widehat \tau}^{0}_{~,kk}) + \kappa \thinspace
{\cal P}_{ij}\Bigl[{\widehat h}^{tt},{\widehat p}^{~tt};{\widehat h}^t,
{\widehat
p}^{~t};{\widehat h},{\widehat p};{\widehat h}^{tr},{\widehat p}^{~tr};
{\widehat
\tau}^0,{\widehat \tau}\Bigr] \eqno(4.38)$$
We use the non-zero modes of ${\widehat \tau}_i$ to perturbatively enforce
${\widehat h}_i' = 0$ by iterating the equation:
$${\widehat \tau}_i = {\widehat h}_i + {\kappa \over \nabla^2} L_{ij}
\partial_k
\thinspace {\cal H}^{\rm NZ}_{jk}\Bigl[{\widehat h}^{tt},{\widehat p}^{~tt};
{\widehat h}^t,{\widehat p}^{~t};{\widehat h},{\widehat p};{\widehat h}^{tr},
{\widehat p}^{~tr};{\widehat \tau}^0,{\widehat \tau}\Bigr] \eqno(4.39)$$
The zero modes of ${\widehat \tau}_i$ are not fixed because they are conjugate
to the global momentum constraints which are not being reduced. We use the
non-zero modes of ${\widehat \tau}^0$ to perturbatively enforce
${\widehat p}^{t
'} = 0$ by iterating the equation:
$${\widehat \tau}^{~0} = - \frac12 {\widehat p}^{~t} - {\kappa \over 2
\nabla^2}
T_{ij} \thinspace {\cal P}^{\rm NZ}_{ij}\Bigl[{\widehat h}^{tt},
{\widehat p}^{~t
t};{\widehat h}^t,{\widehat p}^{~t};{\widehat h},{\widehat p};
{\widehat h}^{tr},
{\widehat p}^{~tr};{\widehat \tau}^0,{\widehat \tau}\Bigr] \eqno(4.40)$$
We will henceforth drop the prime and assume that ${\widehat h}_i = 0 =
{\widehat p}^{~t}$.

There remains the zero mode of ${\widehat \tau}^{0}$. We shall use this to
enforce ${\widehat h}^{tr} = 0$ although the argument for being able to impose
this condition is more subtle. First, note that since the Wheeler-DeWitt
symmetry must be gauge fixed [21], and since the subgroup of constant time
translations is not compact, we do not have the option of declining to enforce
{\it some} zero mode gauge condition. Second, note from $(4.37a)$ that
$h^{tr}$ is
the quantity affected to lowest order by a constant time translation. Let us
label such a transformation by the single parameter ${\widehat z}$:
$${\widehat z} \equiv L^{-3} \int d^3x \thinspace {\widehat \theta}^{~0}({\vec
x}) \eqno(4.41)$$
Whereas the parameter ${\widehat \theta}^{\mu}$ which enforced ${\widehat h}_i
= 0 = {\widehat p}^{~t}$ was of order $\kappa$ we need ${\widehat z}$ to be of
order one. Even so, the fact that $p^{tr}$ is constant to lowest order allows
us
to obtain a perturbative expression for the result of a non-infinitesimal
shift:
$${\widehat h}^{tr'} - {\widehat h}^{tr} = \frac16 {\widehat p}^{tr} + \kappa
\thinspace {\cal Z}\Bigl[{\widehat h}^{tt},{\widehat p}^{~tt};
{\widehat h}^{tr},
{\widehat p}^{~tr};{\widehat z}\Bigr] \eqno(4.42)$$
We see that the desired condition can be imposed formally by iterating the
equation:
$${\widehat z} = - 6 {{\widehat h}^{tr} \over {\widehat p}^{~tr}} - {6 \kappa
\over {\widehat p}^{~tr}} \thinspace {\cal Z}\Bigl[{\widehat h}^{tt},{\widehat
p}^{~tt};{\widehat h}^{tr},{\widehat p}^{~tr};{\widehat z}\Bigr] \eqno(4.43)$$
We include the qualifier ``formally'' because the transformation is obviously
singular when ${\widehat p}^{~tr}$ vanishes. Of course ${\widehat p}^{~tr}$ is
not an independent degree of freedom; and we see from $(4.31)$ that it is
about as
protected from vanishing as it is invariantly possible to get. However, if all
the modes of ${\widehat p}^{~tt}_{ij}$ and all the non-zero modes of ${\widehat
h}^{tt}_{ij}$ vanish then ${\widehat p}^{~tr}$ vanishes, but we can have a
{\it non}-zero ${\widehat h}^{tr}$. In this case both $h_{ij}(t,{\vec x})$ and
$p_{ij}(t,{\vec x})$ are constant in space and time, and no temporal
translation
exists which will enforce ${\widehat h}^{tr'} = 0$.

Our procedure is to go ahead and impose ${\widehat h}^{tr} = 0$ anyway.
Considerable justification for this course derives from the close analogy to
imposing the gauge $q^0(\tau) = t$ for a massless free particle whose position
and momenta are $q^{\mu}(\tau)$ and $p_{\mu}(\tau)$ respectively. In this
system
the constrained variable, $p_0(\tau)$ has an ambiguous sign which necessitates
a
2-component wavefunction. Just as with gravity, the gauge condition conjugate
to
this constrained variable is singular for constant field configurations. This
resolves itself in the quantum theory by the gauge fixed inner product
acquiring
a Faddeev-Popov determinant which endows the troublesome sector of
configuration
space with zero measure [21]:\footnote{*}{We have taken the liberty to correct
an error that appeared in formula (23c) of [21].}
$$\eqalignno{\Bigl\langle \psi \Bigl\vert \thinspace \delta\Bigl(q^0 - t\Bigr)
\thinspace {\rm abs}\Bigl(p_0\Bigr) \thinspace \Bigr\vert \phi \Bigr\rangle &=
-\frac{i}2 \int d^3q \thinspace \Biggl\{\psi^{+*}(t,{\vec x}) \thinspace {\dot
\phi}^+(t,{\vec x}) - {\dot \psi}^{+*}(t,{\vec x}) \thinspace \phi^+
(t,{\vec x})
\Biggr\} \cr
&+ \frac{i}2 \int d^3q \thinspace \Biggl\{\psi^{-*}(t,{\vec x}) \thinspace
{\dot \phi}^-(t,{\vec x}) - {\dot \psi}^{-*}(t,{\vec x}) \thinspace \phi^-(t,
{\vec x})\Biggr\} &(4.44) \cr}$$
We will see at the end of section 6 that Faddeev-Popov gauge fixing endows
gravity with the same sort of inner product. The result obtained for the free
particle has such universal acceptance that we shall henceforth ignore the
completely analogous problem with imposing ${\widehat h}^{tr} = 0$ on constant
field configurations.

It remains to show that we can choose the lapse and shift so as to enforce the
A.D.M. gauge conditions $(4.36)$ for all time. To see this it suffices to
apply the
constraints to evolution equations $(4.20b)$, $(4.21a)$ and $(4.22a)$:
$${\dot p}^t = - 2 \nabla^2 \thinspace n^0 + \kappa \thinspace {\cal N}^0\Bigl[
h^{tt},p^{tt};n^0,n;p^t,h,h^{tr}\Bigr] \eqno(4.45a)$$
$$\nabla^2 \thinspace {\dot h}_i + {\dot h}_{j,ji} = \nabla^2 \thinspace n_i
+ n_{j,ji} + \kappa \thinspace {\cal N}_i\Bigl[h^{tt},p^{tt};n^0,n;p^t,h,h^{tr}
\Bigr] \eqno(4.45b)$$
$${\dot h}^{tr} = \mp L^{-3/2} \thinspace \sqrt{6 E_0} + \kappa \thinspace
{\cal
T}\Bigl[h^{tt},p^{tt};n^0,n;p^t,h,h^{tr}\Bigr] \eqno(4.45c)$$
Since we already have $p^t(0,{\vec x}) = 0 = h_i(0,{\vec x})$ we will have the
A.D.M. condition $(4.36)$ if the weak field lapse and shift are obtained by
iterating
the equations:
$$n^0 = {\kappa \over 2 \nabla^2} \thinspace \thinspace {\cal N}^0_{\rm
NZ}\Bigl[h^{tt},p^{tt};n^0,n;0,0,h^{tr}\Bigr] \eqno(4.46a)$$
$$n_i = - {\kappa \over \nabla^2} L_{ij} \thinspace \thinspace {\cal N}^{
\rm NZ}_j\Bigl[h^{tt},p^{tt};n^0,n;0,0,h^{tr}\Bigr] \eqno(4.46b)$$
Note that these equations only determine the non-zero modes of the lapse and
shift. We propose that the zero modes be left one and zero, respectively, to
all
orders. Of course while $(4.45c)$ --- and the initial condition, $h^{tr}
(0,{\vec
x}) = 0$ --- determines $h^{tr}(t,{\vec x})$, this component field does not
vanish after $t=0$.

\item{(4)}--- The Reduced Canonical Formalism and Its and Correspondence Limit.

In the previous section we succeeded in reducing the theory to the point where
only the transverse-traceless fields survive. It is convenient, however, to
view
the system that results when $p^{tr}$ and ${\widehat \theta}^{~0}$ are not yet
reduced. Because special care must be given to the zero
modes we will do this in $k$--space:
$$\eqalignno{\dot {\widetilde p}^{~tt}_{ij}&= -\frac 12 k^2 \, \widetilde
h^{~tt}_{ij}+\kappa\,\Big[ \widetilde {\cal U}_{ij} (h^{tt},p^{tt})
+\frac{1}{12} \, k^2 \, h^{tr} \, \widetilde h^{~tt}_{ij} -
\frac 13 \, p^{tr} \widetilde p^{~tt}_{ij} \, \Big] +\o2 &(4.47a) \cr \cr
\dot {\widetilde h}^{~tt}_{ij}&=2\, \widetilde p^{~tt}_{ij}
+\kappa \,\Bigg[ \, 2 \, \widetilde {\cal W}_{ij} (h^{tt},p^{tt})
+ \frac 13 \, h^{tr} \, \widetilde p^{~tt}_{ij}
+ \frac 13 \, p^{tr} \, \widetilde h^{~tt}_{ij} \, \Bigg]+\o2 &(4.47b)}$$
The explicit forms of $\widetilde {\cal U}_{ij}$ and $\widetilde
{\cal W}_{ij}$ are:
$$\eqalignno{\widetilde {\cal U}_{ij}\,&=\,{1 \over (2 \pi)^3 L^{3/2}}
\left(T_{in}T_{jr}-\frac12 T_{ij}T_{nr}\right)\int d^3x \, \Bigg\{ e^{-i \vec k
\cdot \vec x}\Bigg[\frac14\,h^{tt}_{lm,n}\,h^{tt}_{lm,r} \cr
&-\frac12\,h^{tt}_{lm,n}\,h^{tt}_{lr,m}-\frac12\,h^{tt}_{ln,m}\,h^{tt}_{lm,r}
-\frac12\,h^{tt}_{nl,m}\,h^{tt}_{rl,m}+\frac12\,h^{tt}_{nl,m}\,h^{tt}_{rm,l}
&(4.48a)\cr
&-\frac12\,h^{tt}_{lm}\,h^{tt}_{nr,lm}-\frac12\,h^{tt}_{nl}\nabla^2 h^{tt}_{lr}
-\frac12\,h^{tt}_{nl}\nabla^2 h^{tt}_{lr}-2\,p^{tt}_{nl} \, p^{tt}_{lr}
\Bigg]\Bigg\} \cr\cr\widetilde {\cal W}_{ij}\,&={1 \over (2 \pi)^3 L^{3/2}}
\, \left(T_{in}T_{jr}-\frac12 T_{ij}T_{nr}\right)
\int d^3x \left[e^{-i \vec k \cdot \vec x}\,
\left(p^{tt}_{nm}\,h^{tt}_{mr}+p^{tt}_{rm}\,h^{tt}_{mn}\right)\right]&(4.48b)
}$$
\noindent and the evolution equations for the zero modes are obtained by
setting $k=0$ {\it after} changing $T_{ij} \to \delta_{ij}$ in the expressions
for $\widetilde {\cal U}_{ij}$ and $\widetilde {\cal W}_{ij}$.

This system is canonical for the same reason that the A.D.M. system is: the
surface
and volume gauges have been chosen so that the variables conjugate to each of
the constrained variables --- that is, $p^t$ for $h^t$ and $h_i$ for $p_i$ ---
remain zero for all time. The act of reducing $p^{tr}$ and $h^{tr}$ spoils
canonicity because $h^{tr}$ does not remain zero after the initial time. The
mechanism is the same as we found in section 3 where the evolution of a
non-zero
longitudinal vector potential in temporal gauge broke the canonicity of scalar
electrodynamics. Because it is {\it only} the trace components which break
canonicity we know that it will be restored if we can transform to variables
$(\widetilde X,\widetilde P)$ with the same evolution
equations except for lack of dependence on $p^{tr}$ and $h^{tr}$.
This transformation will necessarily be time dependent and non-local. The time
dependence arises because the transformation must give $(\widetilde h^{~tt},
\widetilde p^{~tt})$ at $t=0$ so that at this time $(\widetilde X,
\widetilde P)$ obey the same commutation relations obeyed by
$(\widetilde h^{~tt},\widetilde p^{~tt})$; but it must deviate from this
at later times since we wish to eliminate
$(h^{tr},p^{tr})$ from equations $(4.47)$. The non-locality enters merely
because the transformations must depend on $(h^{tr},p^{tr})$ and these are
non-local, as witness equation $(4.31)$.

It is trivial to check that to the order we are working the following
transformations possess the properties mentioned above:

\noindent For $\omega \equiv |\vec k| \neq 0$:
$$\eqalignno{\widetilde P_{ij}\,&=\widetilde p^{~tt}_{ij}
+ \kappa \Bigg\{ \frac 14 p^{tr} \left[{\sin (\omega t) \over \omega} \right]
\left[ \widetilde p^{~tt}_{ij} \cos (\omega t)
+{\omega \over 2} \widetilde h^{~tt}_{ij} \sin (\omega t) \right]
- \frac 1{12} h^{tr} \widetilde p^{~tt}_{ij} \Bigg\}&(4.49a)\cr\cr
\widetilde X_{ij}\,&=\widetilde h^{~tt}_{ij}- \kappa \Bigg\{\frac 14 p^{tr}
\left[{\sin (\omega t) \over \omega} \right] \left[ \widetilde h^{~tt}_{ij}
\cos (\omega t)
-{2 \over \omega} \widetilde p^{~tt}_{ij} \sin (\omega t) \right]
- \frac 1{12} h^{tr} \widetilde h^{~tt}_{ij}\Bigg\}&(4.49b)\cr}$$
while for $\omega = 0$:
$$\eqalignno{\widetilde P_{ij}(0)\,&=\widetilde p^{~tt}_{ij}(0)
+ \kappa \Bigg\{ \frac 14 t p^{tr} \widetilde p^{~tt}_{ij} (0)
- \frac 1{12} h^{tr} \widetilde p^{~tt}_{ij} (0)\Bigg\}&(4.50a)\cr\cr
\widetilde X_{ij}(0)\,&=\widetilde h^{~tt}_{ij}(0)-\kappa\Bigg\{\frac 14 t
p^{tr} \left[ \widetilde h^{~tt}_{ij}(0)-t \widetilde p^{~tt}_{ij}(0) \right]
- \frac 1{12} h^{tr} \widetilde h^{~tt}_{ij}(0)\Bigg\}&(4.50b)\cr}$$
Applying the transformations $(4.49)$ and $(4.50)$
to the evolution equations $(4.47)$ results, as required, in equations
independent of both $h^{tr}$ and $p^{tr}$:
$$\eqalignno{\dot {\widetilde P}_{ij}&= -\frac 12 k^2 \, \widetilde X_{ij}
+\kappa \, \widetilde {\cal U}_{ij} (X,P)+\o2 &(4.51a) \cr \cr
\dot {\widetilde X}_{ij}&=2 \, \widetilde P_{ij}+2\, \kappa \, \widetilde
{\cal W}_{ij} (X,P)
+\o2 &(4.51b)}$$ and again, the equations for the zero modes are obtained by
setting $k=0$ in the above {\it after} changing $T_{ij} \to \delta_{ij}$
in equations $(4.48)$.

As previously mentioned, the variables $\widetilde X$ and $\widetilde P$
are canonical at $t=0$ since at this time they are simply {\it equal}
to $\widetilde h^{tt}$ and $\widetilde p^{tt}$. Furthermore, they will
remain canonical at later times since the evolution equations $(4.51)$ are
just those of the A.D.M. variables.

The Hamiltonian that generates equations $(4.51)$ in terms of the $x$--space
variables is:
$$\eqalignno{H =& \int d^3x\Bigg\{P_{ij}\,P_{ij}+\frac14\,X_{ij,l}\,X_{ij,l}
+ \kappa \, \Bigg[-X_{lm,i}X_{lj,m}X_{ij}-\frac14 X_{lm}X_{ij,lm}X_{ij} \cr
&+\frac12 X_{il,m}X_{jl,m}X_{ij}-\frac12X_{im,l}X_{jl,m}X_{ij} +2\,P_{il}
P_{lj}X_{ij}\Bigg] \Bigg\} &(4.52)}$$
there are two reasons for writing the Hamiltonian in terms of
$(X,P)$ as opposed of $(\widetilde X,\widetilde P)$. The first one is
merely that it is in terms of these variables that the form of $H$ is the
simplest. The second reason is that in this form it is obvious that the
Hamiltonian density implied by $(4.52)$ is local. However, in order to
derive the evolution equations $(4.51)$ while still treating the zero
modes properly, one must work in momentum space. Here the non-zero
bracket relations are:

\noindent For $\vec k,\vec q \neq 0$,
$$\Bigg\{\widetilde {X}_{ij}(\vec k),\widetilde{P}_{lm}(-\vec q)\Bigg\} =
{1 \over 2} \delta_{\vec k,\vec q} \Bigg[\left(T_{il}T_{jm}-\frac12T_{ij}
T_{lm}\right)+\left(T_{im}T_{jl}-\frac12T_{ij}T_{lm}\right) \Bigg]
\eqno(4.53a)$$
\noindent For $\vec k=\vec q=0$
$$\Bigg\{\widetilde{X}_{ij}(0),\widetilde{P}_{lm}(0)\Bigg\}={1 \over 2}
\Bigg[\left(\delta_{il}\delta_{jm}-\frac12\delta_{ij}\delta_{lm}\right)
+\left(\delta_{im}\delta_{jl}-\frac12\delta_{ij}\delta_{lm}\right) \Bigg]
\eqno(4.53b)$$

So far we have succeeded in reducing the theory and extracting the proper
Hamiltonian. We will next prove that in the appropriate
limit the Hamiltonian of equation $(4.52)$ goes to that obtained by following
the A.D.M. procedure. The appropriate limit is that in which a configuration in
$T^3 \times R$ goes to the same configuration in an open space with flat
boundary conditions. Explicitly, the limit in which the two treatments agree
is that in which we take $L\to\infty$ with localized initial perturbations.
We refer to this limit as the {\it open space limit}. It should be obvious
that the proof reduces to showing that in the open space limit
both $p^{tr}$ and $h^{tr}$ vanish; since if this is the case $(X,P)$
become just $(h^{tt},p^{tt})$ respectively and the Hamiltonian $(4.52)$
already has the correct form.

Let us begin then by examining $E_0$ as defined by equation $(4.33)$:
$$E_0 \equiv \int d^3x \thinspace \Biggl\{p^{tt}_{ij} \thinspace p^{tt}_{ij} +
\frac14 h^{tt}_{ij,k} \thinspace h^{tt}_{ij,k}\Biggr\} \eqno(4.33)$$
Note that $E_0$ remains finite in the open space limit, even though
the range of integration increases from $[0,L)$ to $(-\infty,\infty)$.
The reason for this is that localized initial
perturbations guarantees that the integrand above has finite support.

Now let us inspect the evolution equation for $h^{tr}$ (equation $(4.22)$)
together with the constraint equation for $p^{tr}$ (equation $(4.31)$):
$$\dot h^{tr}=- p^{tr} + \kappa L^{-3} \int d^3x \Biggl\{{\cal A}_1
\Bigl[h^{tt},p^{tt};h^{tr},p^{tr}\Bigr]\Biggr\} \eqno(4.54a)$$
$${p^{tr}}^2={6} L^{-3} \, \Biggl\{ E_0 + \kappa \,\int d^3x
{\cal A}_0\Bigl[h^{tt},p^{tt};h^{tr},p^{tr}\Bigr]\Biggr\} \eqno(4.54b)$$
Equations $(4.54)$ are iterative relations for $h^{tr}$ and $p^{tr}$ in terms
of the $tt$ fields {\it after} the gauge has been fixed and the constraints
for $h^t$ and $p_i$ have been enforced. Equation $(4.54a)$ can be integrated
(again iteratively) to give:
$$h^{tr}=\int\limits^{t}_{0} d\tau \Bigg[ - p^{tr} + \kappa L^{-3} \int d^3x
\Biggl\{{\cal A}_1 \Bigl[h^{tt},p^{tt};h^{tr},p^{tr}\Bigr]\Biggr\}\Bigg]
\eqno(4.55)$$
to see that both $h^{tr}$ and $p^{tr}$ vanish in the open space limit we
must examine $L^{-3} \int d^3x {\cal A}_1$ and $L^{-3}\int d^3x
{\cal A}_0$ closer. Let us then explore the $L$ dependance of each of these
two terms separately:

\item {---}${\cal A}_1$ is of second order and higher in the fields;
therefore the highest power of $L$ in $\int d^3x {\cal A}_1$ occurs when the
integral acts on constants (since both $h^{tt}$ and $p^{tt}$ have finite
support). In the open space limit we can then replace $\int d^3x {\cal A}_1$
with $L^3{\cal M}_1$ where ${\cal M}_1$ is at least of
second order in $h^{tr}$ and/or $p^{tr}$.

\item {---}The form of ${\cal A}_0$ is at least of
{\it third} order in the fields (remembering that $E_0$ is independent of $L$
in the limit). Similar considerations as those mentioned for the
case of  ${\cal A}_1$ reveal
that in the open space limit $\int d^3x {\cal A}_0$ can be replaced by $L^3
{\cal M}_0$ where ${\cal M}_0$ is of degree three and higher in
$h^{tr}$ and/or $p^{tr}$.

We can then, in the open space limit, write equations $(4.55)$ and $(4.54b)$
as:
$$
   h^{tr}=\int\limits^{t}_{0} d\tau \Bigg\{ - p^{tr}
             + \kappa {\cal M}_1 \Bigl[h^{tr},p^{tr}\Bigr]\Bigg\}
                \eqno(4.56a)
$$
$$
   {p^{tr}}^2 = L^{-3}  E^{\prime}_0+\kappa\,
    \, {\cal M}^{\prime}_0\Bigl[h^{tr},p^{tr}\Bigr] \eqno(4.56b)
$$
where the primes have been put in to absorb an irrelevant factor of $6$
that would otherwise appear multiplying the right hand side of equation
$(4.56b)$.
At this point it should be obvious to the reader that the perturbative
solutions to equations $(4.56)$ are $h^{tr}=0$ and $p^{tr}=0$. For those that
still have some doubts let us take the $L \to \infty$ limit and re-write
equations $(4.56)$ as:
$$
h^{tr} = \int\limits^{t}_0 d\tau \Bigg\{ -p^{tr} + \kappa \alpha_{nm}
\, \Big({h^{tr}}\Big)^n
\Big({p^{tr}}\Big)^m \Bigg\} \eqno(4.57a)
$$
$$
{p^{tr}}^2 = \kappa \alpha^{\prime}_{n'm'} \, \Big(h^{tr}\Big)^{n'}
\Big(p^{tr}\Big)^{m'} \eqno(4.57b)
$$
with $n+m\geq 2$ and $n'+m' \geq 3$. Each successive iteration of equations
$(4.57)$ brings with it positive powers of $\kappa$. Therefore, to any
order in perturbation in powers of $\kappa$ both $h^{tr}$ and
$p^{tr}$ vanish in the open space limit. Thereby proving the correspondence
between our method and that of A.D.M.

We conclude this section by re-stating the result: Our method of reduction
gives a precise meaning to time and, perhaps more importantly, this time
evolution coincides in the appropriate limit with that obtained by A.D.M. for
open space.

\vskip 1cm
\centerline{\bf 5. Exact Results For Minisuperspace}

In this section we examine two simple models.
The first one is that of pure gravity with a cosmological constant. The
second example deals with the case a gravity coupled to a scalar field.
In both of these models we will truncate the form of the
metric to allow only certain zero modes to be present so as to render
the models exactly solvable.

\item {---} Zero modes of gravity with a cosmological constant.

For the first example we choose the form of the 3-metric to be such as to have
two homogeneous degrees of freedom:$$\eqalignno{ds^2 = & - N^2(\tau) d\tau^2 +
b^{2/3}(\tau) \Bigl[e^{a(\tau)}\,dx^2 \,+\,e^{-a(\tau)}\,dy^2 \,+\, dz^2
\Bigr] &(5.1)}$$the action in canonical form is:$$S=\int d\tau \, \left[
\,p_b\, \dot b\,+\,p_a \,\dot a\,-\, N\,{\cal R}\, \right] \eqno(5.2)$$ with
${\cal R}$ defined as:$${\cal R}\,\equiv\,{\kappa^2 \over 2 b^2}\,p_a^2\,+{2
\,\Lambda \over \kappa^2}\, b\,-\,{3 \kappa^2 \over 8}\,b\,p_b^2 \eqno(5.3)$$

The equations of motion before the reduction is implemented are obtained
by varying the action of $(5.2)$ with respect to $p_a$, $a$, $p_b$ and b
respectively:$$\eqalignno{\dot a\,&=\,N\,{\kappa^2 \over b^2}\,p_a &(5.4a)\cr
\dot p_a\,&=\,0 &(5.4b)\cr\dot b\,&=\,-\,N\,{3 \kappa^2 \over 4}\,b\,p_b
&(5.4c)\cr\dot p_b\,&=\,N\,\left(\,-\,{\kappa^2 \over b^3}\,p_a^2\,+\,{2\,
\Lambda \over \kappa^2}\,-\,{3 \,\kappa^2 \over 8}\,p_b^2\,\right)&(5.4d)}$$
while varying $N$ results in the constraint ${\cal R} = 0$:
$${\kappa^2 \over 2 b^2}\,p_a^2\,+{2\,\Lambda \over \kappa^2}\, b\,-\,{3
\kappa^2 \over 8}\,b\,p_b^2=0 \eqno(5.4e)$$

Since the constraint $(5.4e)$ will be enforced by singling out $p_b$ we wish
to choose the volume gauge by simplifying the equation for $\dot b$ as much
as possible.  The obvious choice is:
$$N\,=\,{1 \over b\,|p_b|} \eqno(5.5)$$
Now that the volume gauge has been fixed we fix the constraint and use equation
$(5.4c)$ to fix $b$ in the following manner\footnote{*}{For simplicity of
exposition we have made a definite choice for the sign of $p_b$, that
corresponding to an expanding space. As in section 4 the wave function really
consists of two components, one for each of the two signs. Note also that we
fix
the surface gauge condition by choosing $b(0)=1$.}:
$$\eqalignno{p_b\,&=\,-\,\sqrt{{16\, \Lambda \over 3\,\kappa^4}\,+\,{4 \over
3\,b^3}\,p_a^2}&(5.6a)\cr \cr b\,&=\,1\,+\,{3\,\kappa^2 \over 4}\,\tau &(5.6b)
\cr}$$

The reduction is now complete. The equations of motion for the {\it physical}
fields become:$$\eqalignno{\dot a\,&=\,{\kappa^2\,p_a\over b(t)^3}\,\Bigg[\,
\sqrt{{16\, \Lambda \over 3\,\kappa^4}\,+\,{4 \over 3\,b(t)^3}\,p_a^2}\,
\Bigg]^{-1}\, &(5.7a)\cr \cr\dot p_a\,&=\,0 &(5.7b)}$$ the above equations
are integrable and the Hamiltonian can be obtained from them:$$H\,=\,{3\,
\kappa^2 \over 4}\,\sqrt{{16\, \Lambda \over 3\,\kappa^4}\,+\,{4 \over 3\,
b(t)^3}\,p_a^2} \eqno(5.8)$$

Before going to the next example we wish to clarify one point: The choice of
$N \neq 1$ signifies that the time evolution implied by equation $(5.8)$ is
not that corresponding to time evolution in flat space (we will see this point
more clearly in the next example when we take the limit $\kappa \to 0$).
$N$ was chosen so as to make
equation $(5.4c)$ exactly solvable. It was by no means a unique choice; for
example had we chosen: $$N={1 \over |p_b|} \eqno(5.9)$$ equation $(5.4c)$
would still be easy to solve but equations $(5.4a-b)$ would have a different
form: $$a'=\kappa ^2 {e ^{3/2 \kappa^2 t} p_a \over p_b} \eqno(5.10)$$
$$p'_a=0$$ where prime denotes differentiation with respect to the new time
parameter $t$. The Hamiltonian then would be:
$$H'={3\,\kappa^2 \over 4}\,e^{-3/4 \kappa^2 t}\,\sqrt{{16\, \Lambda \over
3\,\kappa^4}\,+\,{4 \over 3\,b(t)^3}\,p_a^2} \eqno(5.11)$$This gauge
dependance of the Hamiltonian should come as no surprise since changing
how we gauge fix $N$ changes what we mean by time, thereby changing what
we mean by time evolution. In our next example we will show how despite
this freedom we can make contact with the results obtained in a theory
for which gravity is not dynamical (\ie in the limit $\kappa \to 0$) and
$N=1$ always.

\item {---} Zero mode of gravity coupled to a scalar field

We start by looking at the zero modes of a massive scalar field coupled to
gravity and we allow the $3$--metric to have only one degree of freedom:
$$ds^2=-N^2(\tau) \, d\tau^2 + exp \left[\frac 23\,a(\tau)\right]\,d\vec x^2
\eqno(5.12)$$

The action in canonical form is:
$$ S =\int{d\tau} \left[\pi \dot
{\phi} + p \dot a - N {\cal R} \right]\eqno(5.13)$$ where $(\pi\,{,}\,p)$ are
the variables conjugate to $(\phi\,{,}\,a)$ respectively and: $${\cal R}\,
\equiv\, \left[\,\frac12 \, \pi^2\,-\,\frac 12 \, \alpha^2\,p^2 \right]\,
e^{-a}\,+\,\left[\,\frac12 \, m^2\,\phi^2\right]\,e^a\eqno(5.14)$$ with $
\alpha^2 \equiv 12 \pi G$.

By varying this action we obtain the unreduced equations of motion:
$$\eqalignno{\dot a\, &=\, -\, \alpha^2 \, N e^{-a} \, p \,  &(5.15a) \cr
\dot p\, &=\, N \, \left[\frac12 \, \pi^2\,-\,\frac12\,\alpha^2\,p^2
\right] e^{-a} \, -\,N\,\left[\,\frac12\,m^2\,\phi^2\,\right]\,e^a\,&(5.15b)
\cr
\dot \phi\, &=\,N\, e^{-a} \, \pi \, &(5.15c) \cr \noalign{\hbox{and}}\dot
\pi \,&=\, -\, N \, m^2\, \phi\,e^a\,  &(5.15d)}$$ while variation with
respect to $N$ gives the constraint equation$$\left[\,\frac12 \, \pi^2\,-\,
\frac 12 \, \alpha^2\,p^2 \right]\,e^{-a}\,+\,\left[\,\frac12 \, m^2\,\phi^2
\right]\,e^a=\,0 \, \eqno(5.15e)$$

We must now select a volume gauge condition to impose. Setting $N=1$ will
not do because one would be left with the task of solving equation $(5.15a)$
explicitly for $a$. A gauge choice that simplifies this task is:$$N\,=\,{e^a
\over |p|}\, \eqno(5.16)$$It is obvious that the above choice makes the job
of solving equation $(5.15a)$ a trivial one. Note; however, that $N$ does not
approach $1$ in the limit $\kappa \to 0$. We will have to account for this
when comparing our results to those obtained for flat space.

Having fixed the volume gauge we now proceed to reduce the theory by
enforcing the constraint $(5.15e)$ and fixing the value of $a$ at $\tau=0$
\footnote{*}{We again chose the sign of the constrained variable to give
increasing $a$ for increasing $\tau$}:$$\eqalignno{p\,&=\,-\,{1 \over
\alpha} \sqrt{\pi^2\,+\,m^2 \,\phi^2\,e^{2a}} &(5.17a)\cr a\,&=\,\alpha^2\,
\tau &(5.17b)}$$ where we have chosen $a(0)=0$.

The equations of motion for the remaining variables are, after implementing
reduction: $$\eqalignno{\dot \phi \,&=\,{\alpha\,\pi \over {\sqrt{\pi^2\,+\,
m^2 \,\phi^2\,e^{2a}}}}&(5.18a)\cr\cr\dot \pi \,&=\,{{-\alpha\,\,m^2\, \phi\,
e^{2a}} \over {\sqrt{\pi^2\,+\,m^2 \,\phi^2\,e^{2a}}}} &(5.18b)}$$ The gauge
choice $(5.16)$ leaves the variables canonical. The Hamiltonian obtained by
integrating equations $(5.18)$ is:$$H_\tau\,=\,\alpha\,{\sqrt{\pi^2\,+\,m^2
\,\phi^2\,e^{2\alpha^2\tau}}} \eqno(5.19)$$ the subscript $\tau$ is there
to remind us that this Hamiltonian describes evolution with respect to $\tau$
which in the limit $\alpha \to 0$ does not go to the flat space $t$ simply
because $N$ does not go to $1$ in that limit.

To see how to recover the flat space result in the limit $\alpha \to 0$. Let
us examine the $\tau$ evolution of $\phi$:$$\eqalignno{{\partial \phi \over
\partial \tau}\,&=\,{\partial H_\tau\over \partial \pi} &(5.20a)\cr\cr N\left(
\tau(t)\right)\,{\partial \phi \over \partial t}\,&=\,{\partial H_{\tau(t)}
\over \partial \pi} &(5.20b)\cr\cr{\alpha\, e^{\alpha^2 \tau(t)}\over{\sqrt{
\pi^2\,+\,m^2 \,\phi^2\,e^{2\alpha^2\tau(t)}}}}{\partial \phi \over \partial
t}\,&=\,{\partial \over \partial \pi}\,\left(\,\alpha\,{\sqrt{\pi^2\,+\,m^2
\,\phi^2\,e^{2\alpha^2\tau(t)}}}\right) &(5.20c)\cr}$$ which in the limit
$\alpha \to 0$ can be written as:$$\eqalignno{{\partial \phi \over \partial
t} \to\,& \sqrt{\pi^2\,+\,m^2 \,\phi^2}\, {\partial \over \partial \pi} \,
\sqrt{\pi^2\,+\,m^2 \,\phi^2} &(5.21a)\cr\cr&=\,{\partial \over \partial \pi}
\,\left(\frac12\,\pi^2\,+\,\frac12\,m^2 \,\phi^2\right)&(5.21b)}$$ which of
course is the correct limit.

We end this section by pointing out what we hoped to accomplish
with these two examples. The first example was meant as a simple illustration
of the method for an admittedly simple model. In it we made evident the fact
that the form of the reduced Hamiltonian rests on the choice of the lapse
function (\ie the Hamiltonian is gauge dependent). The second example was
used to show how the paradox of dynamics is resolved by reduction. It was
used to show that the limit of $\kappa \to 0$ gives the same results obtained
for flat space.

\vskip 1cm
\centerline{\bf 6. Correspondence With The Functional Formalism}

The purpose of this section is to show that expectation values and matrix
elements in the reduced canonical theory can be expressed very simply in terms
of the naive functional formalism of the unconstrained theory. The key to
obtaining this result is that {\it reduction affects only the allowed initial
values of Heisenberg operators, not their subsequent time evolutions.} We can
therefore enforce reduction by gauge fixing on the initial value surface and
use the unconstrained Hamiltonian to implement time evolution. Since evolution
is carried out with the known Hamiltonian of the unconstrained theory we get
the
usual functional formalism and we need never go to the trouble of actually
constructing a reduced Hamiltonian or inferring how the original dynamical
variables depend upon the reduced canonical variables. We first derive the
result for a general constrained canonical formalism, then we explain how it
applies to the coupled oscillator model of section 2, to scalar QED in temporal
gauge, and to quantum general relativity with fixed lapse and shift.

We require first a notation for the unconstrained canonical formalism whose
reduction leads to the $v^i(t)$'s of section 2. In a gauge theory this would be
the canonical formalism which results from the imposition of a volume gauge
condition, that is, one set of conditions for each spacetime point. Fixing the
volume gauge is enough to make time evolution unique but it typically leaves a
residual symmetry of gauge transformations of the initial value data. These
residual symmetry transformations are generated by constraints which restrict
the allowed initial value data. It is the act of fixing the residual symmetry
and imposing the constraints which leads to the reduced dynamical system
characterized by the $v^i(t)$'s.

Let us consider the unconstrained system to be described by coordinates
$x^{\alpha}(t)$ and momenta $\pi_{\alpha}(t)$ where  $\alpha = 1,2,\dots,L$,
and
$L = {N+K}$. The variables are canonical, that is, the only non-zero equal time
commutators are:
$$\Bigl[x^{\alpha}(t),\pi_{\beta}(t)\Bigr] = i \delta^{\alpha}_{~\beta}
\eqno(6.1)$$
Time evolution is generated by a possibly time dependent Hamiltonian, $H(x,\pi,
t)$:
$${\dot x}^{\alpha}(t) = -i \Bigl[x^{\alpha}(t),H\Bigl(x(t),\pi(t),t\Bigr)
\Bigr]
\eqno(6.2a)$$
$${\dot \pi}_{\alpha}(t) = -i \Bigl[\pi_{\alpha}(t),H\Bigl(x(t),\pi(t),t\Bigr)
\Bigr] \eqno(6.2b)$$
Equations of this form serve to determine the variables at any time in terms of
the initial values, ${\widehat x}^{\alpha}$ and ${\widehat \pi}_{\alpha}$:
$$x^{\alpha}(t) = X^{\alpha}\Bigl({\widehat x},{\widehat \pi},t\Bigr)
\eqno(6.3a)$$
$$\pi_{\alpha}(t) = \Pi_{\alpha}\Bigl({\widehat x},{\widehat \pi},t\Bigr)
\eqno(6.3b)$$
Finally, we assume the system is subject to a set of $K$ constraints:
$$C_k\Bigl(x(t),\pi(t),t\Bigr) = 0 \qquad , \qquad k = 1,2,\dots,K \eqno(6.4)$$
To accommodate the coupled oscillator model of section 2 we have broken the
usual
convention of forbidding explicit time dependence in the constraints. We do
require, however, that if $C_k\Bigl({\widehat x},{\widehat \pi},0\Bigr) = 0$
for all $k$ then $C_k\Bigl(x(t),\pi(t),t\Bigr) = 0$ as well.

States in the unconstrained formalism can be represented by their wavefunctions
in the basis of position eigenkets at some fixed time:
$$\Bigl\vert \psi ; t\Bigr\rangle = \int d^L\xi \enskip \psi(\xi) \thinspace
\Bigl\vert \xi ; t\Bigr\rangle \eqno(6.5a)$$
$$x^{\alpha}(t) \thinspace \Bigl\vert \xi ; t \Bigr\rangle = \xi^{\alpha}
\thinspace \Bigl\vert \xi ; t \Bigr\rangle \eqno(6.5b)$$
Note that these are Heisenberg states so they do not evolve in time, despite
the
fact that the wavefunction might be given in terms of the eigenkets of the
position operator at any time. The inner product between two such states is the
usual one:
$$\Bigl\langle \psi_2 ; t \Bigl\vert \psi_1 ; t\Bigr\rangle = \int d^Lx \enskip
\psi^*_2(x) \thinspace \psi_1(x) \eqno(6.6)$$

The Heisenberg evolution operator:
$$U(t_2,t_1) \equiv T\Biggl\{ \exp\Biggl[i \int_{t_1}^{t_2} dt \thinspace H
\Bigl(x(t),\pi(t),t\Bigr)\Biggr]\Biggr\} \eqno(6.7a)$$
(the symbol $T$ denotes the ordering convention in which canonical operators at
later times appear to the left of those at earlier times; coordinates stand to
the right of momenta at equal times) enters when we wish to study operators at
different times:
$$x^{\alpha}(t_2) = U(t_2,t_1) \thinspace x^{\alpha}(t_1) \thinspace
U^{\dagger}(t_2,t_1) \eqno(6.7b)$$
or when we wish to specify the wavefunctions in terms of the eigenkets of
operators at different times:
$$\Bigl\vert \xi ; t_2\Bigr\rangle = U(t_2,t_1) \thinspace \Bigl\vert \xi; t_1
\Bigr\rangle \eqno(6.7c)$$
In these cases it is convenient to subsume the evolution operator into a
functional integral. Suppose we wish to study some functional ${\cal O}[x,\pi]$
of the canonical operators defined between times $t_1$ and $t_2 > t_1$. We can
take the matrix element of its time-order product between states specified at
$t_2$ and $t_1$ using the following formula:\footnote{*}{Functional integrals
are sufficiently familiar in particle physics that we shall omit specification
of the detailed skeletonization which would be necessary to make our formulae
well-defined [22]. This could only be done for quantum mechanics and free field
theory in any case. Complete rigor has not yet been obtained for interacting
quantum field theories in four dimensions, although it is completely obvious
how to proceed in regulated perturbation theory.}
$$\eqalignno{\Bigl\langle \psi_2 ; t_2 \Bigl\vert \thinspace T\Bigl({\cal O}
[x&,
\pi]\Bigr) \thinspace \Bigr\vert \psi_1 ; t_1\Bigr\rangle = \Fint \mathop
{\Bigl[
dx(t)\Bigr]}\limits_{\scriptscriptstyle t_2 \geq t \geq t_1} \mathop{\Bigl[
d\pi(t')\Bigr]}\limits_{\scriptscriptstyle t_2 > t' \geq t_1} \thinspace
\psi_2^*\Bigl(x(t_2)\Bigr) \thinspace {\cal O}\Bigl[x,\pi\Bigr] \cr
&\times \exp\Biggl[i\int_{t_1}^{t_2} dt \thinspace \Bigl\{\pi_{
\alpha}(t) \thinspace {\dot x}^{\alpha}(t) - H\Bigl(x(t),\pi(t),t\Bigr)\Bigr\}
\Biggr] \thinspace \psi_1\Bigl(x(t_1)\Bigr) &(6.8) \cr}$$
Note that the symbols ``$x^{\alpha}(t)$'' and ``$\pi_{\alpha}(t)$'' stand for
operators on the left hand side of this equation while they are $\comp$-number
functions in the functional integral to the right. If the Hamiltonian is
quadratic in the momenta then we can perform the Gaussian integrations over
$\pi_{\alpha}(t)$ and pass from the phase space functional formalism to the
more
familiar configuration space version.

If we wish to specify both states at the same time --- say $t_1$ --- then it is
necessary to first evolve forward to some arbitrary point $t_2$, beyond the
final observation in ${\cal O}[x,\pi]$, and then evolve backwards to $t_1$. The
necessary formalism was worked out three decades ago by Schwinger [23] and has
been studied more recently by Jordan [24].\footnote{**}{Although Schwinger and
Jordan assumed that the initial and final states were free vacuum and that
$t_1$
was in the asymptotic past, the generalization to arbitrary states and times is
trivial.} If we denote the fields which implement forward evolution with a
superscript or subscript ``$+$,'' and those which generate backward evolution
with a superscript or subscript ``$-$,'' then the relevant functional integral
is:
$$\eqalignno{\Bigl\langle \psi_2 ; t_1 \Bigl\vert &\thinspace T
\Bigl({\cal O}[x,
\pi]\Bigr) \thinspace \Bigr\vert \psi_1 ; t_1\Bigr\rangle = \Fint \mathop
{\Bigl[
dx_-(t)\Bigr]}\limits_{\scriptscriptstyle t_2 \geq t \geq t_1} \mathop{\Bigl[
d\pi^-(t')\Bigr]}\limits_{\scriptscriptstyle t_2 > t' \geq t_1} \mathop{\Bigl[
dx_+(t)\Bigr]}\limits_{\scriptscriptstyle t_2 \geq t \geq t_1} \mathop{\Bigl[
d\pi^+(t')\Bigr]}\limits_{\scriptscriptstyle t_2 > t' \geq t_1} \thinspace
\delta\Bigl(x_-(t_2) - x_+(t_2)\Bigr) \cr
&\times \psi_2^*\Bigl(x_-(t_1)\Bigr) \thinspace \exp\Biggl[- i\int_{t_1}^{t_2}
dt \thinspace \Bigl\{\pi^-_{\alpha}(t) \thinspace {\dot x}_-^{\alpha}(t) -
H\Bigl(x_-(t),\pi^-(t),t\Bigr)\Bigr\}\Biggr] &(6.9) \cr
&\qquad \times {\cal O}\Bigl[x_+,\pi^+\Bigr] \exp\Biggl[i\int_{t_1}^{t_2} dt
\thinspace \Bigl\{\pi^+_{\alpha}(t) \thinspace {\dot x}_+^{\alpha}(t) -
H\Bigl(x_+(t),\pi^+(t),t\Bigr)\Bigr\}\Biggr] \thinspace \psi_1\Bigl(x(t_1)
\Bigr)
\cr}$$
Note that we must use ``$+$'' variables in the functional ${\cal O}$ if the
functional integral is to give the expectation value of the time-ordered
product; using the ``$-$'' variables would give the anti-time-ordered product.

The unconstrained matrix elements and expectation values we have described are
inadequate in two ways: they are typically divergent for the only interesting
states, and they include information we don't want to know about unphysical
operators. The first problem arises because one enforces the constraints
$(6.4)$
by requiring the states to be annihilated by the $C_k$'s. The inner product
$(6.6)$ between two such states can be divergent on account of the integration
over residual gauge transformations. {\it This is even true when, as in the
case
for gravity, the residual gauge transformation of a coordinate $x^{\alpha}$
involves the momentum $\pi_{\beta}$} [21]. The second problem arises because it
is really operators in the reduced canonical formalism whose matrix elements
and
expectation values we wish to study. These reduced operators have the same time
evolution as the unconstrained ones, but they depend upon $2K$ fewer initial
value operators.

It turns out that by implementing reduction we can also solve the problem with
the inner product. The unconstrained theory is reduced by identifying $K$
residual gauge conditions on an initial value surface which we take to be $t=0$
for definiteness:
$$G^k\Bigl({\widehat x},{\widehat \pi}\Bigr) = 0 \eqno(6.10)$$
These conditions are arbitrary except for the requirement that the
Faddeev-Popov
matrix formed from commutation with the constraints:
$$M_k^{~\ell}\Bigl({\widehat x},{\widehat \pi}\Bigr) \equiv -i \Bigl[C_k\Bigl(
{\widehat x},{\widehat \pi},0\Bigr),G^{\ell}\Bigl({\widehat x},{\widehat \pi}
\Bigr)\Bigr] \eqno(6.11)$$
should have a non-zero determinant.

We can use the conditions $G^k=0$ and $C_{\ell}= 0$ to decompose the $2L$
operators of the unconstrained formalism into two commuting sets of canonical
variables:
$$\Bigl\{{\widehat x},{\widehat \pi}\Bigr\}\longrightarrow
\Bigl\{({\widehat q},
{\widehat p}) \thinspace ; \thinspace ({\widehat c},{\widehat g})\Bigr\}
\eqno(6.12)$$
The $N$ ${\widehat q}^{~a}$'s and $N$ ${\widehat p}_b$'s commute canonically
and
are of course the quantum versions of the reduced canonical variables whose
construction was described in section 2. The $K$ ${\widehat g}^{~k}$'s form
similar conjugate pairs with the $K$ ${\widehat c}_{\ell}$'s. The two sets of
variables commute with one another so that the only non-zero commutators are:
$$\Bigl[{\widehat q}^{~a},{\widehat p}_b\Bigr] = i \delta^a_{~b} \eqno(6.13a)$$
$$\Bigl[{\widehat g}^{~k},{\widehat c}_{\ell}\Bigr] = i \delta^k_{~ \ell}
\eqno(6.13b)$$
The ${\widehat g}^{~k}$'s are pure gauge variables which have the property of
vanishing with the gauge conditions:
$${\widehat g}^{~k} \thinspace \delta^K\Bigl[G\Bigl({\widehat x},{\widehat \pi}
\Bigr)\Bigr] = 0 \eqno(6.14)$$
The ${\widehat c}_{\ell}$'s are constrained variables; the constraint equations
determine them as functions $\kappa_{\ell}
\Bigl({\widehat q},{\widehat p}\Bigr)$
of the reduced canonical variables. The Coulomb potential of electrodynamics
and
the Newtonian potential of gravitation are examples of constrained variables.
When acting on invariant states --- that is, upon states which are annihilated
by the constraint operators --- there is no difference between the ${\widehat
c}_{\ell}$'s and these functions:
$${\widehat c}_{\ell} \thinspace \Bigl\vert \psi_{\rm inv} ; t\Bigr\rangle
= \kappa_{\ell}\Bigl({\widehat q},{\widehat p}\Bigr) \thinspace \Bigl\vert
\psi_{\rm inv} ; t\Bigr\rangle \eqno(6.15a)$$
$$\Bigl\langle \psi_{\rm inv} ; t\Bigr\vert \thinspace {\widehat c}_{\ell}
= \Bigl\langle \psi_{\rm inv} ; t\Bigr\vert \thinspace \kappa_{\ell}\Bigl(
{\widehat q},{\widehat p}\Bigr) \eqno(6.15b)$$

Except for the fact that we have distinguished between the constrained
variables
and the constraints themselves --- that is, we have allowed for the possibility
that $\kappa_{\ell} \neq 0$, and that the reduced canonical variables might not
commute with the constraints --- the decomposition $(6.12)$ is a standard one
in
the theory of constrained quantization [25]. The reduced operators whose matrix
elements and expectation values we really wish to study are:
$$x_r^{\alpha}(t) \equiv X^{\alpha}\Bigl({\widehat x},{\widehat \pi},t\Bigr)
\Bigl\vert_{{\widehat g}=0 \thinspace , \thinspace {\widehat c} = \kappa}
\eqno(6.16a)$$
$$\pi^r_{\alpha}(t) \equiv \Pi_{\alpha}\Bigl({\widehat x},{\widehat \pi},t
\Bigr)
\Bigl\vert_{{\widehat g}=0 \thinspace , \thinspace {\widehat c} = \kappa}
\eqno(6.16b)$$
Note that they have the same time evolution as the unconstrained operators
$(6.3)$.

Now consider the inner product of two states which are annihilated by the
constraints. Since the potential for divergences in the naive inner product
comes from integrating over residual gauge transformations we can avoid the
problem by surface gauge fixing. In the case where the gauge conditions and the
Faddeev-Popov determinant depend only upon the coordinates this is accomplished
by formally inserting unity in the form:
$$1 = \int d^K\theta \thinspace \exp\Bigl[i \theta_k \thinspace C_k\Bigr]
\thinspace \delta^K\Bigl[G\Bigl({\widehat x},{\widehat \pi} \Bigr)\Bigr]
\thinspace {\rm abs}\Biggl\{{\rm det}\Bigl[M_{k \ell} \Bigl({\widehat x},
{\widehat \pi}\Bigr)\Bigr]\Biggr\} \thinspace \exp\Bigl[-i\theta_k
\thinspace C_k\Bigr]
\eqno(6.17)$$
One then changes variables and drops the typically divergent integration over
gauge parameters. Although the question of operator ordering arises forcefully
when either the gauge conditions or the Faddeev-Popov determinant are allowed
to
depend upon non-commuting operators, it is natural to expect that the same
procedure works generally. This expectation is fulfilled in several interesting
cases, including that of perturbative quantum gravity [21]. We therefore
propose
the following reduced inner product:
$$\Bigl\langle \psi_2 ; t_2 \Bigl\Vert \psi_1 ; t_1 \Bigr\rangle_G =
\Bigl\langle \psi_2 ; t_2 \Bigl\vert \thinspace \delta^K\Bigl[G\Bigl({\widehat
x},{\widehat \pi}\Bigr)\Bigr] \thinspace {\rm abs}\Biggl\{{\rm det} \Bigl[M_{k
\ell}\Bigl({\widehat x},{\widehat \pi}\Bigr)\Bigr]\Biggr\} \thinspace
\Bigr\vert \psi_1 ; t_1\Bigr\rangle \eqno(6.18)$$
where it is understood that the issue, if there is one, of operator ordering is
to be resolved on a case-by-case basis. Note that in view of $(6.17)$ this
inner
product is independent of the choice of gauge for states which are annihilated
by the constraints. In fact it amounts to a realization in the unconstrained
formalism of the standard inner product of the reduced theory [25]. Enforcing
this correspondence is an important determinant of the ordering convention for
the gauge fixing paraphernalia; another important condition is the persistence
of Hermiticity when we drop the requirement that the wave functions are
annihilated by the constraints.

The relation we seek between invariant operators in the unconstrained theory
and the corresponding reduced operators holds in the gauge fixed inner product
$(6.18)$. Although invariant operators generally depend upon the
${\widehat g}^{~
k}$'s and the ${\widehat c}_{\ell}$'s in addition to the reduced canonical
variables, when such an invariant operator acts upon an invariant state we can
use the constraints ${\widehat c}_{\ell} = \kappa_{\ell}\Bigl[{\widehat q},
{\widehat p}\Bigr]$:
$${\cal O}_{\rm inv}\Bigl[x,\pi\Bigr] \thinspace \Bigl\vert \psi_{\rm inv} ; t
\Bigr\rangle = {\cal O}_{\rm inv}\Bigl[x,\pi\Bigr] \Bigl\vert_{{\widehat c} =
\kappa} \thinspace \Bigl\vert \psi_{\rm inv} ; t \Bigr\rangle \eqno(6.19a)$$
$$\Bigl\langle \psi_{\rm inv} ; t \Bigr\vert \thinspace {\cal O}_{\rm inv}
\Bigl[
x,\pi\Bigr] = \Bigl\langle \psi_{\rm inv} ; t \Bigr\vert \thinspace {\cal O}_{
\rm inv}\Bigl[x,\pi\Bigr] \Bigl\vert_{{\widehat c} = \kappa} \eqno(6.19b)$$
Note that this is true even when the ${\widehat c}_{\ell}$'s are not themselves
invariant. Once the ${\widehat c}_{\ell}$'s have been eliminated there is no
obstacle to commuting the ${\widehat g}^{~k}$'s through the ${\widehat q}^{~
a}$'s and ${\widehat p}_b$'s to act on the gauge fixing delta function. We
therefore obtain the result that {\it the expectation value or matrix element
of
an invariant operator in the presence of invariant states is the same as the
expectation value or matrix element of the corresponding reduced operator:}
$$\Bigl\langle \psi_{\rm inv} ; t \Bigl\vert \thinspace {\cal O}_{\rm inv}
\Bigl[
x,\pi\Bigr] \thinspace \Bigr\Vert \thinspace \psi'_{\rm inv} ; t'
\Bigr\rangle_G
= \Bigl\langle \psi_{\rm inv} ; t \Bigl\vert \thinspace {\cal O}_{\rm inv}
\Bigl[
x_r,\pi^r\Bigr] \thinspace \Bigr\Vert \thinspace \psi'_{\rm inv} ; t' \Bigr
\rangle_G \eqno(6.20)$$

To reach the functional formalism we consider the action of the gauge fixing
paraphernalia on the ket to be a state in the unconstrained theory:
$$\delta^K\Bigl[G\Bigl({\widehat x},{\widehat \pi}\Bigr)\Bigr] \thinspace
{\rm abs}\Biggl\{{\rm det}\Bigl[M_{k \ell}\Bigl({\widehat x},{\widehat \pi}
\Bigr)\Bigr]\Biggr\} \thinspace \Bigl\vert \psi'_{\rm inv} ; t' \Bigl\rangle
\equiv \Bigl\vert \psi' ; t' \Bigl\rangle \eqno(6.21)$$
and apply the functional formulae $(6.8)$ and $(6.9)$. It is clearly convenien
for
this purpose to either assume that $t'=0$ or else to fix the residual gauge on
the surface $t=t'$. The relevant formulae for matrix elements and expectation
values are, respectively:
$$\eqalignno{\Bigl\langle \psi_2 ; t_2 \Bigl\vert \thinspace T
\Bigl({\cal O}[x&,
\pi]\Bigr) \thinspace \Bigr\Vert \psi_1 ; t_1\Bigr\rangle_G = \Fint \mathop{
\Bigl[dx(t)\Bigr]}\limits_{\scriptscriptstyle t_2 \geq t \geq t_1} \mathop{
\Bigl[d\pi(t')\Bigr]}\limits_{\scriptscriptstyle t_1 > t' \geq 0} \thinspace
\psi_2^*\Bigl(x(t_2)\Bigr) \thinspace {\cal O}\Bigl[x,\pi\Bigr] \cr
&\times \exp\Biggl[i\int_{t_1}^{t_2} dt \thinspace \Bigl\{\pi_{\alpha}(t)
\thinspace {\dot x}^{\alpha}(t) - H\Bigl(x(t),\pi(t),t\Bigr)\Bigr\}\Biggr]
&(6.22a) \cr
&\qquad \times \delta^K\Bigl[G\Bigl(x(t_1),\pi(t_1)\Bigr)\Bigr] \thinspace {\rm
abs}\Biggl\{{\rm det}\Bigl[M_{k \ell}\Bigl(x(t_1),\pi(t_1)\Bigr)\Bigr]\Biggr\}
\thinspace \psi_1\Bigl(x(t_1)\Bigr) \cr}$$
$$\eqalignno{\Bigl\langle \psi_2 ; t_1 \Bigl\vert &\thinspace T
\Bigl({\cal O}[x,
\pi]\Bigr) \thinspace \Bigr\Vert \psi_1 ; t_1\Bigr\rangle_G = \Fint \mathop{
\Bigl[dx_-(t)\Bigr]}\limits_{\scriptscriptstyle t_2 \geq t \geq t_1} \mathop{
\Bigl[d\pi^-(t')\Bigr]}\limits_{\scriptscriptstyle t_2 > t' \geq t_1} \mathop{
\Bigl[dx_+(t)\Bigr]}\limits_{\scriptscriptstyle t_2 \geq t \geq t_1} \mathop{
\Bigl[d\pi^+(t')\Bigr]}\limits_{\scriptscriptstyle t_2 > t' \geq t_1}
\thinspace
\delta\Bigl(x_-(t_2) - x_+(t_2)\Bigr) \cr
&\times \psi_2^*\Bigl(x_-(t_1)\Bigr) \thinspace \exp\Biggl[- i\int_{t_1}^{t_2}
dt \thinspace \Bigl\{\pi^-_{\alpha}(t) \thinspace {\dot x}_-^{\alpha}(t) -
H\Bigl(x_-(t),\pi^-(t),t\Bigr)\Bigr\}\Biggr] \cr
&\qquad \times {\cal O}\Bigl[x_+,\pi^+\Bigr] \exp\Biggl[i\int_{t_1}^{t_2} dt
\thinspace \Bigl\{\pi^+_{\alpha}(t) \thinspace {\dot x}_+^{\alpha}(t) -
H\Bigl(x_+(t),\pi^+(t),t\Bigr)\Bigr\}\Biggr] &(6.22b) \cr
&\qquad \qquad \times \delta^K\Bigl[G\Bigl(x(t_1),\pi(t_1)\Bigr)\Bigr]
\thinspace
{\rm abs}\Biggl\{{\rm det}\Bigl[M_{k \ell}\Bigl(x(t_1),\pi(t_1)\Bigr)\Bigr]
\Biggr\} \thinspace \psi_1\Bigl(x(t_1)\Bigr) \cr}$$
The reader should be aware that these expressions may require modification on
the  initial value surface to account for whatever operator ordering
prescription is imposed upon the paraphernalia of surface gauge fixing. If both
the time-ordered product of ${\cal O}[x,\pi]$ and the two states are invariant
then relation $(6.20)$ shows that $(6.22a)$ and $(6.22b)$ give the matrix
element and
expectation value, respectively, of the reduced operator $T\Bigl({\cal O}[x_r,
\pi^r]\Bigr)$. If ${\cal O}[x,\pi]$ is invariant {\it before} time-ordering
then
the necessary ordering corrections are the simple ones of the unconstrained
theory, not the potentially complicated ones of the reduced theory. That is, we
first convert the reduced operator ${\cal O}[x_r,\pi^r]$ into the analogous
unconstrained operator, ${\cal O}[x,\pi]$, inside the matrix element or
expectation value. Then we write it as the time-ordered product, $T\Bigl({\cal
O}[x,\pi]\Bigr)$, plus ordering corrections, and we use $(6.22)$ to evaluate
the
matrix element or expectation value of the time-ordered product. Since the
ordering corrections are also operators in the unconstrained theory we time
order {\it them} and apply $(6.22)$ again, repeating the procedure as often as
is
necessary to reduce everything to functional integrals.

If the states and the operator in $(6.22)$ are invariant then these expressions
are independent of the residual gauge condition $G^k=0$. However, even when the
wavefunctions or the functional ${\cal O}[x,\pi]$ are not manifestly invariant
the right hand sides of $(6.22a)$ and $(6.22b)$ still represent the matrix
elements
or expectation values of some invariant operator in the presence of some
invariant states. That this must be so follows from the fact that the gauge has
been completely fixed and {\it any} quantity becomes invariant when it is
defined in a unique gauge. To find {\it which} invariants one employs the
residual gauge condition to ``invariantize'' the wavefunctions and the operator
[21,26].

It is to some extent pointless to expend effort in order to discover a
manifestly invariant state or operator when the only way physical information
can be extracted from these objects is by taking gauge fixed inner products.
The
aesthetic advantage to using manifestly invariant wavefunctions and operators
is
that then matrix elements and expectation values are independent of the choice
of gauge. If we understand physics in a particular gauge --- as we often do ---
then this is not much of an advantage. The practical advantage is that manifest
invariance allows us, through relation $(6.20)$, to compute the matrix element
or
expectation value of a reduced operator ${\cal O}[x_r,\pi^r]$ using the same
matrix element or expectation value of the unconstrained operator ${\cal O}[x,
\pi]$. {\it We can therefore avoid the need to ever construct the reduced
Hamiltonian} which, as we have seen, can be a formidable task. Of course we can
always invariantize non-invariant states and operators, but then the
non-locality
and complicated field dependence this typically entails makes it difficult to
compute the time ordering corrections. However, we emphasize that the process
is
simple enough to carry out perturbatively --- witness section 4 --- and the
very
fact that only operator ordering corrections are needed to relate ${\cal O}[x,
\pi]$ to ${\cal O}[x_r,\pi^r]$ inside gauge fixed inner products shows that we
are well and truly free of the formal paralysis described as the paradox of
second coordinatization.

We turn now to a survey of how the general formalism manifests itself in the
three dynamical systems we have studied. For the coupled oscillator model of
section 2 the $x^{\alpha}(t)$'s are $q_1(t)$ and $q_2(t)$ while the $\pi_{
\alpha}(t)$'s are $p_1(t)$ and $p_2(t)$. The evolution functions $X^{\alpha}
\Bigl({\widehat x},{\widehat \pi},t\Bigr)$ and $\Pi_{\alpha}\Bigl({\widehat x},
{\widehat \pi},t\Bigr)$ can be read off from $(2.34)$. The unconstrained
Hamiltonian is:
$$H = \frac1{2m} p_1^2 + \frac1{2m} p_2^2 + \frac12 m \omega^2 \thinspace
\Bigl(
\frac54 \thinspace q_1^2 + q_1 \thinspace q_2 + \frac54 \thinspace q_2^2\Bigr)
\eqno(6.23)$$
Because it is quadratic in the momenta we can convert the canonical functional
integrals of $(6.22)$ into the more familiar, configuration space functional
integrals.

Recall that since the coupled oscillator model is not really a gauge theory we
imposed ${\widehat p}_2 = 0$ as an ersatz ``constraint.'' We might therefore
identify the single constraint by inverting $(2.34)$ to give ${\widehat p}_2$
in
terms of the time evolved variables:
$$\eqalign{C = \frac12 \Bigl[p_1(t) &+ p_2(t)\Bigr] \thinspace \cos
\Bigl(\frac32
\omega t\Bigr) + \frac12 \Bigl[- p_1(t) + p_2(t) \Bigr] \thinspace \cos\Bigl(
\frac12 \omega t\Bigr) \cr
&+ \frac34 m \omega \thinspace \Bigl[q_1(t) + q_2(t)\Bigr] \thinspace
\sin\Bigl(
\frac32 \omega t\Bigr) + \frac14 m \omega \thinspace \Bigl[- q_1(t) + q_2(t)
\Bigr] \thinspace \sin\Bigl(\frac12 \omega t\Bigr) \cr} \eqno(6.24)$$
The most general invariant operator is a function of ${\widehat q}_1$,
${\widehat p}_1$ and ${\widehat p}_2$. We can of course express these initial
value operators in terms of the evolved operators. The result for ${\widehat
p}_2$ is just $(6.24)$; the results for ${\widehat q}_1$ and ${\widehat p}_1$
are:
$$\eqalign{{\widehat q}_1 = \frac12 \Bigl[q_1(t) &+ q_2(t)\Bigr] \thinspace
\cos\Bigl(\frac32 \omega t\Bigr) + \frac12 \Bigl[q_1(t) - q_2(t) \Bigr]
\thinspace \cos\Bigl(\frac12 \omega t\Bigr) \cr
&+ \frac1{3 m \omega} \thinspace \Bigl[p_1(t) + p_2(t)\Bigr] \thinspace
\sin\Bigl(\frac32 \omega t\Bigr) + \frac1{m \omega} \thinspace \Bigl[- p_1(t)
+ p_2(t)\Bigr] \thinspace \sin\Bigl(\frac12 \omega t\Bigr) \cr} \eqno(6.25a)$$
$$\eqalign{{\widehat p}_1 = \frac12 \Bigl[p_1(t) &+ p_2(t)\Bigr] \thinspace
\cos\Bigl(\frac32 \omega t\Bigr) + \frac12 \Bigl[p_1(t) - p_2(t) \Bigr]
\thinspace \cos\Bigl(\frac12 \omega t\Bigr) \cr
&+ \frac34 m \omega \thinspace \Bigl[q_1(t) + q_2(t)\Bigr] \thinspace
\sin\Bigl(
\frac32 \omega t\Bigr) + \frac14 m \omega \thinspace \Bigl[q_1(t) - q_2(t)
\Bigr]
\thinspace \sin\Bigl(\frac12 \omega t\Bigr) \cr} \eqno(6.25b)$$
``Invariant'' states are those whose wavefunctions are independent of $q_2$ in
the $t=0$ basis of position eigenkets. Since invariant operators are
independent
of ${\widehat q}_2$ it is clear why any ${\widehat p}_2$'s can also be
neglected
when acting on an invariant state.

Since the residual gauge condition is $G = {\widehat q}_2$ the associated
Faddeev-Popov matrix is:
$$M = - 1 \eqno(6.26)$$
The single canonical pair $({\widehat q},{\widehat p})$ is of course just
$({\widehat q}_1,{\widehat p}_1)$; the pure gauge variable is
${\widehat g}^{~k}
\longrightarrow {\widehat q}_2$, and the constrained variable is
${\widehat c}_{
\ell} \longrightarrow {\widehat p}_2$. Note that the unconstrained operators
$q_i(t)$ and $p_i(t)$ are not invariant because they depend upon ${\widehat q
}_2$. This means that products of them do not give products of the analogous
reduced operators in matrix elements and expectation values. For example, it is
straightforward to show that in this gauge:
$$\eqalign{\Bigl\langle &\psi_{\rm inv} ; t \Bigl\vert \thinspace q_1(s)
\thinspace q_1(s') \thinspace \Bigr\Vert \psi'_{\rm inv} ; t' \Bigr\rangle_G =
\Bigl\langle \psi_{\rm inv} ; t \Bigl\vert \thinspace q^r_1(s) \thinspace
q^r_1(s') \thinspace \Bigr\Vert \psi'_{\rm inv} ; t' \Bigr\rangle_G \cr
&- \frac{i}{m \omega} \thinspace \Bigl[\frac13 \sin\Bigl(\frac32 \omega s\Bigr)
- \sin\Bigl(\frac12 \omega s\Bigr)\Bigr] \Bigl[\frac12 \cos\Bigl(\frac32 \omega
s'\Bigr) - \frac12 \cos\Bigl(\frac12 \omega s'\Bigr)\Bigr] \thinspace
\Bigl\langle \psi_{\rm inv} ; t \Bigr\Vert \psi'_{\rm inv} ; t' \Bigr\rangle_G
\cr} \eqno(6.27)$$
The terms on the extreme right represent unphysical and gauge dependent
information that is included in the expectation value or matrix element of the
unconstrained, non-invariant operator.

For scalar QED in temporal gauge the $x^{\alpha}(t)$'s are the fields $\phi(t,
{\vec x})$, $\phi^*(t,{\vec x})$ and $A_i(t,{\vec x})$; the $\pi_{\alpha}(t)$'s
are $\pi(t,{\vec x})$, $\pi^*(t,{\vec x})$ and $E_i(t,{\vec x})$. Owing to the
interaction, there is no closed form expression for the evolution functions
$X^{\alpha}\Bigl({\widehat x},{\widehat \pi},t\Bigr)$ and $\Pi_{\alpha}\Bigl(
{\widehat x},{\widehat \pi},t\Bigr)$. The unconstrained Hamiltonian is some
ordering of $(3.6)$. Since it is quadratic we can convert the canonical
functional
formalism into the configuration space version.

The constraint --- of which there is one for each space point ${\vec x}$ --- is
some ordering of $(3.8)$. A typical invariant operator is:
$$\eqalign{{\cal O}_{\rm inv} = \phi(t,{\vec x}) \thinspace &\exp\Biggl[ i e
\thinspace \int d^3x' \thinspace G\Bigl({\vec x}; {\vec x}{~'}\Bigr) \thinspace
\partial_i \thinspace A_i(t,{\vec x})\Biggr] \cr
&\times E_j(u,{\vec y}) \thinspace \pi(v,{\vec z}) \thinspace \exp\Biggl[- i e
\thinspace \int d^3z' \thinspace G\Bigl({\vec z}; {\vec z}{~'}\Bigr) \thinspace
\partial_k \thinspace A_k(v,{\vec z})\Biggr] \cr} \eqno(6.28)$$
where $G\Bigl({\vec x}; {\vec x}{~'}\Bigr)$ is the Green's function first used
in $(3.13)$.
The exponential term associated with each of the charged fields creates the
associated longitudinal Coulomb field. A typical invariant wavefunctional is
that of the free vacuum which, up to a normalization factor, is:
$$\eqalign{\Psi_{\rm inv}\Bigl[A,\phi,\phi^*\Bigr] \propto \exp\Biggl[- \int
d^3x \thinspace &\Phi^*({\vec x}) \thinspace \sqrt{-\nabla^2} \thinspace
\Phi({\vec x})\Biggr] \cr
&\times \exp\Biggl[- \frac14 \int d^3x F_{ij}({\vec x}) \thinspace {1 \over
\sqrt{-\nabla^2}} \thinspace F_{ij}({\vec x})\Biggr] \cr} \eqno(6.29a)$$
$$\Phi({\vec x}) \equiv \phi({\vec x}) \thinspace \exp\Biggl[i e \int d^3x'
\thinspace G\Bigl({\vec x}; {\vec x}{~'}\Bigr) \thinspace \partial_i \thinspace
A_i({\vec x})\Biggr] \eqno(6.29b)$$
The apparent coupling between the vector potential and the charged fields is a
gauge fiction; it disappears when the instantaneous Coulomb gauge condition
$(3.11)$ is used. This is an example of why it is pointless to struggle to
achieve
manifest invariance in an expression which must ultimately rest against a gauge
fixing delta functional. Note as well that $(6.28)$ and $(6.29)$ are only
invariant
with respect to the residual, time independent gauge symmetry.

For instantaneous Coulomb gauge, $(3.11)$, the Faddeev-Popov matrix is:
$$M\Bigl({\vec x};{\vec y}\Bigr) = - \frac{\partial}{\partial x^i} \frac{
\partial}{\partial y^i} \thinspace \delta^3\Bigl({\vec x} - {\vec y}\Bigr)
\eqno(6.30)$$
Note that it has a non-singular determinant on the appropriate function space.
The most natural identification of canonical, gauge and constrained operators
on
the initial value surface is:
$${\widehat q}^{~a} \longrightarrow \Bigl\{{\widehat A}^T_i({\vec x}),
{\widehat
\phi}({\vec y}), {\widehat \phi}^*({\vec z})\Bigr\} \eqno(6.31a)$$
$${\widehat p}_a \longrightarrow \Bigl\{{\widehat E}^T_i({\vec x}), {\widehat
\pi}({\vec y}), {\widehat \pi}^*({\vec z})\Bigr\} \eqno(6.31b)$$
$${\widehat g}^k \longrightarrow A^L_i({\vec x}) \eqno(6.31c)$$
$${\widehat c}_k \longrightarrow E^L_i({\vec x}) \eqno(6.31d)$$
Note that although these initial operators do commute canonically --- on the
appropriate function space --- we saw in section 3 that their time evolved
versions do not. If we assume that $t > u > v > 0$ then the invariant operator
$(6.28)$ is also time ordered and we can write:
$$\eqalignno{\Bigl\langle \Psi_{\rm inv} ; t &\Bigl\vert \thinspace T\Bigl(
{\cal O}_{\rm inv}\Bigr) \thinspace \Bigr\Vert \Psi_{\rm inv} ; 0\Bigr\rangle_G
= \Fint \mathop{\Bigl[dA_{\mu}(s,{\vec w})\Bigr] \Bigl[\phi(s,{\vec w})\Bigr]
\Bigl[d\phi^*(s,{\vec w})\Bigr] \thinspace \delta\Bigl[A_0(s,{\vec w})\Bigr]}
\limits_{t \geq s \geq 0 \thinspace , \thinspace {\vec w} \in R^3} \cr
&\times \Psi^*_{\rm inv}\Bigl[A(t,{\vec w}), \phi(t,{\vec w}), \phi^*(t,{\vec
w})\Bigr] \thinspace \phi(t,{\vec x}) \thinspace \exp\Bigl[ i e \thinspace
\int d^3x' \thinspace G\Bigl({\vec x}; {\vec x}{~'}\Bigr) \thinspace \partial_i
\thinspace A_i(t,{\vec x})\Bigr] \cr
&\qquad \times F_{0j}(u,{\vec y}) \thinspace {\dot \phi}^*(v,{\vec z})
\thinspace \exp\Biggl[- i e \thinspace \int d^3z' \thinspace G\Bigl({\vec z};
{\vec z}{~'}\Bigr) \thinspace \partial_k \thinspace A_k(v,{\vec z})\Biggr]
&(6.32) \cr
&\qquad \qquad \times \exp\Biggl[i \int_0^t ds \int d^3w \thinspace {\cal L}
\Biggr] \thinspace \delta\Bigl[\partial_{\ell} \thinspace A_{\ell}(0,{\vec w})
\Bigr] \thinspace \Psi_{\rm inv}\Bigl[A(0,{\vec w}), \phi(0,{\vec w}),
\phi^*(0,
{\vec w})\Bigr] \cr}$$
where expression $(3.1)$ defines the Lagrangian ${\cal L}$. Note that we have
integrated out the momenta and that we have introduced $A_0$ as an integral
over
the volume gauge fixing delta functional. Since the operator $(6.28)$ is
invariant, as are the states, it follows that this functional integral
represents the matrix element of the corresponding reduced operator. The
interested reader can worked out examples from standard QED in ref. [26].

For quantum general relativity at fixed lapse and shift the $x^{\alpha}(t)$'s
are the 3-metrics, $\gamma_{ij}(t,{\vec x})$; the $\pi_{\alpha}(t)$'s are their
conjugate momenta, $\pi^{ij}(t,{\vec x})$. It is no more possible than for
scalar QED to give explicit forms for the evolution functions, $X^{\alpha}
\Bigl(
{\widehat x},{\widehat \pi},t\Bigr)$ and $\Pi_{\alpha}\Bigl({\widehat x},
{\widehat \pi},t\Bigr)$. The Hamiltonian is some operator ordering of the
classical result:
$$H\Bigl[\gamma,\pi\Bigr](t) = \int d^3x \thinspace N^{\mu}\Bigl[\gamma,\pi
\Bigr](t,{\vec x}) \thinspace {\cal H}_{\mu}\Bigl[\gamma,\pi\Bigr](t,{\vec x})
\eqno(6.33)$$
Note that the lapse and shift may depend upon time and also upon the dynamical
variables; in fact dependence upon the dynamical variables is necessary
classically in order to have a chance of avoiding the evolution of coordinate
singularities. Although no one has ever exhibited a gauge which is classically
free of coordinate singularities its existence seems obvious if a sufficient
amount of field dependence and non-locality is permitted in the lapse and
shift.
In any case we shall assume that such a gauge exists.

Note that if the lapse and shift {\it do} depend upon the $\pi^{ij}$'s then the
canonical action may not be quadratic in the momenta and we would not be able
to
reach the usual configuration space functional formalism. Although
inconvenient,
there is no inconsistency in this. It could happen as well in scalar QED if we
had allowed $A_0$ to depend upon $E_i$, $\pi$ or $\pi^*$. Note also that even
if
the lapse and shift depend only upon $\gamma_{ij}$, integrating out the momenta
will give a field dependent measure factor. If the lapse depends {\it
non-locally} upon $\gamma_{ij}$ then this measure factor can be non-local, and
it
may therefore give non-zero contributions even in dimensional regularization.
Again, this is not specific to gravity; we could have made it happen as well
for
scalar QED by permitting $A_0$ to depend non-locally upon the fields.

There are four constraints for each space point ${\vec x}$. Without regard to
operator ordering they are:\footnote{*}{It is futile to pay much attention to
ordering so long as the problem with renormalization precludes being able to
take the unregulated limit of inner products [27].}
$${\cal H}_0 = {\kappa^2 \over \sqrt{\gamma}} \thinspace \Bigl(\gamma_{ik}
\thinspace \gamma_{j \ell} - \frac12 \gamma_{ij} \thinspace \gamma_{k \ell}
\Bigr) \thinspace \pi^{ij} \thinspace \pi^{k \ell} - {1 \over \kappa^2} \Bigl(
R - 2 \Lambda\Bigr) \thinspace \sqrt{\gamma} \eqno(6.34a)$$
$${\cal H}_i = - 2 \thinspace \gamma_{ij} \thinspace \pi^{jk}_{~~; k}
\eqno(6.34b)$$
It is reasonably straightforward to construct a basis of operators which
transform as scalar densities under an arbitrary diffeomorphism. The method is
to evaluate Riemann tensors at the ends of geodesic segments emanating from a
central origin, with the indices of each Riemann tensor defined in the local
inertial frame field of the origin as obtained by parallel transport along the
connecting geodesic [28]. If the coordinate time generated by $(6.33)$ can be
extended infinitely then full diffeomorphism invariance results when the origin
is integrated over $R \times M^3$. Since such operators are invariant under all
diffeomorphisms they are necessarily invariant under those of the residual
symmetry group which preserve the fixed lapse and shift.

No one has obtained closed form expressions for normalizable invariant {\it
states}, but one can of course construct them order by order in perturbation
theory. It is the fact that the asymptotic functional formalism does this
automatically which explains the observation in [26] and [28] that all of
Einstein's equations are obeyed in regulated perturbation theory. Note also
that
there is a considerable phenomenological advantage to building invariant states
perturbatively: one knows what they mean this way. If given a solution of the
constraints it is difficult to tell what level of excitation it represents
since
on a closed spatial manifold all states are degenerate with zero energy.
However, if this solution is perturbatively related to the invariantized free
vacuum then one can assume --- at least as a first guess --- that it represents
empty space. Similar comments apply to the wavefunctionals built up from other
Fock space states.

In section 4 we reduced the theory in two steps. In the first step we enforced
the constraints that fixed $h^t$ and $p_i$ and surface gauge fixed their
conjugate variables. Explicitly:

\noindent The $C_k$'s are:
$$\eqalignno{
	h^t&=0+\o1 &(6.35a)\cr
	p_i&=0+\o1 &(6.35b)
	    }
$$
while the $G_k$'s are:
$$\eqalignno{
	{\widehat p}^t & =0 &(6.36a)\cr
	{\widehat h}_i & =0 &(6.36b)
	    }
$$
The Faddeev-Popov determinant arising from this first step of reduction is just
a $\comp$-number to lowest order. This is not the case for the second step. In
the second step we enforced the constraint on $p^{tr}$ and surface gauge fixed
$h^{tr}$ using:
$$
{\left(p^{tr}\right)}^2 = 6 \, E_0 + \o1 \eqno(6.37a)
$$
$$
{\widehat h}^{tr}=0 \eqno(6.37b)
$$
We see now that the Faddeev-Popov matrix for this, the second step is
$2\, \widehat p^{tr}$. The choice of variables on the initial value surface
goes as
follows:
$$
\widehat q^{~a} \longrightarrow  \widehat h^{tt}  \eqno(6.38a)
$$
$$
\widehat p_a \longrightarrow  \widehat p^{tt}  \eqno(6.38b)
$$
$$
\widehat g^k \longrightarrow \Big\{ \widehat p^t , \widehat h^i ,
\widehat h^{tr} \Big\} \eqno(6.38c)
$$
$$
\widehat c_k \longrightarrow \Big\{ \widehat h^t , \widehat p^i ,
\widehat p^{tr} \Big\} \eqno(6.38d)
$$
We will see next how the inner product breaks up in two parts. One part for
negative $p^{tr}$ and one for positive, representing and expanding and a
contracting space respectively.

Let us look at the inner product defined in $(6.18)$:
$$\eqalignno{
\Bigl\langle \psi_2 ; t_2 \Bigl\Vert \psi_1 ; t_1 \Bigr\rangle_G &=
\Bigl\langle \psi_2 ; t_2 \Bigl\vert \thinspace \delta^K\Bigl[G\Bigl({\widehat
x},{\widehat \pi}\Bigr)\Bigr] \thinspace {\rm abs}\Biggl\{{\rm det} \Bigl[M_{k
\ell}\Bigl({\widehat x},{\widehat \pi}\Bigr)\Bigr]\Biggr\} \thinspace
\Bigr\vert \psi_1 ; t_1\Bigr\rangle &(6.18) \cr
&\simeq \Bigl\langle \psi_2 ; t_2 \Bigl\vert \thinspace \delta \Bigl[\widehat
h^{tr} \Bigr] {\rm abs} \, \Biggl\{ 2\,p^{tr} \Biggr\}
\Bigr\vert \psi_1 ; t_1\Bigr\rangle &(6.39)
}
$$
Where we used $\simeq$ because we are disregarding overall multiplicative
factors and we are working to lowest order in $\kappa$. As mentioned in section
4 each wavefunction is divided in two parts depending on the action of
$p^{tr}$ upon them:
$$
\Bigr\vert \psi ; t\Bigr\rangle = \Bigr\vert \psi^{+} ; t
\Bigr\rangle + \Bigr\vert \psi^{-} ; t\Bigr\rangle \eqno(6.40)
$$
The inner product of $(6.39)$ can then be written as:
$$\eqalignno{
\Bigl\langle \psi_2 ; t_2 \Bigl\Vert \psi_1 ; t_1 \Bigr\rangle_G &=
\Bigl\langle \psi_2 ; t_2 \Bigl\vert \Bigg[ \Theta \Bigl(p^{tr}\Bigr) \, \delta
\Bigl(h^{tr} \Bigr) \, p^{tr} \, \Theta \Bigl(p^{tr}\Bigr) +
\Theta \Bigl(p^{tr}\Bigr)\,p^{tr}\,\delta \Bigl(h^{tr} \Bigr) \, \Theta \Bigl(
p^{tr}\Bigr) \cr &- \Theta \Bigl(-p^{tr}\Bigr) \, \delta
\Bigl(h^{tr} \Bigr) \,p^{tr}\, \Theta \Bigl(-p^{tr}\Bigr) -
\Theta \Bigl(-p^{tr}\Bigr) \, p^{tr} \, \delta
\Bigl(h^{tr} \Bigr) \, \Theta \Bigl(-p^{tr}\Bigr) \Bigg]
\Bigr\vert \psi_1 ; t_1\Bigr\rangle\cr & &(6.41)
}$$
Where we have chosen a Hermitian ordering. Using $(6.40)$ equation $(6.41)$
becomes:
$$\eqalignno{
\Bigl\langle \psi_2 ; t_2 \Bigl\Vert \psi_1 ; t_1 \Bigr\rangle_G &=
\Bigl\langle \psi^+_2 ; t_2 \Bigl\vert \Bigg[ \delta \Bigl(h^tr\Bigr) \,
|p^{tr}| + |p^{tr}| \, \delta \Bigl(h^{tr}\Bigl)\Bigg]\Bigr\vert \psi^+_1 ; t_1
\Bigr\rangle \cr
&+\Bigl\langle \psi^-_2 ; t_2 \Bigl\vert \Bigg[ \delta \Bigl(h^tr\Bigr) \,
|p^{tr}| + |p^{tr}| \, \delta \Bigl(h^{tr}\Bigl)\Bigg]\Bigr\vert \psi^-_1 ; t_1
\Bigr\rangle &(6.42)}$$
We see that, as previously promised, the inner product breaks up into two
parts. Both of these parts are present quantum mechanically; however,
classically either $\psi^+=0$ {\it or} $\psi^-=0$.

\vskip 1cm
\centerline{\bf 7. Conclusions}

The context of our analysis is that of a gauge theory in which the ability to
perform local, time dependent transformations has been fixed but there is still
a residual gauge symmetry characterized by how it acts on the initial value
surface. Examples of such theories include scalar electrodynamics in temporal
gauge, which was studied in section 3, and general relativity with fixed lapse
and shift, which was studied in section 4. The initial value problem has a
unique solution in this setting --- which it does not in the fully invariant
theory --- because residual symmetry transformations cannot change the time
evolved dynamical variables without also changing their initial values. We
assume that a canonical formalism describes this initial value problem; we
assume as well that the residual gauge symmetry is generated in the usual way
by
functions of the dynamical variables which are constrained to vanish as a
consequence of the Euler-Lagrange equations of the (volume) gauge fixed
variables. This is what we mean by the ``unconstrained theory,'' and its
generic
dynamical variables are the $x^{\alpha}(t)$'s and $\pi_{\beta}(t)$'s of section
6.

The generic reduced theory is obtained by fixing the residual gauge freedom,
imposing the constraints, and identifying a subset, $\Bigl\{v^i(t)\Bigr\}$, of
the original dynamical variables which gives a complete and minimal
representation of physics. What this means is that the constraints, the
residual
gauge conditions and the values of the $v^i(t)$'s at any time $t$ uniquely
determine the $x^{\alpha}(t)$'s and the $\pi_{\beta}(t)$'s. The reduced
dynamical variables inherit their time evolution and their equal time bracket
(or commutation) algebra from the unconstrained theory. This is true even when
the unconstrained Hamiltonian vanishes after reduction; it is even true if {\it
no} Hamiltonian exists which generates the time evolution of the $v^i(t)$'s.

Section 2 described a standard construction of the last century for identifying
canonical variables $q^a(t)$ and $p_b(t)$ in the reduced theory, and for
finding
the non-zero Hamiltonian which generates their time evolution. The
identification
of a reduced canonical formalism is not unique. On the classical level we can
vary it by performing canonical transformations; with suitable attention to
operator ordering there is a similar class of transformations on the quantum
level. What keeps {\it physics} unchanged is the fact that measurement theory
is
based upon the $x^{\alpha}(t)$'s and the $\pi_{\beta}(t)$'s, considered as
functions of the reduced canonical variables. As our identification of the
reduced canonical variables changes, the functional dependence of the unreduced
canonical variables changes so as to keep the $x^{\alpha}(t)$'s and
$\pi_{\beta}
(t)$'s the same.

An immediate consequence of the multiplicity of reduced canonical formalisms is
that the associated Hamiltonians have no physical significance independent of
whatever meaning attaches to the reduced canonical variables whose evolution
they generate. Although this may seem strange, quantum field theorists are
familiar with the phenomenon through the ``interaction representation.'' This
is
a field redefinition in which any perturbatively well defined interacting
quantum field theory is transformed into the corresponding free quantum field
theory. We do not conclude that all perturbatively well defined quantum field
theories are free because we insist that physics be inferred from the original
fields.

The construction of a reduced canonical formalism is irrelevant to most issues
in classical physics. This is because we infer physics from the $x^{\alpha}(t)
$'s and $\pi_{\beta}(t)$'s, and we may as well solve for these variables
directly in the unconstrained theory, starting from initial value data which
obey the constraints. The principle motivation for erecting a reduced canonical
formalism is as a prelude to applying canonical quantization. This is
especially
relevant to systems, such as gravity on a spatially closed manifold, for which
reduction causes the unconstrained Hamiltonian to vanish. As has been noted, we
did not invent the construction discussed in section 2; this was done by the
classical physicists of the previous century [5,6,10-15]. Our contribution is
rather to propose that quantum gravity should be defined by canonically
quantizing a reduced canonical formulation of whatever is the correct theory of
gravity.

We constructed explicit reduced canonical formalisms for the coupled oscillator
model of section 2, and for scalar electrodynamics in section 3. Although no
complete construction seems possible for general relativity we showed in
section
4 how to do it perturbatively around a flat background on $T^3 \times R$. We
also gave explicit constructions for a handful of minisuperspace truncations in
section 5. Our inability to give an explicit, non-perturbative construction for
the complete theory of general relativity does not pose a practical barrier ---
even for the study of non-perturbative phenomena --- because of the relation we
were able to obtain in section 6 between the relatively simple quantum
mechanics
of the unconstrained theory and that of the reduced theory. Equation $(6.20)$
asserts that the matrix elements or expectation values of invariant functionals
of the reduced operators are {\it equal} to the matrix elements or expectation
values of the same functionals of the unconstrained operators in the presence
of
invariant states. Explicit functional integral representations exist for
unconstrained matrix elements and expectation values. These representations are
as simple to evaluate as it seems possible for anything to be in an interacting
quantum field theory, and they have at least the potential for extension beyond
perturbation theory. {\it Thus it is not really necessary to explicitly
construct a reduced canonical formalism; it suffices to know that such a
construction exists and that we can study it using the far simpler formalism of
the unconstrained theory.}

It is worth reviewing how the formulation we propose for canonical quantum
gravity avoids the four correspondence paradoxes mentioned in section 1. Our
resolution to the paradox of second coordinatization is that the lapse and the
shift determine what is meant by time evolution in quantum gravity the same way
they do for classical gravity.\footnote{*}{A tangential point concerns the
order
of gauge fixing. Our volume gauge condition determines the lapse and shift as
functionals of $\gamma_{ij}(t,{\vec x})$, $\pi^{ij}(t,{\vec x})$, and possibly
also of time and space; we then impose a surface gauge condition upon the
$\gamma_{ij}$'s and $\pi^{ij}$'s, and use the constraints to determine the
evolution of some components of the $\gamma_{ij}$'s and $\pi^{ij}$'s. Many
researchers [3,7,8,9] prefer to impose a volume gauge condition upon $\gamma_{i
j}(t,{\vec x})$ and  $\pi^{ij}(t,{\vec x})$; they then use the constraint
equations, with some surface gauge condition, to solve for the lapse and shift.
An example of our method in scalar electrodynamics is fixing $A_0(t{\vec x}) =
0$ as the volume gauge condition, and then using the constraints and the
surface
condition, $\partial_i A_i(0,{\vec x}) = 0$, to determine the longitudinal
field
components. An example of the other procedure would be fixing
$\partial_i A_i(t,
{\vec x}) = 0$ as the volume gauge, and then using the constraint equation plus
the freedom to perform time dependent, harmonic gauge transformations, to
determine $A_0(t,{\vec x})$. Our method gives the best chance of defining a
successful evolution in gravity because it allows one to adjust the rate of
evolution in response to what the fields are doing. In this way we can avoid
coordinate singularities, and also --- by merely slowing down before they form
---  the evolution of physical singularities. Of course if the other method
gives a successful evolution it can always be transcribed into our form by
merely regarding the derived solutions for the lapse and shift as volume gauge
conditions.} It is neither necessary, nor even particularly desirable, to try
doing the job twice by identifying some other variable as ``time'' and then
attempting to interpret the Wheeler-DeWitt constraint as a Schr\"odinger
evolution
equation. Our method works because the Heisenberg field operators depend upon
the time fixed by the lapse and the shift, whether or not we restrict the
initial value data by fixing the residual gauge and imposing the constraints.
The conventional method will only work if one of the Heisenberg operators is an
{\it invertible} function of time, and the method will only produce tractable
results if this time dependence is relatively simple. Gravity is a formidably
complex dynamical system and it may somewhere harbor such an operator; it is
certain though that no one has found it yet, despite years of search.

A key point of our proposal is that one should infer physics the same way in
the
quantum theory as in the classical theory: by studying the reduced variables,
$x_r^{\alpha}(t)$ and $\pi^r_{\beta}(t)$. The fact that these variables are not
manifestly invariant before gauge fixing is irrelevant. {\it Any} quantity can
be made gauge invariant by defining it in a particular gauge and this is just
what fixing the residual gauge freedom does. No classical relativist would
demur from modeling planetary orbits using the geodesics of the Schwarzschild
metric just because the latter was reported in Schwarzschild coordinates. Our
formalism permits the same indifference on the quantum level. Correspondence in
the limit that $\hbar$ vanishes is manifest through relation $(6.20)$ and the
functional expressions $(6.22)$ for unconstrained matrix elements and
expectation
values.

Our resolution to the paradox of dynamics is that proper correspondence should
pertain in the limit that $G$ vanishes if that the reduced canonical variables
are taken to include the canonical variables of the pure matter theory. In this
case the Hamiltonian of the reduced canonical formalism for gravity plus matter
will go over into that of free gravitation around some background, plus the
pure matter Hamiltonian for that background. The same correspondence will
obviously apply for inner products. We have no general proof that a suitable
choice exists for the reduced canonical variables but, if it does, then the
result follows automatically from the correspondence limit of the evolution
equations.\footnote{*}{That the field equations obey the correct correspondence
limit is not in doubt.} We feel it is plausible that such a choice exists, and
we have given an explicit example in the minisuperspace truncations of section
5.

Our resolution to the paradox of topology is that a correspondence limit should
exist between the non-zero Hamiltonians of infinite, spatially open manifolds
[3,4] and the reduced Hamiltonians of spatially closed manifolds. This
correspondence will exist when the reduced canonical variables are chosen to
include the canonical variables of the open space theory, and when the
coordinate volume of the closed space is taken to infinity in such a way that
the initial value data obey the asymptotic condition of the open space. The
idea
is that in such a situation causality prevents the closed topology from
reaching
around the universe to influence the localized initial disturbances. As with
the
paradox of dynamics, the desired correspondence follows immediately from the
field equations provided that a suitable choice can be made for the reduced
canonical variables. We have no general proof that such a choice exists but we
did show in section 4 that it does for gravity on the manifold $T^3 \times R$.

We comment that if the aforementioned topological correspondence limit holds
generally then the $2+1$ dimensional constructions of Moncrief [7], Hosoya and
Nakao [8] and of Carlip [9] would be special cases of the general formalism
described in section 2. The same would be true of the $3+1$ dimensional
constructions of A.D.M.[3], and of Deser and Abbott [4]. Indeed, the method of
section 2 seems to provide the long sought unifying principle needed to define
energy on a space of arbitrary topology.

We do not avoid the paradox of stability by appealing to the fact that reduced
Hamiltonians have non-trivial spectra. As has been noted, what the reduced
Hamiltonian is depends upon how we identify the reduced canonical variables, so
its spectrum for an arbitrary identification has no intrinsic significance. Nor
are reduced Hamiltonians typically conserved. They do not therefore fill the
usual role of constraining the way in which a theory can excite its various
degrees of freedom. In fact for gravity on a spatially closed manifold the only
generally conserved energy functional is the unconstrained Hamiltonian which
vanishes upon reduction. This means that all states really are degenerate and
hence that the universe is liable to evaporate into pairs in the manner of
section 1.

That we are here is a consequence --- assuming that our spatial manifold is
closed --- of causality and of the weakness of the gravitational interaction.
The $H=0$ constraint is not met, as is sometimes supposed, by a vast reduction
in the number of possible states compared with gravity on an open space. The
perturbative analysis of section 4 shows that there are at least as many
positive energy graviton modes on $T^3 \times R$ as on $R^3 \times R$; in fact
the $H=0$ constraint is enforced by the global negative energy mode, $p^{tr}$.
We conjecture that this is the case generally; that is, positive energy modes
can only be excited by corresponding excitations of a global negative energy
mode. But a global mode by definition pervades the spatial manifold, so
exciting
it requires a similarly extensive process. On large manifolds causality imposes
a formidable barrier to such excitation. There is an additional barrier in the
fact that the global mode can only be excited gravitationally. Since
gravitational interactions are typically very weak in our current universe they
must proceed slowly. Note that neither barrier would apply to a strongly
gravitating system of small physical volume. One consequence of our work is the
prediction that such systems ought to be unstable.

The barrier causality imposes against instability becomes absolute in the limit
that the coordinate volume goes to infinity while the initial value data are
only locally disturbed from a vacuum solution. In this limit the global mode
decouples, and both its (negative) energy and the (positive) energy of the
local
modes become separately conserved. This is how we can approach the conserved
Hamiltonians [3,4] of spatially open manifolds. The gravitational barrier
becomes similarly infinite in the limit that $G$ vanishes. In this limit the
negative mode again decouples --- along with all the other gravitational modes
--- and the energies of matter and of each gravitational mode become separately
conserved. This is how we can approach the conserved Hamiltonians of pure
matter
theories.

We should comment as well on a venerable argument sometimes used to deny the
possibility of non-zero Hamiltonians for gravity on closed spatial manifolds.
The
argument begins with the observation that any such Hamiltonian would have to be
the integral of a local function of the metric and its first derivative. If it
is to have physical significance such a Hamiltonian must also be an invariant,
but there are no invariant functions of the metric and its first derivative.
Therefore the Hamiltonian of any theory with dynamical gravity must be zero
when
the spatial manifold is closed. We evade this argument by the simple device of
permitting the Hamiltonian to be non-invariant. It generates time evolution in
a
particular coordinate system --- the one provided by the fixed lapse and shift
--- so it can and should be a gauge dependent object. The physical energy, on
the other hand, must really be an invariant, and we have seen that it {\it is}
zero. There is no contradiction between these two facts because the non-zero
Hamiltonian that generates time evolution for a reduced dynamical system need
not be the physical energy, neither must it be necessarily conserved. We
cobbled
together a simple example of this with the coupled oscillator system of section
2, and the phenomenon is generic in gravity.

We close with an admonition to those who would prefer to continue searching for
a conventional solution to the ``problem of time'' --- i.e., the identification
of some component of the metric as physical time and the interpretation of the
Wheeler-DeWitt constraint as a Schr\"odinger evolution equation. Though our
frank
admission that time is a gauge choice and our frequent appeal to perturbation
theory in implementing this gauge choice may seem ugly, the construction of
section 2 is not intrinsically tied to perturbation theory --- witness the
functional representations of section 6 --- and it has the great advantage that
one can tailor the coordinate system to the state and the operator under study.
There is no reason why a ``magic bullet'' should even exist for identifying
physical time, much less that it should be a simple functional of the metric.
Nor is it in accord with our experience in other branches of physics to assume
that a single calibration should govern all conceivable measurements. We do not
insist, for example, that condensed matter experimentalists use the same
thermometers in the microkelvin regime as in the heart of an exploding, high
field magnet. Nor do we deny the validity or utility of measuring temperature
just because different techniques are employed in different regimes. What we
insist upon is rather that any two methods should agree in those environments
for which they both apply. There is no reason to impose a more stringent
standard in the vastly more varied and extreme environments imaginable to a
theorist in quantum gravity.

\vskip 1cm
\centerline{ACKNOWLEDGEMENTS}

We have profitted from conversations and correspondence
with A. O. Barvinsky, S.
Carlip, S. Deser and A. Higuchi. This work was partially supported by DOE
contract DOE-DE-FG05-86-ER40272 and by a GAANN fellowship from the Department
of
Education.

\references
\doublespace

[1] P. A. M. Dirac, {\sl Proc. R. Soc. London} {\bf A246} (1958) 333.

[2] K. Kucha{\v r}, in {\it Conceptual Problems of Quantum Gravity}, ed. A.
Ashtekar and J. Stachel (World Scientific, Singapore, 1991); \hfill\break
C. J. Isham, ``Canonical Quantum Gravity and the Problem of Time,'' Imperial
College preprint TP/91-92/25.

[3] R. Arnowitt, S. Deser and C. Misner, in
{\it Gravitation: An Introduction to
Current Research}, ed. L. Witten (Wiley, New York, 1962).

[4] L. F. Abbott and S. Deser, {\sl Nucl. Phys.} {\bf B195} (1982) 76.

[5] S. Lie, {\sl Archiv for Math. og Natur.} {\bf II} (1877) 10.

[6] G. Koenigs, {\sl Comptes Rendus} {\bf CXXXI} (1895) 875.

[7] V. Moncrief, {\sl J. Math. Phys.} {\bf 30} (1989) 2907; {\bf 31} (1990)
2978.

[8] A. Hosoya and K. Nakao, {\sl Class. Quantum Grav.} {\bf 7} (1990) 163;
{\sl Prog. Theor. Phys.} {\bf 84} (1990) 739.

[9] S. Carlip, ``Time in $2 + 1$ Dimensional Quantum Gravity,'' talk given at
the Banff Conference on
Gravitation, Banff, Alberta, Canada, August 12-24, 1990.

[10] J. F. Pfaff, {\sl Abhandl. Akad. der Wiss.} (1814-15) 76.

[11] A. R. Forsyth, {\it Theory of Differential Equations, pt. I. Exact
Equations and Pfaff's Problem}
(Cambridge Univ. Press, 1890).

[12] E. Goursat, {\it Le\c cons sur le Probl\`eme de Pfaff}
(Hermann,1922).

[13] J. A. Schouten and W. v. d. Kulk, {\it Pfaff's Problem and its
Generalizations}
(Clarendon Press, 1949).

[14] V. I. Arnold, {\it Mathematical Methods of Classical Mechanics}
(Springer-Verlag, New York, 1978) pp. 230-232.

[15] E. T. Whittaker, {\it A Treatise On the Analytical Dynamics of Particles
and Rigid Bodies}, 4th ed. (Dover, New York, 1944), pp. 275-276.

[16] P. A. M. Dirac, {\sl Can. J. Math.} {\bf 2} (1950) 129.

[17] D. Brill and S. Deser, {\sl Commun. Math. Phys.} {\bf 32} (1973) 291.

[18] A. E. Fischer and J. E. Marsden, in {\it General Relativity: An Einstein
Centenary Survey}, ed. S. W. Hawking and W. Israel (Cambridge University Press,
Cambridge, 1979).

[19] A Higuchi, {\sl Class. Quantum Grav.} {\bf 8} (1991) 2023.

[20] M. B. Green, J. H. Schwarz and E. Witten, {\it Superstring Theory}, vol. I
(Cambridge University Press, Cambridge, 1987).

[21] R. P. Woodard, {\sl Class. Quantum Grav.} {\bf 10} (1993) 483.

[22] R. P. Feynman and A. R. Hibbs, {\it Quantum Mechanics and Path Integrals}
(McGraw-Hill, New York, 1965).

[23] J. Schwinger, {\sl J. Math. Phys.} {\bf 2} (1961) 407; {\it Particles,
Sources and Fields} (Addison-Wesley, Reading, MA, 1970).

[24] R. D. Jordan, {\sl Phys. Rev.} {\bf D33} (1986) 444.

[25] L. D. Faddeev, in {\it Methods in Field Theory}, ed. R. Balian and J.
Zinn-Justin (Amsterdam: North-Holland, 1976).

[26] N. C. Tsamis and R. P. Woodard, {\sl Class. Quantum Grav.} {\bf 2} (1985)
841.

[27] N. C. Tsamis and R. P. Woodard, {\sl Phys. Rev.} {\bf D36} (1987) 3641.

[28] N. C. Tsamis and R. P. Woodard, {\sl Ann. Phys.} {\bf 215} (1992) 96.

\endreferences

\bye